\documentclass[12pt]{article}
\usepackage{yfonts}
\DeclareMathAlphabet{\scr}{U}{rsfs}{m}{n}
\usepackage{supertabular}
\usepackage{amsmath}
\usepackage{amssymb}
\usepackage{bbold}
\usepackage[sans]{dsfont}
\usepackage{graphicx}
\usepackage{epsfig}
\usepackage[dvips]{color}

\newcommand{\cleqn}{\setcounter{equation}{0}}
\newcommand{\newc}{\newcommand}
\newc{\eps}{\epsilon}
\newc{\lam}{\lambda}
\newc{\Lam}{\Lambda}
\newc{\ra}{\rightarrow}
\newc{\lra}{\leftrightarrow}
\newc{\wtilde}{\widetilde}
\newc{\ie}{{\it i.e.}}
\newc{\eg}{{\it e.g.}}
\newc{\rpv}{\not\!\! M_p}
\newc{\lsim}{\stackrel{<}{\sim}}
\newc{\beq}{\begin{equation}}
\newc{\eeq}{\end{equation}}
\newc{\beqn}{\begin{eqnarray}}
\newc{\eeqn}{\end{eqnarray}}
\newc{\PLB}{\emph{Phys.Lett.}{\bf{B}}}
\newc{\NPB}{\emph{Nucl.Phys.}{\bf{B}}}
\newc{\mcal}{\mathcal}
\newc{\bsym}{\boldsymbol}
\newc{\nonum}{\nonumber}
\newc{\ol}{\overline}
\definecolor{Red}{cmyk}{0,1,1,0}
\definecolor{luhn}{rgb}{1,0,0.5}
\definecolor{thor}{rgb}{0,0.6,1}
\begin{document}
\setlength{\baselineskip}{.6cm}
\title{\hfill ~\\[-30mm]
       \hfill\mbox{\small UFIFT-HEP-07-4}\\[6mm]
       \textbf{Baryon Triality and Neutrino Masses\\
    from an Anomalous Flavor $\bsym{U\!(1)}$ }} \date{}
\author{\\Herbi K. Dreiner$^1$,\footnote{ E-mail: {\tt
      dreiner@th.physik.uni-bonn.de}}~~ Christoph Luhn$^{1,2}$,\footnote{
    E-mail: {\tt luhn@phys.ufl.edu}}\\
  Hitoshi Murayama$^{3,4}$,\footnote{ E-mail: {\tt
      murayama@hitoshi.berkeley.edu}}~~ Marc Thormeier$^5$\footnote{
    E-mail: {\tt marc.thormeier@cea.fr}}
  \\
  \\
  \emph{$^1$\small{}Physikalisches Institut der Universit\"at Bonn,}\\
  \emph{\small Nu\ss allee 12, 53115 Bonn, Germany}\\[4mm]
 \emph{$^2$\small{}Institute for Fundamental Theory, Department of Physics,}\\
  \emph{\small University of Florida, Gainesville, FL 32611, USA}\\[4mm]
  \emph{$^3$\small{}Theoretical Physics Group,}\\\emph{\small Ernest
    Orlando Lawrence     Berkeley National Laboratory,}\\
  \emph{\small University of California, Berkeley, CA 94720, USA}\\[4mm]
  \emph{$^4$\small{}Department of Physics,}\\\emph{\small University of
    California, Berkeley, CA 94720, USA}\\[4mm]
  \emph{$^5$\small{}Service de Physique Th\'eorique, CEA-Saclay,}\\
  \emph{\small Orme des Merisiers, 91191 Gif-sur-Yvette Cedex, France}}

\maketitle

\begin{abstract}
\noindent
We construct a concise $U(1)_X$ Froggatt-Nielsen model in which baryon
triality, a discrete gauge $\mathbb{Z}_3$-symmetry, arises from $U(1)_
X$ breaking. The proton is thus stable, however, $R$-parity is
violated. With the proper choice of $U(1)_X$ charges we can obtain
neutrino masses and mixings consistent with an explanation of the
atmospheric and solar neutrino anomalies in terms of neutrino
oscillations, with no right-handed neutrinos required. The only mass
scale apart from $M_{\mathrm{grav}}$ is
$m_{\mathrm{soft}}$.

\end{abstract}





\vfill
\newpage

\section{Introduction} 
The Standard Model of particle physics (SM) has 19 free parameters,
excluding the neutrino sector. 13 of these parameters arise through
the trilinear Higgs couplings: nine masses and  three mixing angles, among 
which one finds a hierarchy, and one phase. The Froggatt-Nielsen (FN) scenario is
an elegant mechanism, which nicely explains these hierarchies 
in terms of a spontaneously broken gauge
symmetry \cite{Froggatt:1978nt}. In the supersymmetric SM\footnote{For 
a discussion of the difference between the minimal
SSM (MSSM) and the SSM we refer to Ref.~\cite{Dreiner:2005rd}.} (SSM),
the renormalizable superpotential has 48 additional unknown new complex
parameters. If we also consider 
non-renormalizable operators with mass-dimensionality five and six, there 
are correspondingly more. 
The SSM as such is inconsistent with the experimental lower bound on the
proton lifetime and requires an additional symmetry beyond the SM
gauge symmetry. From the low-energy point of view, the simplest
possibility is to impose a new discrete symmetry. However, global discrete
symmetries are inconsistent with quantum gravity \cite{Krauss:1988zc}.
It is possible to embed the discrete symmetry in a gauge symmetry,
which is spontaneously broken at a high energy scale
\cite{Krauss:1988zc,Banks:1989ag}. In this case, when requiring the
original gauge symmetry to be anomaly-free, and demanding a viable
low-energy superpotential, only three consistent discrete symmetries
remain: matter parity\footnote{Matter parity, [{\it cf.} 
Eq.~(\ref{Mptrafo})] acts on the superfields, while $R$-parity,
$R_p\equiv (-1)^{2S+3B+L}$, is defined on the components of the
superfields. Mostly this difference is irrelevant, because both
symmetries allow and forbid exactly the same operators in the
Lagrangian. So the two terms are often used synonymously.}, baryon
triality and proton hexality\footnote{Proton hexality is a discrete
gauge anomaly free $\mathbb{Z}_6$-symmetry and has recently been
introduced in Ref.~\cite{Dreiner:2005rd}. It is defined by the
transformation in Eq.~(\ref{P6trafo}).} \cite{Dreiner:2005rd,
Ibanez:1991hv,Ibanez:1991pr}. In Ref.~\cite{Dreiner:2003yr} a
realistic supersymmetric FN model was constructed which conserved
matter parity to all orders. It is the purpose of this paper to
construct a realistic FN model with a low-energy baryon-triality
symmetry to all orders.

As stated above, in the SM the FN mechanism explains the hierarchy of 
low-energy quark and charged lepton masses as well as the CKM mixings. 
But there is more to FN than just this:
\begin{itemize}

\item It was soon realized that the FN idea can be nicely combined with 
the SSM. It is then equally well applicable to the
other superpotential coupling constants which arise, like the
trilinear $R$-parity violating and/or higher dimensional operators,
see for example Ref.~\cite{Dreiner:2003hw}. 

\item Furthermore in gravity mediated supersymmetry breaking, the FN 
model can go hand in hand with the Giudice-Masiero/Kim-Nilles (GM/KN)
mechanism \cite{{Giudice:1988yz}, {Kim:1994eu}}. This allows for the
natural generation of a $\mu$-term and the other dimensionful
bilinears without having to introduce their corresponding mass scales
by hand.

\item In addition, if the family symmetry is a $U(1)_X$, the
FN scenario can plausibly be conjectured to come from an underlying
string theory. The scale of spontaneous $U(1)_X$ breaking is then
naturally just below $M_{\mathrm{grav}}=2.4\times10^{18}~\mathrm {GeV}$ 
\cite{Dine:1986zy,Dine:1987bq,Atick:1987gy,Dine:1987gj},\footnote{That
  the Green-Schwarz mechanism can soak up the anomalies requires a modified
  dilatonic K\"ahler potential,   
which induces a $U(1)_X$ Fayet-Iliopoulos term. This 
contribution to the $D$-term is then responsible
for  $U(1)_X$ to be broken, even in the case of 
having just one flavon. For more details see~\cite{Dreiner:2003yr}.} if
$U(1)_X$ is anomalous in the sense of Green and Schwarz (GS)
\cite{Green:1984sg}.
\end{itemize}
Our aim here is to build a supersymmetric FN model with the
above-mentioned characteristics, \textit{i.e.} with a hierarchical
prediction for all allowed superpotential couplings and with the
correct $U(1)_X$ breaking scale. We shall employ a 
single\footnote{There are models with more than one flavon,
see \emph{e.g.} Ref.~\cite{Jack:2003pb}. However, in the context of 
obtaining a residual discrete symmetry through the spontaneous 
breaking of a $U(1)$, several flavon fields would unnecessarily 
complicate our business. We therefore choose the simplest possibility 
of only one flavon. Furthermore, as pointed out in 
Ref.~\cite{Binetruy:1996xk},  if one works with
a vectorlike pair of flavons, the $D$-flat direction is spoiled, ``leading 
to large hierarchy among the vacuum expectation values'' (just before 
their Subsection~2.1).} 
$U(1)_X$ breaking chiral superfield~$A$, the so-called flavon. We refer to
Refs.~\cite{Dreiner:2003yr, Dreiner:2003hw} and references therein for
a review of the details on these models. In addition, we aim for some
novel features:
\begin{itemize}
\item To have proton longevity, we want as a $U(1)_X$-remnant
  the discrete gauge $\mathbb{Z}_3$-sym\-metry baryon triality
  \cite{Ibanez:1991hv,Ibanez:1991pr}, defined by the transformation in
  Eq.~(\ref{B3trafo}), to arise by virtue of the $X$-charges.
\item To have the particle content as minimal as possible, we do
  not introduce right-handed neutrinos, instead we get the
  phenomenologically viable neutrino masses and mixings from matter
  parity
violating but nevertheless baryon
  triality conserving coupling constants. Apart from minimality,
  there is a practical reason not to introduce right-handed neutrinos
  in this case: We do not want linear superpotential terms like
  $\ol{N^i}$, in order for the right-handed sneutrinos not to acquire
  a VEV to eat up this tree-level tadpole term, as this would
  constitute further flavon fields.
\end{itemize}
These two points are complementary to the model presented in
Ref.~\cite{Dreiner:2003yr}, where compactness was also the guiding
principle. 
There, the $U(1)_X$ gauge charges automatically lead to conserved
matter parity at all orders and non-vanishing neutrino masses required
the introduction of right-handed neutrinos. It was however assumed
that one of the right-handed neutrino superfields mimics the flavon
superfield.  Inspired by \emph{three} generations of SSM superfields,
we introduced \emph{three} generations of SSM singlets, two of them
being right-handed neutrinos and one constituting the flavon. Having
effectively only two right-handed neutrinos then results in a massless
lightest neutrino. The proton had a sufficiently long lifetime by
suppressing the matter parity allowed, but baryon triality forbidden,
operators $QQQL$ and $\ol{UUDE}$. However, the predicted lifetime is
not too far from the current experimental bound \cite{Harnik:2004yp,
Larson:2004ji}. In addition, the choice of two right-handed neutrinos
may seem a bit arbitrary. Without matter parity, the lepton doublets
and the down-type Higgs doublet mix, hence effectively one could
also advocate \emph{four} generations of SSM singlets.

Here, compared to \cite{Dreiner:2003yr}, we stabilize the proton
completely, and simultaneously we are even more compact concerning the
ingredients. However, as baryon triality allows for matter parity
violation, the lightest supersymmetric particle decays. Therefore it
does not provide a natural solution to the origin of dark matter. In
our model, we thus have to stick to other possible candidates such as
{\it e.g.} axions or axinos \cite{Chun:1999cq,Choi:2001cm,Chun:2006ss}.

Before going on we state the renormalizable superpotential  
of the SSM to fix our notation:
\beqn
W&=& 
G^{(E)}_{ij}\, H^DL^i \overline{E^j} \,+\, 
G^{(D)}_{ij}\, H^DQ^i \overline{D^j}\,+\, 
G^{(U)}_{ij}\, H^UQ^i \overline{U^j}\,+\,
\mu_0 \, H^DH^U\nonumber\\
&~& +\: 
\frac{1}{2}\, \lambda_{ijk}\, L^i  L^j \overline{E^k} \,+\, 
\lambda^\prime_{ijk} \, L^i Q^j \overline{D^k} \,+\, 
\frac{1}{2} \, \lambda^{\prime\prime}_{ijk} \,\overline{U^i}\overline{D^j}\overline{D^k}\,+\, 
\mu_i \, L^i H^U\,.
\label{superpot}
\eeqn
$SU(3)_C$ and $SU(2)_W$ indices are suppressed, $i,j,k$ are generation
indices, $Q^k$, $\ol{U^k}$, $\ol{D^k}$, $L^k$, $\ol{E^k} $, $H^D,H^{U}
$ denote superfields: quark doublets, $u$-type antiquark singlets, $d
$-type antiquark singlets, lepton doublets, right-handed antielectron
singlets and two Higgs doublets, respectively. $\mu_0,\mu_i$ are the
dimensionful bilinear parameters. The 
$G^{(E)}_{ij},\,G^{(D)}_{ij},\,G^{(U)}_{ij}$
and $\lam_{ijk},\,\lam^\prime_{ijk},\,\lam^{\prime\prime}_{ijk}$ are 
coupling constants. Denoting the
scalar component of a superfield by a tilde, the soft supersymmetry
breaking Lagrangian density is (see, {\it e.g.},
Ref.~\cite{Grossman:1998py})
\beqn
- \mathcal{L}_{\mathrm{soft}} &=&   
[\boldsymbol{M_{\tilde{Q}}^2}]_{ij}\widetilde{Q^i}^\dagger\widetilde{Q^j} \,+\,
[\boldsymbol{M_{\tilde{U}}^2}]_{ij}\widetilde{U^i}\widetilde{U^j}^\ast \,+\,
[\boldsymbol{M_{\tilde{D}}^2}]_{ij}\widetilde{D^i}\widetilde{D^j}^\ast \,+\,
[\boldsymbol{M_{\tilde{E}}^2}]_{ij}\widetilde{E^i}\widetilde{E^j}^\ast
\notag \\ 
&& +\,[\boldsymbol{M_{\tilde{L}}^2}]_{\alpha\beta}
\widetilde{L^\alpha}^\dagger\widetilde{L^\beta} \,+\,
{M_{{H^U}}^2}\widetilde{H^U}^\dagger\widetilde{H^U} \,+\, \left( b_\alpha \,
  \widetilde{L^\alpha} \widetilde{H^U} \,+\, \mathrm{h.c.}\right) \notag \\ 
&& +\, \left(\frac{1}{2} \,[\boldsymbol{A_E}]_{\alpha \beta k} \,
\lambda_{\alpha 
\beta k}\,\widetilde{L^\alpha} \widetilde{L^\beta} \widetilde{E^k}^\ast \,+\,
[\boldsymbol{A_D}]_{\alpha jk} \, \lambda^\prime_{\alpha jk} \,
\widetilde{L^\alpha} \widetilde{Q^j} \widetilde{D^k}^\ast \right)
\,+\,\mathrm{h.c.}\notag\\ 
&& +\, \left( [\boldsymbol{A_U}]_{ij} \, G^{(U)}_{ij} \, \widetilde{H^U}
\widetilde{Q^i} \widetilde{U^j}^\ast \,+\, 
[\boldsymbol{A_{UDD}}]_{ijk} \, \lambda^{\prime\prime}_{ijk} \,
\widetilde{U^i}^\ast \widetilde{D^j}^\ast \widetilde{D^k}^\ast \right)
\,+\,\mathrm{h.c.}\notag \\
&& +\, \frac{1}{2} \, \left( M_1 \tilde{B}\tilde{B} \,+\, M_2 \tilde{W}^a
  \tilde{W}^a \,+\, M_3 \tilde{G}^b \tilde{G}^b \right) \,+\, \mathrm{h.c.}
\label{softEl}
\eeqn
The supersymmetry breaking bino, wino, and gluino mass terms are given
in the last line, with $a=1,2,3$ and $b=1,...,8$.  Note that we have
applied the compact notation $L^\alpha = (L^0 \equiv H^D , \,L^i)$,
$\alpha=0,1,2,3$, since the lepton doublets $L^i$ and the down-type
Higgs doublet $H^D$ have identical quantum numbers.

In supersymmetric theories, the presence of 
$\mathcal{L}_{\mathrm{soft}}$ is a potential
source for flavor changing neutral currents (FCNCs). 
Even worse, there is one unresolved drawback to mention which is generic
to FN models employing the GM/KN mechanism
(for details see Ref.~\cite{Dreiner:2003yr}), inducing non-universal 
and $U(1)_X$-charge dependent contributions to the sparticle soft squared
masses. This potentially causes problems with low-energy FCNCs, and 
is common to all FN models.  In order to
suppress these FCNCs, it is necessary to theoretically provide for a
specific structure of the soft supersymmetry breaking terms (\emph{e.g.} 
an approximate alignment of the quark and squark mass matrices, \emph{cf}. 
Refs.~\cite{Nir:1993mx,Dudas:1995eq}).
It is beyond the
scope of this work to detail the subtleties of such issues. We simply
expect (or hope?) that the problem of low-energy FCNCs will be solved
together with an as yet non-existing proper model of supersymmetry
breaking.

Our paper is structured as follows: In Section~\ref{B3} we review the
importance of discrete gauge symmetries; then we show how to obtain
matter parity, baryon triality and proton hexality from a
Froggatt-Nielsen $U(1)_X$. The next three sections are the core of
this paper.  Section~\ref{bases} discusses our choice of basis.
Section~\ref{pheno} combines the requirements of anomaly cancellation,
the phenomenological constraints on the $X$-charges of quarks and
charged leptons and the conditions of achieving baryon
triality. Sect.~\ref{NeutrinoSector} then deals with the neutrino
sector in detail. In Sect.~\ref{viableSets} we finally arrive at six
distinct sets of $X$-charges, so that the quintessential result is
given by Table~\ref{allsets}.  Sect.~\ref{summarY} concludes. The
Apps.~\ref{bpb3-app}, \ref{2by2}+\ref{vevcond}, \ref{symconsid},
\ref{diagonalizeM}, \ref{top-down} complement Sect.~\ref{B3}
(comparing baryon \emph{parity} and baryon \emph{triality}),
Sect.~\ref{bases} (remarks on supersymmetric zeros + eliminating
sneutrino VEVs), Sect.~\ref{nuMatrix} (analyzing the symmetries in the
$LL\ol{E}$ loop contribution to the neutrino mass matrix),
Sect.~\ref{mixingconstraints} (diagonalization of the neutrino mass
matrix) and Sect.~\ref{viableSets} (an explicit example of how a
set of $X$-charges produces low-energy physics), respectively.





\cleqn

\section{\label{B3}Low-Energy Discrete Symmetries 
from Froggatt-Nielsen  Charges}
\subsection{Discrete Gauge Symmetries}
The SM Lagrangian is invariant under Poincar\'e transformations as
well as $SU(3)_C\times SU(2)_W\times U(1)_Y$ local gauge
transformations.  The most general supersymmetric SM Lagrangian with
one additional Higgs doublet leads to unobserved exotic processes, in
particular rapid proton decay, inconsistent with the experimental
bounds \cite{Dreiner:1997uz}. In the low-energy effective Lagrangian,
this problem is resolved by introducing a global discrete
multiplicative symmetry, which prohibits a subset of the
superpotential interactions. Prominent examples are matter parity,
$M_p$, (or equivalently $R$-parity), baryon triality, $B_3$, as well as
the recently introduced proton hexality, $P_6$ \cite{Dreiner:2005rd}.
There is however a problem when embedding a global discrete symmetry
in a unified theory at the gravitational scale, such as string theory:
it is typically violated by quantum gravity effects (worm holes)
\cite{Krauss:1988zc}. This is avoided if the discrete symmetry of the
low-energy effective Lagrangian is the remnant of a local gauge
symmetry, $G$, which is spontaneously broken, in our case at or near
the {gravitational} scale. Since $G$ is unaffected by quantum gravity
effects, after the spontaneous breakdown of $G$ also the residual
discrete symmetry remains intact \cite{Krauss:1988zc,Banks:1989ag,
  Banks:1991xj}. Such a discrete symmetry is denoted a ``discrete
\emph{gauge} symmetry'' (DGS) \cite{Wegner:1984qt}. In the following,
we only treat Abelian DGSs, originating in an Abelian local gauge
group $G\equiv U(1)_X$.  For examples see
\cite{Dreiner:2005rd,Martin:1992mq}.

In order to obtain a consistent quantum field theory, we demand that
the underlying local gauge theory $G$ is anomaly-free. In general, we
include the possibility that the anomalies of the original gauge
symmetry are canceled by the Green-Schwarz mechanism (GS)
\cite{Green:1984sg}. Thus either 
\begin{itemize}
\item[\emph{1)}] the low-energy DGS is a remnant of an anomaly-free
  local gauge symmetry, in which case the DGS is anomaly-free in the
  sense of Ib\'a\~nez and Ross \cite{Ibanez:1991hv}, or
\item[\emph{2)}] the DGS is a remnant of a local gauge symmetry whose
  anomalies are canceled by the GS mechanism. In this case the DGS
  can be either
\begin{itemize}
\item[\emph{a)}] anomaly-free in the sense of Ib\'a\~nez and Ross or
\item[\emph{b)}] GS-anomalous, {\it i.e.} the DGS anomalies are
  canceled via a discrete version of the GS mechanism
  \cite{Ibanez:1992ji}.\footnote{We emphasize that not every discrete symmetry
  is an anomaly-free DGS, \emph{e.g.} baryon parity, $B_p$, see also
  App.~\ref{bpb3-app}.}
\end{itemize}\end{itemize}
The model we construct in this paper belongs to class \emph{2a)}, {\it
  i.e.} the $U(1)_X$ gauge anomalies are canceled by the GS
mechanism; however, the low-energy DGS satisfies the anomaly
cancellation conditions of Ib\'a\~nez and Ross, without the GS
mechanism.

In Ref.~\cite{Ibanez:1991pr}, the family-independent $\mathbb{Z}_2$
and $\mathbb{Z}_3$ DGSs were determined, which are anomaly-free
without invoking the existence of extra light particles besides those
of the SSM and, in addition, which leave the MSSM superpotential (and
possibly more) invariant. This investigation was generalized to
$\mathbb{Z}_N$ DGSs with arbitrary $N$ in Ref.~\cite{Dreiner:2005rd}.
Taking into account the need for neutrino masses, only three
family-independent DGSs survive: matter parity [Eq.~(\ref{Mptrafo})],
baryon triality [Eq.~(\ref{B3trafo})] and proton hexality
[Eq.~(\ref{P6trafo})], which we discuss in more detail below.

In Ref.~\cite{Dreiner:2003yr}, several of the authors constructed a
phenomenologically viable FN model with a low-energy matter parity
DGS. In the following subsection, we wish to first derive the
necessary and sufficient conditions on the MSSM $X$-charges for baryon
triality and proton hexality to arise as a DGS from a family-dependent
local gauge symmetry. Afterwards, in Sect.~\ref{NeutrinoSector} we
construct phenomenologically viable FN models with a low-energy {\it
  baryon triality} DGS. We postpone the construction of such a model
with a low-energy proton hexality to Ref.~\cite{Dreiner:2007vp}.

\subsection{Baryon Triality Arising from $\bsym{U(1)_X}$}
Consider a general product of MSSM left-chiral superfields
$\Phi^a\;\in\; \{Q^k,\ol{U^k},\ol{D^k}, L^k,\ol{E^k},$ $H^D,H^{U} \}$ and
their charge conjugates $\ol{\Phi^a}$,
\beq \label{R}
R\equiv\prod_{a,b}{(\Phi^a)}^{\alpha_a}\,(\ol{\Phi^b})^
{\ol{\alpha_b}}\,.
\eeq 
In general, such an operator can appear in the K\"ahler potential or,
if the $\ol{\alpha_b}$ vanish, in the superpotential. Imposing a
discrete symmetry forbids some of these SM-invariant operators. We now
wish to obtain a specific low-energy discrete symmetry by an
appropriate $U(1)_X $ gauge charge assignment. We fix the gauge charge
normalization such that the flavon superfield $A$ has $U(1)_X$ charge
$X_A=-1$. It is then obvious that only those operators with an
\textit{integer} overall $X$-charge, $X_{\mathrm{total}}$, are allowed
after the breaking of $U(1)_X$. We obtain further constraints on $X_{
  \mathrm{total}}$ by requiring $SU(3)_C\times SU(2)_W\times U(1)_Y$
gauge invariance of the given operator, as well as by demanding that
the renormalizable MSSM superpotential operators are necessarily
allowed. We thus have the conditions on $X_{\mathrm{total}}$ for an
operator to be allowed or forbidden. We then make the connection with
the corresponding discrete symmetry, originating from the MSSM
$X$-charges, stating the necessary and sufficient conditions thereof.

We shall denote the ``combined'' multiplicity of each superfield in a
given operator by $n_{\Phi^a}\equiv \alpha_a-\ol{\alpha_a}$.  Thus for
example the term $Q^1\ol{Q^2}\ol{U^1}\ol{D^1}\ol{D^2}$ has $n_{Q^1}=1,
n_{Q^2}=-1$, $n_{\ol{U^1}}=n_{\ol{D^1}}=n_{\ol{D^2}}=1$.  The total
$X$-charge of a general product, $R$, of superfields
$\Phi^a,\,\ol{\Phi^b}$ can then be expressed as
\begin{eqnarray}\label{xc}
X_{\mathrm{total}}&=& n_{H^D}~
X_{H^D}+n_{H^{U}}~
X_{H^{U}} ~+~\sum_in_{Q^i}~X_{Q^i}~+~\sum_in_{\ol{D^i}}~ 
X_{\ol{D^i}}\nonumber\\
&~&~+~\sum_in_{\ol{U^i}}~ X_{\ol{U^i}}~+~\sum_i~ n_{L^i}~
X_{L^i}~+~\sum_in_{\ol{E^i}}~X_{\ol{E^i}} .~~~~~~~~
\end{eqnarray}
The coefficients $n_{\ldots}$ and charges $X_{\ldots}$ above are not all
mutually independent:
\begin{itemize}
\item Since each product $R$ should be $SU(3)_C\times SU(2)_W \times
  U(1)_Y$ gauge invariant, the $n_{\ldots}$ are subject to the
  conditions (for the first equation see for example Chapter 10 in
  Ref.~\cite{Georgi:1982jb})
\begin{eqnarray}
\sum_i n_{Q^i}-\sum_in_{\ol{D^i}}-\sum_in_{\ol{U^i}}&=&
3\mathcal{C}, \label{n-eq1}\\
n_{H^D}+n_{H^{U}}+\sum_in_{Q^i}+\sum_in_{L^i}&=&2\mathcal{W},
\label{n-eq2}\\
Y_{H^D}~n_{H^D}+~Y_{H^{U}}~
n_{H^{U}}~+~Y_Q~\sum_in_{Q^i}~+~Y_{\ol{D}}~
\sum_in_{\ol{D^i}}& &\nonumber\\
+~Y_{\ol{U}}~\sum_in_{\ol{U^i}}~+~Y_L~\sum_i n_{L^i}~+~
Y_{\ol{E}}~\sum_in_{\ol{E^i}}&=&0.\label{n-eq3}
\end{eqnarray}
Here $\mathcal{C}$ is an integer, $\mathcal{W}$ is an integer which is
non-negative for terms in the superpotential. $Y_{...}$ denotes the
hypercharge of the corresponding field. For the MSSM fields we have:
$Y_{H^D}=-3Y_Q$, $Y_{H^{U}}=3Y_Q$; $Y_{L}=-3Y_Q$,
$Y_{\ol{E}}=6Y_Q$; $Y_{\ol{U}}=-4Y_Q$, $Y_{\ol {D}}=2Y_Q$.  Solving
Eqs.~(\ref{n-eq1})-(\ref{n-eq3}) for $n_{Q^1},n_{\ol{D^1}},$
and $n_{\ol{E^1}}$ we obtain
\beqn  
n_{Q^1} &=& \phantom{-3 \mathcal{C} +}\,\;2 \mathcal{W}-(n_{H^{D}} + 
n_{H^{U}})- (n_{Q^2} + n_{Q^3})- \sum_{i}n_{L^i} \,,
\label{above-1} \\
n_{\ol{D^1}} &=& -3 \mathcal{C}+ 2\mathcal{W}-(n_{H^D}+ 
n_{H^{U}}) -(n_{\ol{D^2}} + n_{\ol{D^3}})- \sum_{i}n_{L^i} 
-\sum_{i}n_{\ol{U^i}}\,, \label{above-2} \\
n_{\ol{E^1}} &=&\phantom{-3} \mathcal{C}-\phantom{2}\mathcal{W}+
\phantom{(}n_{H^D} \phantom{+ n_{H^{U}})}\;\,-\,
(n_{\ol{E^2}} + n_{\ol{E^3}} ) + \sum_{i}n_{L^i} +
\sum_{i}n_{\ol{U^i}}.\label{above-3} 
\eeqn
\item Since we assume that after the breaking of $U(1)_X$ all
  renormalizable MSSM superpotential operators are allowed, the
  corresponding gauge invariant products $R$ must have non-fractional
  powers of the flavon superfield $A$, \ie\ we require
\begin{itemize}
\item[1.] {The renormalizable} superpotential terms $H^{U} Q^i \ol{U^j}
$, $H^D Q^i \ol{D^j}$, $H^D L^i \ol{E^j}$, and $H^DH^{U}$ have an overall 
integer $X$-charge.
\end{itemize} 
This corresponds to the conditions
\beqn 
X_{H^{U}}+X_{Q^1}+X_{\ol{U^1}}&=&\mbox{integer},\nonumber\\
X_{H^D}+X_{Q^1}+X_{\ol{D^1}}&=&\mbox{integer},\nonumber\\
X_{H^D}+X_{L^1}+X_{\ol{E^1}}&=&\mbox{integer},\nonumber
\eeqn
\beqn
X_{Q^{2,3}}-X_{Q^1}&=&\mbox{integer},\nonumber\\
X_{L^{2,3}}-X_{L^1}&=&\mbox{integer},\nonumber\\
X_{\ol{D^{2,3}}}-X_{\ol{D^{1}}}&=&\mbox{integer},\nonumber\\
X_{\ol{U^{2,3}}}-X_{\ol{U^{1}}}&=&\mbox{integer},\nonumber\\
X_{\ol{E^{2,3}}}-X_{\ol{E^{1}}}&=&\mbox{integer},\nonumber\\[2mm]
X_{H^D}+X_{H^{U}}&=&\mbox{integer.}~~~~\label{cond0}
\eeqn 
We leave it open at the moment which other gauge invariant terms shall
also have an overall integer $X$-charge. 
\end{itemize}
With the help of Eq.~(\ref{cond0}), we can express all $X$-charges in
terms of $X_{L^1}$, $X_{Q^1}$, $X_{H^D}$, and unknown
integers. Inserting this and Eqs.~(\ref{above-1})-(\ref{above-3}) in
Eq.~(\ref{xc}) we get for the total $X$-charge
\begin{eqnarray}
X_{\mathrm{total}}&=&
\mathcal{C}\cdot\big[3X_{Q^1}+X_{L^1}+2(X_{H^D}-X_{L^1})
\big]~\nonumber\\
&~&~+~\Big(n_{H^D}~-~\mathcal{W}~+~\sum_in_{\ol{U^i}}  
\Big)\cdot(X_{H^D}-X_{L^1})~+~
\mbox{integer}.~~~~~\label{x=i+f}
\end{eqnarray}
If we now require \emph{no} remnant DGS at low energy whatsoever, {\it
  i.e.}  if {\it all} renormalizable and non-renormalizable terms
which are $SU(3)_C\times SU(2)_W\times U(1) _Y$ gauge invariant are
allowed, then we must have an overall integer $X$-charge, and thus 
$X_{H^D}-X_{L^1}$ and $3X_{Q^1}+X_ {L^1}$ must be integers. However,
we wish to determine the necessary constraints on the gauged
$X$-charges in order to obtain a remnant discrete matter parity,
baryon triality, or proton hexality DGS arising from the $U(1)_X$. Our
treatment here does not rely on the absence or cancellation of
anomalies and is thus equally applicable to $\mathbb{Z}_{N}
$-symmetries other than $M_p$, $B_3$ and $P_6$, {\it e.g.}  the
$\mathbb{Z}_{2} $-symmetry baryon parity, $B_p$. (For the definition
of baryon parity and an investigation of its phenomenological
difference with respect to baryon triality see App.~\ref{bpb3-app}.)

Under the respective DGSs, the MSSM left-chiral superfields transform
as follows

$\bullet$ Matter parity\footnote{An alternative but at low-energies
  physically equivalent definition of matter parity is
\beqn
\begin{array}{rccl}
\Big\{Q^i,L^i\Big\} &\longrightarrow&
\phantom{e^{2\pi \mathrm{i}/2}~}&
\Big\{Q^i,L^i\Big\},\\[2mm]
\Big\{\ol{U^i},\ol{D^i},
\ol{E^i},H^D, H^{U} \Big\} &\longrightarrow&
e^{2\pi \mathrm{i}/2}& \Big\{\ol{U^i},\ol{D^i},\ol{E^i},H^D, 
H^{U}\Big\}.\nonumber
\end{array}
\eeqn 
See Ref.~\cite{Dreiner:2005rd} for details. With this definition
it is easy to see that proton hexality is the direct product of matter
parity and baryon triality.} ($M_p$)
\beqn
\begin{array}{rccl}
\Big\{H^D, H^{U}\Big\} &\longrightarrow&
\phantom{e^{2\pi \mathrm{i}/2}~}&
\Big\{H^D, H^{U}\Big\},\\[2mm]
\Big\{Q^i,\ol{U^i},\ol{D^i},L^i,
\ol{E^i} \Big\} &\longrightarrow&
e^{2\pi \mathrm{i}/2}& \Big\{Q^i,\ol{U^i},\ol{D^i},L^i,\ol{E^i}\Big\},
\end{array} \label{Mptrafo}
\eeqn 
\begin{itemize}
\item Baryon triality ($B_3$)
\beqn
Q^i\phantom{\Big\}}&
\longrightarrow&\phantom{e^{4\pi \mathrm{i}/3}}~
\phantom{\Big\{}Q^i,
\nonumber\\
\Big\{H^{U},\ol{D^i} \Big\}
&\longrightarrow&e^{2\pi \mathrm{i}/3}~\Big\{H^{U},\ol{D^i}\Big\},  
\nonumber \\
\Big\{H^D, \ol{U^i}, L^i,\ol{E^i} \Big\}
&\longrightarrow&
e^{4\pi \mathrm{i}/3}~\Big\{ H^D, \ol{U^i}, L^i,
\ol{E^i} \Big\}, \label{B3trafo}
\eeqn
\item Proton hexality ($P_6$), \textit{cf.} Ref.~\cite{Dreiner:2005rd},
\beqn
Q^i\phantom{\Big\}}&
\longrightarrow&\phantom{e^{4\pi \mathrm{i}/6}}~\;\:\!\phantom{\Big\{}Q^i,
\nonumber\\
\Big\{H^D,\ol{U^i},\ol{E^i} \Big\}
&\longrightarrow&e^{2\pi \mathrm{i}/6}~\;\:\!\Big\{H^D,\ol{U^i},
\ol{E^i} \Big\},\nonumber \\
\phantom{\Big\{} L^i \phantom{\Big\}}
&\longrightarrow&e^{8\pi \mathrm{i}/6}~\;\:\!\phantom{\Big\{} L^i\,,
\nonumber \\
\Big\{H^{U}, \ol{D^i}  \Big\}
&\longrightarrow&e^{10\pi \mathrm{i}/6}~\Big\{ H^{U}, 
\ol{D^i} \Big\}. \label{P6trafo}
\eeqn
\end{itemize}
(None of these three symmetries has a domain wall problem, since the
discrete charges of the two Higgs superfields are opposite to each
other, for details see Ref.~\cite{Dreiner:2005rd}.) In other words,
under $M_p$, $B_3$, and $P_6$ transformations a general product of MSSM
superfields is respectively multiplied by
\beqn 
&&\bullet~~\left(
  e^{2\pi\mathrm{i}/2}\right)^{\sum_in_{Q^i}~+~\sum_in_{\ol{U^i}}~+~
  \sum_in_{\ol{D^i}}~+~\sum_in_{L^i}~+~\sum_in_{\ol{E^i}}},
\\\nonumber\\
&&\bullet~~\left( e^{2\pi\mathrm{i}/3}
\right)^{n_{H^{U}}~+~\sum_in_{\ol{D^i}} ~+~2~n_{H^D}
~+~2\sum_in_{\ol{U^i}}~+~2\sum_in_{L^i}~+~2\sum_in_{\ol{E^i}}},\\
\nonumber\\
&&\bullet~~   \left( e^{2\pi \mathrm{i}/6} \right)^{n_{H^D} 
~+~ \sum_in_{\ol{U^i}} 
~+~ \sum_in_{\ol{E^i}} 
~+~ 4 \sum_in_{L^i} 
~+~ 5~   n_{H^{U}}  
~+~5  \sum_in_{\ol{D^i}}}\,.
\eeqn
Thus in turn we may write for [$M_p/B_3/P_6$]
\beqn\label{rrpp}
&&\hspace{-0.8cm} \begin{tabular}{lrcl}
$\bullet$&$\!\!\sum_i n_{Q^i}\;+\;\sum_in_{\ol{U^i}}\;+\;\sum_i n_{\ol{D^i}}
\;+\;\sum_i n_{L^i} \;+\;\sum_i n_{\ol{E^i}}$
&$\!=\!$&
$2\mathcal{I}_M\:+\:\iota_M,$ \\ \\
$\bullet$& $\!\!n_{H^{U}}\;+\;\sum_in_{\ol{D^i}}\;+\;2\;n_{H^D}
\;+\;2\sum_in_{\ol{U^i}}\;+\;2\sum_in_{L^i}\;+\;2
\sum_in_{\ol{E^i}}$
&$\!=\!$&
$3\mathcal{I}_B\;+\;\iota_B,$ \\ \\
$\bullet$& $\!\!n_{H^D} \;+\; \sum_in_{\ol{U^i}} \;+\; \sum_in_{\ol{E^i}} 
\;+\; 4\; \sum_in_{L^i} \;+\; 5\;   n_{H^{U}} \;+\;5\;  
\sum_in_{\ol{D^i}}$ 
&$\!=\!$&
$6\mathcal{I}_P \;+\;\iota_P \, .$
\end{tabular} \nonumber \\ 
\eeqn
$\mathcal{I}_M$, $\mathcal{I}_B$, and $\mathcal{I}_P$ are integers;
$\iota_M$ is $0$ or $1$ if matter parity is conserved or broken,
$\iota_B$ is $0$ or $1,2$ if baryon triality is conserved or broken,
and $\iota_P$ is $0$ or $1,\ldots,5$ if proton hexality is conserved or
broken.  With Eqs.~(\ref{above-1})-(\ref{above-3}) we get from
Eq.~(\ref{rrpp}) that
\beqn\label{alfa1}
M_p:~~n_{H^D} &=& 3\mathcal{W}-\iota_M-2(\mathcal{C} 
+\mathcal{I}_M  + n_{H^{U}}) +\sum_{i} n_{\ol{U^i}},\\
B_3:~~~~~~\;\!\mathcal{C} &=& 3\Big(-\mathcal{I}_B + n_{H^D}  +  
\sum_i n_{L^i}+\sum_{i} n_{\ol{U^i}}\Big)  - \iota_B,  \\ 
\label{alfa3}
P_6:~~\,n_{H^D} &=&  3\mathcal{W} - \sum_i n_{\ol{U^i}} 
- \frac{14\, \mathcal{C} + 6\mathcal{I}_P+\iota_P}{3} ,
\eeqn
respectively. We  now require  
\begin{itemize}
\item[$1^\prime\!.$] \textit{All} $SU(3)_C\times SU(2)_W \times U(1)_Y$
  gauge invariant terms which conserve the discrete symmetry
  $[M_p$/$B_3$/$P_6]$ each have an overall integer $X$-charge. This
  requirement is a generalization of Point~1. above Eq.~(\ref{cond0}).
 
\item[2.]  All $SU(3)_C\times SU(2)_W \times U(1)_Y$ gauge invariant
  terms which do not conserve the discrete symmetry $[M_p/B_3/P_6]$
  each have an overall fractional $X$-charge. It follows that all
  superfield operators which violate $[M_p/B_3/P_6]$ are forbidden
  even after the spontaneous breaking of $U(1)_X$. $[M_p/B_3/P_6]$
  is thus conserved exactly, \ie\ to all orders.
\end{itemize}
For any $SU(3)_C\times SU(2)_W\times U(1)_Y$ invariant operator $R$
[\textit{cf.} Eq.~(\ref{R})], which violates $[M_p$/$B_3$/$P_6]$ one
has respectively that $[R^2$/$R^3$/$R^{6\,}]$ conserves $[M_p$/$B_3
$/$P_6]$. From Point~$1^\prime\!.$ above, we find that the $X$-charge
of each of the latter operators, namely $[2\cdot X_{\mathrm{total}}(R)
$/$3\cdot X_{\mathrm{total}}(R)$/$6\cdot X_{ \mathrm{total}}(R)]$
respectively, is integer.  Point~2.  demands that $X_{\mathrm{total}}
(R)$ is fractional.  \emph{It follows that all superfield operators
  which violate $[M_p/B_ 3/P_6]$ have an $X_{\mathrm{total}}$ of the
  form $~[\frac{1}{2}\,+$\,integer$\big/$$\frac{1~\mathrm{or}~2}{3}
\,+$\,integer$\big/$$\frac{ 1,2,3,4~\mathrm{or}~5}{6}\,+$\,integer}].
Bearing this in mind, we plug Eqs.~(\ref{alfa1}) - (\ref{alfa3}) into
Eq.~(\ref{x=i+f}) to eliminate $[n_{H^D}/\mathcal{C}/n_{H^D}]$,
respectively.
\begin{itemize}
\item We first treat $M_p$. In this case
\beqn\label{mattparit}
X_{\mathrm{total}}&=&\mathcal{C}\cdot(3X_{Q^1} +  X_{L^1})\\&~&
+~\Big[2 \Big( \mathcal{W}   
-\mathcal{I}_M - n_{H^{U}}+\sum_{i} n_{\ol{U^i}} \Big) - 
\iota_M\Big]
\cdot (X_{H^D}-X_{L^1})~+~\mbox{integer}.\nonumber
\eeqn
Now consider an operator $R$ forbidden by $M_p$, {\it i.e.} with
$\iota_M=1$. $X_{\mathrm{total}}$ must then be $\frac{1}{2}$+integer.
Choosing a forbidden operator where $\mathcal{C}=0$, and $\mathcal{W}=
\mathcal{I}_M+n_{H^{U}}-\sum _{i} n_{\ol{U^i}}$, we obtain the
condition
\beqn
\label{cond1} X_{H^D}~-~X_{L^1}~\stackrel{!}{=}\,
  - \frac{1}{2}~+~\mbox{integer}.  
\eeqn
We insert this into the expression for $X_{\mathrm{total}}$:
\beq
X_{\mathrm{total}}~=~\mathcal{C}\cdot(3X_{Q^1}+X_{L^1}
)+\frac{\iota_M}{2}+~\mbox{integer}\,.
\eeq

\ \ \ Now for the terms which are allowed by $M_p$, \emph{i.e.} which
have $\iota_M=0$ (and thus $X_{\mathrm{total}}$ is integer). Here we
get another condition on the $X$-charges of the MSSM superfields when
choosing an operator for which $\mathcal{C}=1$,
\beq\label{cond2}
3X_{{Q^1}}~+~X_{L^1}~\stackrel{!}{=}~\mbox{integer}\,.
\eeq
To check consistency, we plug Eqs.~(\ref{cond1}) and (\ref{cond2}) 
into Eq.~(\ref{mattparit}); we thus find that
\beq
X_{\mathrm{total}}~=~\frac{\iota_M}{2}~+~\mbox{integer}\,.
\eeq
In Refs.~\cite{{Dreiner:2003yr},{Harnik:2004yp},{Larson:2004ji}} the 
implications of Eqs.~(\ref{cond0}), (\ref{cond1}), and (\ref{cond2})
in combination  with a viable phenomenology  were studied 
in detail.

\item Next we treat $B_3$. We get
\begin{eqnarray}
X_{\mathrm{total}} \!\!&  \!\!=  \!\!& \!\!
\Big[3\Big( \! n_{H^D}-\mathcal{I}_B   +  
 \!\sum_i n_{L^i}+ \!\sum_{i} n_{\ol{U^i}}\Big)  - 
\iota_B\Big] \! \!\cdot  \!\!\Big[3X_{Q^1}+X_{L^1}+2(
X_{H^D}-X_{L^1} \!)\Big] \nonumber\\
&~& +~\Big( \! n_{H^D}~-~\mathcal{W}~+~ \!\sum_in_{\ol{U^i}}  
\Big) \!\cdot (X_{H^D}-X_{L^1} )~+~
\mbox{integer}.\label{xtb3}
\end{eqnarray}
Considering an allowed operator, \ie\ with $\iota_B=0$ (thus
$X_{\mathrm{total}}$ is integer) and for which also $\mathcal{I}_B=n_
{H^D} + \sum_i n_{L^i}+\sum_i n_{\ol{U^i}}$ we arrive
at
\beq
X_{\mathrm{total}}~=~\big(n_{H^D}-\mathcal{W}+ \sum_i
n_{\ol{U^i}}
\big)\cdot(X_{H^D}-X_{L^1})~+~\mbox{integer}\,.
\eeq
If we furthermore require that $\mathcal
{W}=n_{H^D}+ \sum_i n_{ \ol{U^i}} +1$ we obtain the condition
\beqn\label{cond11}
X_{H^D}~-~X_{L^1}~\stackrel{!}{=}~\mbox{integer} 
\eeqn
[to be compared with Eq.~(\ref{cond1})].  We insert this into the
expression for $X_{\mathrm{total}}$, getting 
\beqn
X_{\mathrm{total}}= \Big[3\big(n_{H^D} - \mathcal{I}_B +
\sum_i n_{L^i}+ \sum_i n_{\ol{U^i}} \big) - \iota_B\Big]
\cdot(3X_{Q^1}+X_{L^1})~+~\mbox{integer}.~~~~
\eeqn 
\ \ \ Now considering a forbidden operator for which $\mathcal{I}_B=n_
{H^D} +\sum_i n_{L^i} + \sum_i n_{\ol{U^i}}$, we arrive at
\beqn
X_{\mathrm{total}}~=~-~\iota_B\cdot(3X_{Q^1}+X_{L^1})~+~\mbox{integer}\,.
\eeqn 
Setting $\iota_B=1$ (thus $X_{\mathrm{total}}$ being
$\frac{1~ \mathrm{or}~2}{3}$+integer) we get
\beq\label{cond22}
3X_{Q^1}+X_{L^1}~\stackrel{!}{=}\,-\frac{b}{3}+~\mbox{integer}\,,
\eeq
with $b\in\{1,2\}$ [to be compared with Eq.~(\ref{cond2})]. This is
compatible with $\iota_B=2$ also requiring $X_{\mathrm{total}}$ not to
be an integer.  To check consistency, we plug Eqs.~(\ref{cond11}) and
(\ref{cond22}) into Eq.~(\ref{xtb3}); this gives
\beq
X_{\mathrm{total}}~=\,\frac{b\cdot\iota_B}{3}~+~\mbox{integer}\,.
\eeq


\item For $P_6$ we find that
\beqn
X_{\mathrm{total}}&=&\mathcal{C}\cdot(3X_{Q^1} +  X_{L^1}) \label{xtp6} 
\nonumber \\
&~&+~\left[2 (\mathcal{W} -  \mathcal{C} - \mathcal{I}_P)
- \frac{ 2  \mathcal{C} + 
\iota_P}{3} \right]
\! \cdot \! (X_{H^D}-X_{L^1})\:+\:\mbox{integer}\,.
\eeqn
Before continuing it is important to point out that $\frac{2\mathcal
{C} + \,\iota_P}{3}$ has to be an integer due to Eq.~(\ref{alfa3}).
Hence, when deriving the conditions on the $X$-charges for $P_6
$~conservation, we are restricted to consider only those operators
for which $2\mathcal{C} + \iota_P $ is a multiple of three. Thus we
have that $\iota_P=0,3\big/1,4\big/2,5$ requires $\mathcal{C}=0\big/1
\big/2~\mathrm{mod}~3$, respectively. Defining $\mathcal{J}\equiv\frac
{2\mathcal{C}+\,\iota_P}{3} \in \mathbb{Z}$, we have
\beqn
2\mathcal{C} &=& - \iota_P + 3 \mathcal{J}.\label{vergesseneBED}
\eeqn
Returning to Eq.~(\ref{xtp6}), 
consider $\iota_P=3$ (already the square of such an operator is
$P_6$ invariant, therefore we have $X_{\mathrm{total}}=\frac{1}{2}+\mathrm
{integer}$ in this case) and $\mathcal{C}=\mathcal{W}=\mathcal{I}_P=0
$; this leads to the condition\footnote{This condition has already been 
stated in Ref.~\cite{Dreiner:2005rd}, below Eq.~(6.9). 
But in its PRD-version we unfortunately 
made a typo by including a wrong factor of "3",  
which we have however corrected in the newest arXiv-version.}
\beqn
X_{H^D} - X_{L^1} &\stackrel{!}{=}& - \,\frac{1}{2}+\mathrm{integer}\,,
\label{Pcond11} 
\eeqn
[to be compared with Eq.~(\ref{cond1})]. Inserting this into
$X_{\mathrm{total}}$ of Eq.~(\ref{xtp6}) we get
\beqn
X_{\mathrm{total}}&=&\mathcal{C}\cdot(3X_{Q^1} + X_{L^1}) ~+~
\frac{2 \, \mathcal{C} + \iota_P}{6} ~+~\mbox{integer}\,.  
\eeqn 
Next consider $\iota_P=\mathcal{C}=1$. For $\iota_P=1$ we need $X_{
\mathrm{total}}=\frac{p}{6} + \mathrm{integer}$, with $p=1,5$. $p=2,3
,4$ are not allowed as these have common prime factors with 6. This
would lead to a term in the Lagrangian whose square or cube is $P_6$
invariant contrary to the assumption that $\iota_P$ is 1. This way the
following condition is obtained
\beqn 
3X_{Q^1} + X_{L^1} &
\stackrel{!}{=}& -\,\frac{3-p}{6}~+~\mbox{integer}\notag \\ 
&\equiv& -\,\frac{\widetilde{p}}{3}~+~\mathrm{integer},   \label{Pcond22} 
\eeqn 
with $\widetilde{p}=\pm 1$ [to be compared with Eq.~(\ref{cond2})]. 
Plugging Eqs.~(\ref{Pcond11}) and (\ref{Pcond22}) into
Eq.~(\ref{xtp6}) we get
\beq
X_{\mathrm{total}}~=\frac{\iota_P}{6}~+~  (1-\widetilde{p})\cdot
\frac{2\mathcal{C}}{6}  ~+~\mbox{integer}\,. 
\eeq
Recalling the condition in Eq.~(\ref{vergesseneBED}), we can rewrite this as
\beqn
X_{\mathrm{total}}&=&\frac{\iota_P}{6}~+~  (1-\widetilde{p})\cdot
\frac{-\iota_P}{6} ~+~  (1-\widetilde{p})\cdot
\frac{3\mathcal{J}}{6} ~+~\mbox{integer}\,\notag \\
&=&\frac{\widetilde{p} \cdot \iota_P}{6}~+~  (1-\widetilde{p})\cdot
\frac{\mathcal{J}}{2}~+~ \mbox{integer}\,. 
\eeqn
As $(1-\widetilde{p})$ is always an even number, we finally arrive at
\beqn
X_{\mathrm{total}}&=& \frac{\widetilde{p} \cdot \iota_P}{6}
~+~\mbox{integer}\,.  
\eeqn
\end{itemize}
As a summary, in addition to Eq.~(\ref{cond0}), depending on the
desired remnant low-energy discrete symmetry, we need to impose the
following conditions on the $X$-charges:
\beqn & X_{H^D}-X_{L_1}~=~\left\{\begin{array}{lll}
\mbox{integer} \\
\mbox{integer}-\mbox{${m}/{2}$}\\
\mbox{integer} \\
\mbox{integer} \\
\mbox{integer} -\mbox{${1}/{2}$}
\end{array}\right.,~~~~~~3X_{Q^1}+X_{L^1}~=~\left\{\begin{array}{lll}
\mbox{integer} \\
\mbox{integer}  \\
\mbox{integer}-\mbox{${b^{\,\prime}}/{2}$}\\
\mbox{integer}-\mbox{${b}/{3}$}  \\
\mbox{integer}-\mbox{${\widetilde{p}}/{3}$} 
\end{array}\right., & \nonumber \\\nonumber \\
& ~~ ~~\Longrightarrow~~ ~~
X_{\mathrm{total}}~=~\left\{\begin{array}{lll}
\mbox{integer} \\
\mbox{integer}+\mbox{${m\cdot\iota_M}/{2}$,}  \\
\mbox{integer}+ \,\mbox{${b^{\,\prime}\cdot\iota_B^{\,\prime}}/{2}$} \\
\mbox{integer}+ \,\mbox{${b\cdot\iota_B}/{3}$} \\
\mbox{integer}+ \,\mbox{${\widetilde{p}\cdot\iota_P}/{6}$}
\end{array}\right.,& 
\label{condns}
\eeqn
with $m,b^{\,\prime}=1, ~b\in\{1,2\},$ and $\widetilde{p}\in\{-1,1\}$. We also have $
\iota_M,\iota_B^ \prime\in\{0,1\}, ~\iota_B\in\{0,1,2\}$, and $\iota
_P\in\{0,1,2,3,4,5\}$. The five cases in Eq.~(\ref{condns}) correspond
to having {\it all} terms, only $M_p$ terms, only $B_p$ terms (see
App.~\ref{bpb3-app}), only $B_3$ terms, or only $P_6$ terms allowed by
virtue of the $X $-charges, respectively. Note that in
Ref.~\cite{Dreiner:2003yr} it was shown that Eq.~(\ref{cond0})
together with the coefficients $ \mathcal{A}_{CCX}$ and $\mathcal{A}_
{WWX}$ of the $SU(3)_C$-$SU(3)_C $-$U(1)_X$ and $SU(2)_W$-$SU(2)_W
$-$U(1)_X$ anomalies and the condition of Green-Schwarz anomaly
cancellation require
\beq\label{irgendeinscheisslabel}
3X_{Q^1}+X_{L^1}=\frac{\mbox{integer}}{\mathcal{N}_g}, 
\eeq
where $\mathcal{N}_g$ symbolizes the number of generations. With
$\mathcal{N}_g=3$ all possibilities listed above except the anomalous
$B_p$ are compatible with Eq.~(\ref{irgendeinscheisslabel}).

\cleqn


\section{\label{bases}Sequence of Basis 
Transformations}

The Froggatt-Nielsen charges determine the structure of the theory
just below the gravitational scale $M_{\mathrm{grav}}$. The low-energy
theory emerges after the successive breakdown of the $U(1)_X$ gauge
symmetry, supersymmetry, and then $SU(2)_W\times U(1)_Y$. The
hierarchy of the fermion mass spectrum is given in terms of powers of
the ratio $\eps\equiv\frac{\langle A\rangle}{M_{\mathrm {grav}}}$ of
the vacuum expectation value (VEV) of the $U(1)_X$ flavon field, $A$,
and the gravitational scale. Within a string-embedded FN framework
this expansion parameter originates in the Dine-Seiberg-Wen-Witten
mechanism \cite{Dine:1986zy,Dine:1987bq,Atick:1987gy,Dine:1987gj},
leading to a value of about $\eps \sim 0.2$ (see {\it e.g.}
Ref.~{\cite{Dreiner:2003yr}}).  Neglecting $\mathcal{O}(1)$
renormalization flow effects and imposing that $B_3$ arises from the
$X$-charges as described in Section~\ref{B3}, we obtain a Lagrangian
which is $B_3$ invariant but $M_p$ violating ($\not\!\!\!M_p$).  The
resulting kinetic terms obtained from the K\"ahler potential are
non-canonical, {\it i.e.}  they are not diagonal in generation space
and not properly normalized.  Furthermore, the sneutrino vacuum
expectation values are in general non-zero. It is usually more
convenient to formulate $\not\!\!M_p$ theories where the neutrino
masses are induced radiatively in a basis with vanishing sneutrino
VEVs and the down-type fields rotated to their mass bases
\cite{Grossman:1998py}.\footnote{For a pedagogical review of the different
contributions to the neutrino masses, expressed in a basis independent way,
see Ref.~\cite{Rakshit:2004rj}.} Therefore, we apply the sequence of basis
transformations, depicted in the diagram below, and study its effects
on the FN-generated coupling constants.  The numbers in brackets refer
to the explanations of each step, below.
\vspace{4mm}
\begin{center}
\begin{tabular}{cl}
\fbox{$\phantom{\Big|}$Type of Basis$\phantom{\Big|}$} & \fbox{
$\phantom{\Big|}$Redefinition of Chiral Fields$\phantom{\Big|}$} \\[7mm]
Froggatt-Nielsen basis \\
(denoted by the subscript FN) \\[3mm] 
$\Big\downarrow$ & 
$\begin{array}{l} \mbox{(1)~Non-unitary~transformation~of} \\
 \mbox{$\;\;\;\;\;\;Q^i$, $\ol{D^i}$, $\ol{U^i}$, 
$L^\alpha$, $\ol{E^i}$, $H^D$, $H^U$}\\
\end{array}$ 
\\[6mm]
Basis with canonical  \\
K\"ahler potential \\ \\
$\Big\downarrow$ & 
$\begin{array}{l} \mbox{(2)~Unitary~transformation~of~$L^\alpha$} 
\end{array}$ \\[6mm]
Basis without \\
sneutrino VEVs \\[3mm]
$\Big\downarrow$ & 
$\begin{array}{l}\mbox{(3) Unitary~transformation~of}\\ 
~~~~~ \mbox{$Q^i$,~$\ol{D^i}$,~$L^i$,~and~$\ol{E^i}$} 
\end{array}$\\[6mm]
Mass basis of down-type quarks\\
and charged leptons
\end{tabular}
\end{center}
\vspace{0mm}
In the third step we only rotate the $L^i$, not the $L^\alpha$. The
transformations of the $\ol{U^i}$ do not affect any of the terms we
are interested in, so that we do not further consider them. After the
above transformations, we again find an FN structure for the coupling
constants in the new basis. Working backwards, it is then possible to
deduce phenomenologically viable $X$-charge assignments from the
experimentally observed masses and mixings of quarks and leptons:
\begin{enumerate}
\item \emph{Canonicalization of the K\"ahler potential (CK):} The
  K\"ahler potential for $n$ species of superfields
  $\Phi_{\mathrm{FN}}^i$ ($i=1,...,n$) with equal gauge quantum
  numbers in the FN basis, \ie\ which can mix, is canonicalized by the $n\times n$
  {\it non-unitary} matrix $\bsym{C^{(\Phi)}}$, with the texture (see
  {\it e.g.}  Ref.~\cite{Dreiner:2003yr})
\beqn\label{defofckmatrix}
{C^{(\Phi)}}_{ij} & \sim & \eps^{|X_{\Phi^i} - X_{\Phi^j}|}\,.  
\eeqn
In terms of the canonicalized superfields $\Phi^i \equiv{C^{(\Phi)}}
_{ij} ~\Phi_{\mathrm{FN}}^j$, the kinetic operators are given in
their standard diagonal and normalized form. The interaction coupling
constants ${c_{\,\mathrm{FN}}}_{\,i}$ also change correspondingly
through the basis transformation, \emph{e.g.} for a trilinear
interaction of superfields $\Phi^i$, $\Psi^j$ and $\Theta^k$
\beqn 
{c_{\,\mathrm{FN}}}_{\,ijk} \:\Phi_{\mathrm{FN}}^i 
\Psi_{\mathrm{FN}}^j\Theta_{\mathrm{FN}}^k &=& c_{\,ijk} \: 
\Phi^{i} \Psi^{j}\Theta^{k}\,, 
\eeqn 
with
\beqn 
c_{ijk} &\equiv& [{\bsym{C^{(\Phi)}}}^{-1}]_{i'i} \,
[{\bsym {C^{(\Psi)}}}^{-1}]_{j'j} \,
[{\bsym{C^{(\Theta)}}}^{-1}]_{k'k} \;
{c_{\,\mathrm{FN}}}_{\,i'j'k'}\,.  \label{coupling}
\eeqn 
Note that each index transforms separately. In the following, while
discussing the general FN power structure, we focus on one index for
notational simplicity, \ie\ we suppress additional indices that might
be attached to the coupling constants
\beqn
c_{\,i} &\equiv& [{\bsym{C^{(\Phi)}}}^{-1}]_{ji} 
\;{c_{\,\mathrm{FN}}}_{\,j} \;.
\eeqn
The generalization to $n$ indices is trivial. Considering
superpotential couplings which are free of supersymmetric
zeros,\footnote{The problems connected with having supersymmetric
  zeros in the Yukawa mass matrices are discussed in
  App.~\ref{2by2}.} we have $~{c_{\,\mathrm{FN}}}_{\,i} \propto
\eps^{X_{\Phi^i}}$. Under the above transformations, we obtain
\cite{Binetruy:1996xk}
\beqn 
c_{\,i} & \propto &
\eps^{|X_{\Phi^j} - X_{\Phi^i}|} \; \eps^{X_{\Phi^j}} \;\; \sim \;\;
\eps^{X_{\Phi^i}}.  
\eeqn
Coupling constants which are not generated by FN alone but involve  a 
combination of FN and Giudice-Masiero/Kim-Nilles mechanism (see {\it e.g.}  
Ref.~\cite{Dreiner:2003yr}) are treated slightly differently:
\begin{itemize}
\item Later we will \emph{e.g.} assume that the bilinear
  superpotential terms $\mu_\alpha L^\alpha H^U$ are due to the GM/KN
  mechanism. Therefore the corresponding coupling constants have the
  $X$-charge dependence ${c_ {\,\mathrm{FN}}}_{\,i} \propto
  \eps^{-X_{\Phi^i}}$. In this case, the canonicalization of the
  K\"ahler potential yields
\beqn c_{\,i} & \propto & \eps^{|X_{\Phi^j} - X_{\Phi^i}|} \; \eps^{-
  X_{\Phi^j}} \;\; \sim \;\; \eps^{- X_{\Phi^i}}.  
\eeqn 
\item We also deal with the case where on the one hand the MSSM
  operators $H^D L^i\overline{E^j}$, $H^D Q^i\overline{D^j}$ are
  required to have overall positive integer $X$-charges, whereas the
  corresponding $\not\!\!M_p$-operators with $L^0\equiv H^D
  \rightarrow L^i$ ($i=1,2,3$) replaced, \emph{i.e.} $L^iL^j
  \overline{E^k}$, $L^iQ^j\overline{D^k}$, have overall negative integer 
  $X$-charges.  This assumption implies $X_{L^0} > X_{L^i}$. Due to
  the GM/KN mechanism, the supersymmetric zeros of the coupling
  constants with negative overall $X$-charge are actually not zero.
  However, for trilinear couplings the resulting terms are suppressed
  by a factor of $\mathcal{O}(\frac{m_{\mathrm{soft}}}{M_{\mathrm
  {grav}}})$ and therefore effectively absent; \emph{e.g.}
\beqn
X_{H^D}+X_{Q^2}+X_{\ol{D^2}}=2~~
&\stackrel{\mathrm{pure~FN}}{=\!\!=\!\!=\!\!=\!\!\Longrightarrow}&~\eps^2, 
\\
\mbox{but} ~~~~~
X_{L^1}+X_{Q^2}+X_{\ol{D^2}}=-3~
&\stackrel{\mathrm{GM/KM}}{=\!\!=\!\!=\!\!=\!\!\Longrightarrow}&
~\frac{m_{\mathrm{soft}}}
{M_{\mathrm{grav}}}\cdot\eps^3\ll\eps^3\,,
\eeqn 
where on the right we show the power of $\eps$ of the corresponding
coupling. So for the coupling constants we effectively have ${c_{\,
\mathrm{ FN}}}_{\, \alpha}\propto(\eps^{X_{L^0}},0,0,0)_\alpha$. 
But thanks to the canonicalization of the kinetic terms, these ``quasi
supersymmetric zeros'' are filled in so that
\beqn 
c_{\,\alpha} = [{\bsym{C^{(L)}}}^{-1}]_{0 \alpha} \;{c_{\,
    \mathrm{FN}}}_{\,0} \;\; \propto \;\; \eps^{|X_{L^0} -
  X_{L^\alpha}|} \; \eps^{ X_{L^0}} \;\; \sim \;\; \eps^{2 X_{L^0} -
  X_{L^\alpha}}.\label{CKlqd} 
\eeqn 
We can apply a similar consideration to operators which contain
$\eps^{ab}L^\alpha_a L^\beta_b$, where $a,b\in\{1,2\}$ are $SU(2)$
doublet indices. As the symmetric part of the corresponding coupling
constant ${c_{\,\mathrm{FN}}}_{\,\alpha \beta}$ cancels
automatically, it can be taken antisymmetric without loss of
generality. Now when constructing a viable model, we choose the $X
$-charges such that the terms $\eps^{ab}L^i_a L^j_b$, with $i,j=1,2,
3$, are forbidden by a negative integer total $X$-charge, whereas
$\eps^{ab} L^i_a L^0_b$ and $\eps^{ab}L^0_a L^j_b$ are allowed. In
this special case we find\footnote{This can be seen as follows:
$c_{\alpha \beta} = [{\bsym{C^{(L)}}}^{\!-1}]_ {\alpha^\prime
\alpha}[{\bsym{C^{(L)}}} ^{\!-1}]_{\beta^\prime
\beta}\,{c_{\,\mathrm{FN}}}_{\,\alpha^
\prime \beta^\prime}$. The assumption ${c_{\,\mathrm{FN}}}_{
\,ij}=0$ together with the condition of antisymmetry, ${c_{\,
\mathrm{FN}}}_{\,\alpha 0} = -{c_{\,\mathrm{FN}}}_{\,0 \alpha}$,
leads to
\beqn 
c_{\alpha \beta} &=& [{\bsym{C^{(L)}}}^{\!-1}]_{0 \alpha}
  [{\bsym{C^{(L)}}}^{\!-1}]_{j \beta} \cdot
  {c_{\,\mathrm{FN}}}_{0 j}+ [{\bsym{C^{(L)}}}^{\!-1}]_{i
    \alpha} [{\bsym{C^{(L)}}}^{-1}]_{0 \beta}\cdot
  {c_{\mathrm{FN}}}_{i 0}\,, \notag \\
  &=&[{\bsym{C^{(L)}}}^{\!-1}]_{0 \alpha}
  [{\bsym{C^{(L)}}}^{\!-1}]_{j \beta}\cdot {c_{\mathrm{FN}}}_{\,0
    j} - (\alpha\lra\beta).\notag 
\eeqn 
The $\eps$-structure is then given by 
\beqn
c_{\alpha \beta} &\propto&
\eps^{|X_{L^0}- X_{L^\alpha}|} \, \eps^{|X_{L^j}-X_{L^\beta}| } \,
\eps^{X_{L^0}+X_{L^j}} - (\alpha \lra\beta)\notag \\
&\propto& \eps^{2 X_{L^0}-X_{L^\alpha} +
  X_{L^\beta}} - (\alpha \lra\beta)\,.\notag
\eeqn
}
\beqn 
c_{\,\alpha \beta} &=&
[{\bsym{C^{(L)}}}^{-1}]_{0 \alpha} \;
[{\bsym{C^{(L)}}}^{-1}]_{j \beta} \;{c_{\,\mathrm{FN}}}_{\,0
  j}\; - \; (\alpha \lra \beta) \label{CKlle} \nonumber\\
&\propto & \eps^{ 2 X_{L^0} - X_{L^\alpha} + X_{L^\beta}} \; - \;
(\alpha \lra \beta).  
\eeqn 
\end{itemize}

After the canonicalization of the K\"ahler potential, all
superpotential coupling constants of the fields $\Phi^i$ will
therefore include a factor of either $\eps^{\,X_{\Phi^i}}$ or $\eps
^{\,-X_{\Phi^i}}$.

\item \emph{Rotating away the sneutrino VEVs:} Next we perform a
  unitary transformation on the superfields $L^\alpha$ ($\alpha=0,1,2,
  3$) in order to eliminate the sneutrino VEVs. The four vacuum
  expectation values $\upsilon_\alpha$ of the scalar component fields
  in $L^\alpha$ are determined by the minimization conditions for the
  neutral scalar potential. If we make the well-motivated
  \cite{Banks:1995by} assumption of an FN structure in the soft
  supersymmetry breaking terms (for details see App.~\ref{vevcond}),
  we find
\beqn 
\upsilon_\alpha & \propto & \eps^{-X_{L^\alpha}}.  
\eeqn 
We eliminate the sneutrino VEVs $\upsilon_i$ ($i=1,2,3$) by the
unitary matrix which in Ref.~\cite{Dreiner:2003hw} was used to rotate
away the bilinear superpotential terms $L^iH^U$. In our case it has
the texture\footnote{Replacing $\mu \rightarrow \upsilon_0$ and $K_i
  \rightarrow \upsilon_i$ in Eq.~(4.10) of Ref.~\cite{Dreiner:2003hw},
  we have $K = \sqrt{{\upsilon_i^\ast} \upsilon_i}$ and $\mathcal{M} =
  \sqrt{{\upsilon_\alpha^\ast} \upsilon_\alpha}$. For the matrix we
  then have
\beqn
{U^{\mathrm{VEVs}}}_{0j} &=& \frac{|\upsilon_0|}{\mathcal
    {M}} \cdot \frac{{\upsilon_j^\ast}}{{\upsilon_0^\ast}} \sim
  \eps^{X_{L^0}-X_{L^j}}\,,\quad
{U^{\mathrm{VEVs}}}_{i0} =-
  \frac{|\upsilon_0|}{\mathcal{M}} \cdot \frac{\upsilon_i}{\upsilon_0}
  \sim \eps^{X_{L^0}-X_{L^i}}\,,\notag\\
{U^{\mathrm{VEVs}}}_{ij} &=&
  \delta_{ij} + \frac{\upsilon_i {\upsilon_j^\ast}}{K^2} \cdot \left(
    \frac{|\upsilon_0|}{\mathcal{M}} -1 \right) \approx \delta_{ij} -
  \frac{\upsilon_i {\upsilon_j^\ast}}{2 |\upsilon_0|^2} \sim
  \delta_{ij} + \eps^{2 X_{L^0} - X_{L^i} - X_{L^j}}\,.\notag
\eeqn
In the penultimate step we applied the approximation $K\ll \mathcal{M}
\approx |\upsilon_0|$.}
\beqn 
\bsym{U^{\mathrm{VEVs}}} & \sim
& \begin{pmatrix} 1 & \eps^{\,X_{L^0}-X_{L^j}} \\
  \eps^{\,X_{L^0}-X_{L^i}} & \delta_{ij} + \eps^{\,2 X_{L^0} - X_{L^i}
    - X_{L^j}} \end{pmatrix} .\label{vevrot} 
\eeqn
Accordingly, all coupling constants involving $L^\alpha$ also have to
be transformed.  However, as $[{\bsym{U^{\mathrm{VEVs}}} }^\dagger]_{
\beta \alpha} \; \eps^{\,\pm X_{L^\beta}} \; \sim \;\eps^{\,\pm X_{L^
\alpha}}$, their $\eps$-structure remains unchanged.

\item \emph{Rotation of the quarks and charged leptons into their mass
    bases:} In a third step, the down-type quark\footnote{As we apply
    the basis transformations equally on both components of the
    $SU(2)_W$ superfield doublets $Q^i$, we can diagonalize {\it
    either} the up- {\it or} the down-type quark mass matrix. The
    latter is more appropriate for our purpose because, in the context
    of radiatively generated neutrino masses, only down-type loops
    contribute to the neutrino mass matrix and the computations are
    simpler in this basis \cite{Grossman:1998py}. After $SU(2)_W\times
    U(1)_Y$ breaking, we rotate the left- and right-handed up-type
    quark superfields $U_L$ and $\ol{U}$ into their mass basis.}  and
    charged lepton mass matrices are diagonalized by the unitary
    transformations $\bsym{U^{(Q)}}$, $\bsym{U^{(\ol{D})}}$,
    $\bsym{U^{(L)}}$, and $\bsym{U^{(\ol{E})}}$ of the corresponding
    superfields. Their $\eps$-power structure is given by, see also
    Ref.~\cite{Hall:1993ni},
\beqn 
{U^{(Q)}}_{ij} \; \sim \;
  \eps^{|X_{Q^i} - X_{Q^j}|}\,, && {U^{(\ol{D})}}_{ij} \; \sim \;
  \eps^{|X_{\ol{D^i}} - X_{\ol{D^j}}|}\,,
  \nonumber \\
  {U^{(L)}}_{ij} \; \sim \; \eps^{|X_{L^i} - X_{L^j}|}\,, &&
  {U^{(\ol{E})}}_{ij} \; \sim \; \eps^{|X_{\ol{E^i}} -
    X_{\ol{E^j}}|}\,.\label{massbasisQL} 
\eeqn 
Here we have to assume a decreasing $X$-charge for increasing
generation index.\footnote{The diagonalization matrices $\bsym{U^
{(...)}}$ have the structure of Eq.~(\ref{massbasisQL}) only if the
$X$-charges of the left- and the right-chiral superfields are ordered
in the same way. Demanding further that the third generation is the
heaviest and the first the lightest, we are restricted to decreasing
$X$-charge for increasing generation index.} The transformations of
Eq.~(\ref{massbasisQL}) diagonalize the down-type mass
matrices. However, they do not alter the $\eps$-structure of the
up-type Yukawa couplings and other renormalizable or
non-renormalizable coupling constants, with one significant exception.
The $\eps$-structure of the down-type Yukawa mass matrices is
obviously changed drastically in its off-diagonal entries by the
transition to the mass basis. Through our transformations it is
possible that supersymmetric zeros are filled proportional to the
down-type mass matrices. Diagonalizing the down-type mass matrix then
also diagonalizes these coupling constants in the corresponding two
indices. In our model, we encounter such a proportionality when
generating interactions as in Eq.~(\ref{CKlqd}), for example the
trilinear $\not\!\!\!M_p$ terms $\lam^\prime_{ijk} \, L^i Q^j
\ol{D^k}$. Together with the mass terms $G^{(D)}_{jk}\,
H^DQ^j\ol{D^k}~\equiv~\lam^\prime_{0jk} \, L^0 Q^j \ol{D^k}$ for
the down-type quarks we find
\beqn 
\lam^\prime_{ijk} & = & \frac{
  [{\bsym{C^{(L)}}}^{-1}\cdot{\bsym{U^{\mathrm{VEVs}}}}^\dagger]_{0 l} }{\ 
  [{\bsym{C^{(L)}}}^{-1} \cdot{\bsym{U^{\mathrm{VEVs}}}}^\dagger]_{0 0} } \;
[{\bsym{U^{(L)}}}^\dagger]_{l i} \; \lam^\prime_{0jk}\,,
\label{rpvbyCK}
\eeqn
with the coupling constants $\lam^\prime_{\alpha jk}$ now given in
the basis of diagonal down-type mass matrices. Analogously, we have
for the superpotential terms $\frac{1}{2} \, \lam_{ijk}\,L^i L^j\ol{E^k}$, 
together
with ${\lam}_{0jk} \, L^0 L^j\ol{E^k}$,
\beqn 
\lam_{ijk} & = & \frac{ [{\bsym{C^{(L)}}}^{-1} \cdot
  {\bsym{U^{\mathrm{VEVs}}}}^\dagger]_{0 l} }{\ 
  [{\bsym{C^{(L)}}}^{-1} \cdot
  {\bsym{U^{\mathrm{VEVs}}}}^\dagger]_{0 0} } \;
[{\bsym{U^{(L)}}}^\dagger]_{l i} \; \lam_{0jk} \; - \; (i
\lra j).\label{rpvbyCKlept} 
\eeqn 
Here we have neglected the second antisymmetrizing contribution of
Eq.~(\ref{CKlle}) when expressing $\lam_{0jk}$ in terms of
${\lam_{\,\mathrm{FN}}}_{\, 0j'k'}$ as it is suppressed by a factor
of $\eps^{\,2 \, (X_{L^0} - X_{L^j})}$. Hence, both types of trilinear
$\not\!\!M_p$ coupling constants are proportional to the corresponding
Yukawa mass matrices,\footnote{In Ref.~\cite{Kim:2005} quite generally
models for radiatively generated neutrino masses are studied in which
as it so happens \emph{1)} baryon triality is (accidentally) conserved
and \emph{2)} the trilinear $\not\!\!M_p$ coupling constants are
proportional to the mass matrices of the down-type quarks and charged
leptons. Our models belong to this category, with both \emph{1)} and
\emph{2)} arising by virtue of the $X$-charges.  The 5$^{th}$ charge
assignment in Table~\ref{allsets} is presented in
Ref.~\cite{Kim:2005}, as an example.}  which are diagonal in our
basis.
\end{enumerate}
Table~\ref{sequence} summarizes the FN structure of some important
superpotential coupling constants at different steps in the sequence
of basis transformations. We omitted the up-type quark Yukawa coupling
constants in Table~\ref{sequence}, as they have the standard FN
structure which does not change under the sequence of basis
transformations.
\begin{table}[t]
\begin{tabular}{||c||c|c|c||} 
\hline \hline & $\frac{\mu_\alpha}{m_{3/2}}$ & $\lam^\prime_{\alpha jk}$  
& $\lam_{\alpha \beta k}$\\ 
\hline \hline $(0)\phantom{\Big{|}}$ & $\eps^{-X_{L^\alpha}-
X_{H^U}}$   & only $\lam^\prime_{0 jk} \sim 
\eps^{X_{L^0} +X_{Q^j} + X_{\ol{D^k}}}$  &  only $\lam_{0 jk} 
\sim \eps^{X_{L^0} +X_{L^j} + X_{\ol{E^k}}}$  \\ 
\hline  $(1)\phantom{\Big{|}}$ &  $\eps^{-X_{L^\alpha}-X_{H^U}}$  
& $ \eps^{2 X_{L^0}- X_{L^\alpha} + X_{Q^j} + X_{\ol{D^k}}}$  &  
$\eps^{2 X_{L^0} - X_{L^\alpha} + X_{L^\beta}+ X_{\ol{E^k}}}$ $-$ 
$(\alpha \lra \beta)$  \\  
\hline  $(2)\phantom{\Big{|}}$ &  $\eps^{-X_{L^\alpha}-X_{H^U}}$  
& $ \lambda^\prime_{0jk} \sim  \delta_{jk} \; \eps^{ X_{L^0}+ X_{Q^k} + X
_{\ol{D^k}}}$, &   $\lambda_{0jk} \sim \delta_{jk} \;\eps^{X_{L^0} + 
X_{L^k}+ X_{\ol{E^k}}}$ ,\\
 &  $\phantom{\Big{|}}$ &  $\lambda^\prime_{ijk} \sim \eps^{X_{L^0} - 
X_{L^i}} \lambda^\prime_{0jk}$,  &   $\lambda_{ijk} \sim \eps^{X_{L^0} 
- X_{L^i}} \lambda_{0jk} - (i \lra j)$  \\ \hline \hline  
\end{tabular}   \caption{\small FN structure of superpotential couplings at 
  various stages of the basis transformations: Before $(0)$ and after $(1)$ 
  the canonicalization of the K\"ahler potential, and finally in the mass 
  basis of the down-type quarks and charged leptons $(2)$.}\label{sequence}
\end{table}

\noindent We now state a first set of constraints on the $X$-charges,
required for our model, which we shall make more quantitative in the
subsequent sections:
\begin{itemize}
\item By choosing positive integer $X$-charges 
for all trilinear MSSM interactions we avoid troubles in the fermionic 
mass spectrum associated with supersymmetric zeros 
(see App.~\ref{2by2}). 
\item The generalized $\mu$-problem ($\mu_\alpha \,L^\alpha
  H^D$) is solved by the Giudice-Masiero/Kim-Nilles
  mechanism.
\item In order to avoid too heavy neutrino masses \cite{
    {Allanach:2003eb}}, we require $\frac{\mu_i}{\mu_0}\sim\eps^
  {X_{L^0} - X_{L^i}} < 1 $, \ie\ $X_{L^0} > X_{L^i}$.
\item If the trilinear $\not\!\!M_p$ interactions are only suppressed
  by powers of $\eps$ comparable to the trilinear MSSM terms, they are
  in disagreement with the experimental bounds \cite{Allanach:1999ic,
  Dreiner:2003yr}. Therefore we choose to generate trilinear
  $\not\!\!M_p$ terms by the canonicalization of the K\"ahler
  potential as described above, {\it cf.} Eq.~(\ref{CKlqd}).  This
  mechanism has also been employed in\footnote{The authors of
  Ref.~\cite{Mira:2000gg} construct their model such that $X_{
  \ol{U^i}\ol{D^j}\ol{D^k}} <0$, so that it is GM/KN-suppressed if
  $X_{\ol{U^i}\ol{ D^j}\ol{D^k}}$ is integer.  However, with
  $X_{L^i}+X_{ H^U}$ required to be integer and working with
  $\Delta^L_{21}=\Delta^L_{31}=0$, $z=1$, their proposed $X$-charge
  assignment in fact also accidentally generates $B_3$, so that
  $\ol{U^i}\ol{D^j}\ol {D^k}$ is not only highly suppressed but absent
  altogether.}  Ref.~\cite{Mira:2000gg}.  The term $\ol{UDD}$ is
  forbidden altogether by $B_3$.
\end{itemize}

\noindent In the following we study phenomenological constraints on
the $X$-charges arising from the fermionic mass spectrum. In our
basis, the down-type mass matrices are diagonal. Therefore the CKM
matrix is obtained solely from the diagonalization of the up-type
quark mass matrix, and exhibits the $\eps$-structure ${U^{\mathrm{C
      KM}}}_{ij} \sim \eps^{|X_{ Q^i}-X_{Q^j}|}$. Furthermore, we need
to specify and diagonalize the neutrino mass matrix.






\section{\label{pheno}Non-Neutrino Constraints 
on the $\bsym{X}$-Charges}
\cleqn
In the previous section, we translated our model from the scale of
$U(1)_X$ breaking down to the electroweak scale. The FN charges of the
MSSM superfields are now directly con\-nected to the low-energy
fermionic mass spectrum. For our model, we require the $X$-charges to
reproduce phenomenologically acceptable quark masses and mixings as
well as charged lepton masses. Furthermore, we demand the GS anomaly
cancellation conditions \cite{{Maekawa:2001uk},{Dreiner:2003yr}} to be
satisfied (apart from the constraints listed at the end of the
previous section, which we will not implement right away but
later). We then find that all 17 $X$-charges can be expressed in terms
of only six real numbers (see Table~1 in Ref.~\cite{Dreiner:2003yr}):
\beqn 
x =0,1,2,3 \;,\;\;\;\, & y = -1,0,1 \;,\;\, & \;\;\; z = 0,1 \;,\nonumber\\
\Delta^L_{31} \equiv X_{L^3} - X_{L^1} \;,\;\; & \Delta^L_{21} \equiv
X_{L^2} - X_{L^1}\;, & \;\;\; X_{L^1} \;.\label{parameters} 
\eeqn
$\Delta^L_{31}$ and $\Delta^L_{21}$ are necessarily integer whereas
$X_{L^1}$ is arbitrary.\footnote{$\Delta^L_{31}$ and $\Delta^L_{21}$
actually do not have to be integers, if it weren't for the sake of
Eq.~(\ref{cond0}).  Furthermore, with $\Delta^L_{31}$ and $\Delta^L_{
21}$ being fractional Eq.~(\ref{defofckmatrix}) would not hold.} For
phenomenological reasons $x,y,z$ can only take on the shown integer
values.  As we choose to generate the $\mu$-term via the GM/KN
mechanism, we take $z\equiv -X_{H^U} -X_{H^D}=1$ throughout this
article.  $x$ is related to the ratio of the Higgs VEVs by $\eps^x\sim
\frac{m_b}{m_t} \tan \beta$, with $\tan\beta=\big{|}\frac{\upsilon_u}{
\upsilon_0}\big{|}$. Recall, the sneutrino VEVs are rotated away, so $
|\upsilon_0|=|\upsilon_d| \equiv\sqrt{{\upsilon_\alpha^\ast}~ \upsilon
_\alpha}$. $y~ $parameterizes all phenomenologically viable CKM 
matrices. Our preferred choice is $y=0$, resulting in ${U^{\mathrm{CKM
}}}_ {12}\sim \eps$, ${U^{\mathrm{CKM}}}_{13}\sim\eps^3$, and ${U^{
\mathrm{CKM}}}_{2 3}\sim\eps^2$, see Ref.~\cite{Dreiner:2003yr}.

Assuming a string-embedded FN framework, the parameter $\eps$
originates solely in the Dine-Seiberg-Wen-Witten mechanism
\cite{{Dine:1986zy},{Dine:1987bq}, {Atick:1987gy},{Dine:1987gj}}.
Thus it is a derived quantity which depends on $x$ and $z$ (for
details see Ref.~\cite{Dreiner:2003yr}). Taking $z=1$ and $x=0,1,2,3$
we get $\eps$ within the interval $\eps\in[0.186,0.222]$.\footnote{
  The parameterization of the mass ratios of the SM fermions in terms
  of $\eps$ is based on $\eps=0.22$, so that working with other values
  for $\eps$ is strictly speaking slightly inconsistent.}

Our goal is to construct a conserved $B_3$ model. As discussed in
detail in Sect.~\ref{B3}, this leads to additional constraints, which
are best expressed in terms of the new parameters $\Delta^H,\,\zeta$
\beqn 
\Delta^H \equiv X_{L^1}-X_{L^0}\;,&&
3\zeta+b\equiv \Delta^L_{21}+\Delta^L_{31}-z\,.
\label{B_3condsa}
\eeqn 
Here the parameter $b=1,2$ is as introduced in Eq.~(\ref{cond22}).
Rewriting Eqs.~(\ref{cond11}) and (\ref{cond22}), the latter by making
use of Table~1 in Ref.~\cite{Dreiner:2003yr}, we see that demanding
$B_3$ conservation is equivalent to demanding\footnote{
  \label{rpv-footnote} The corresponding conditions for conserved
  $M_p$ are \beqn \Delta^H \equiv X_{L^1}-X_{L^0}-\mbox{$\frac{1}
    {2}$}\;\in\;\mathbb{Z}\;, && \zeta\equiv\frac{1}{3}(
  \Delta^L_{21}+\Delta^L_{31}-z)\;\in\;\mathbb{Z}.\nonumber \eeqn}
\beq
 \Delta^H,\,\zeta \;\in\; \mathbb{Z}\,. \label{B_3conds} 
\eeq
We now replace $X_{L^1}$ and $\Delta^L_{21}$ in favor of $\Delta^H$,
$\zeta$, and $b=1,2$ and we arrive at the constrained $X$-charges of
Table \ref{Table2}. This is the equivalent of Table~2 in Ref.
\cite{Dreiner:2003yr} for the case of $B_3$ instead of $M_p$.  Note
that the parameters $\zeta$ and $b$ appear in Table~\ref{Table2} only
in the combination $3\zeta + b$.

\begin{table}[t!]
\begin{center}
\begin{tabular}{|rcl|}
\hline
  & & \\
$~~~\phantom{\Big|}X_{H^D}$
 &$=$&$\frac{1}{5~(6+x+z )}~\Big(6y + x ~(2x + 12 + z 
- 2\Delta^{\!H}) $\\
       & &$  - z~(4 + 3 \Delta^{\!H}) - 
  2 ~(3 + 6 \Delta^{\!H} - \Delta^L_{31})-\frac{2}{3}~(6+x+z)(3 \zeta+ b)
\Big)~~~$\\
$~\phantom{\Big|}X_{H^U}$
 &$=$&$-z-\phantom{\Big|}X_{H^D}  $\\
$~\phantom{\Big|}X_{Q^1}$&$=$&$\frac{1}{3}\Big(
      10 - X_{H^D} + x + 2y + z - 
\Delta^{\!H} - \frac{1}{3}(3 \zeta + b) \Big)$\\
$~\phantom{\Big|}X_{Q^2}$&$=$&$X_{Q^1}-1-y $\\
$~\phantom{\Big|}X_{Q^3}$&$=$&$X_{Q^1}-3-y $\\
$~\phantom{\Big|}X_{\ol{U^1}}$&$=$&$
X_{H^D}-X_{Q^1}+8+z $\\
$~\phantom{\Big|}X_{\ol{U^2}}$&$=$&$
X_{\ol{U^1}}-3+y $\\
$~\phantom{\Big|}X_{\ol{U^3}}$&$=$&$
X_{\ol{U^1}}-5+y $\\
$~\phantom{\Big|}X_{\ol{D^1}}$&$=$&$-X_{
H^D}-X_{Q^1}+4+x $\\
$~\phantom{\Big|}X_{\ol{D^2}}$&$=$&$X_{
\ol{D^1}}-1+y $\\
$~\phantom{\Big|}X_{\ol{D^3}}$&$=$&$
X_{\ol{D^1}}-1+y $\\
$~\phantom{\Big|}X_{L^1}$&$=$&$
X_{H^D}+\Delta^{\!H}$\\
$~\phantom{\Big|}X_{L^2}$&$=$&$X_{L^1} -
\Delta^L_{31}+ z+ (3\zeta + b) $\\
$~\phantom{\Big|}X_{L^3}$&$=$&$X_{L^1}
+\Delta^L_{31}$\\
$~\phantom{\Big|}X_{\ol{E^1}}$&$=$&$-X_{
H^D}        {+~4} - X_{L^1} + x + z $\\
$~\phantom{\Big|}X_{\ol{E^2}}$&$=$&$X_{\ol{E^1}}
 -2 - 2z + \Delta^L_{31} - (3\zeta  + b)    $ \\ 
$~\phantom{\Big|}X_{\ol{E^3}}$&$=$&$
X_{\ol{E^1}} -4 - \phantom{2}z - \Delta^L_{31}             $ \\ 
  & &\\ \hline
\end{tabular}
\caption{\label{Table2}\small The constrained $X$-charges with an acceptable
  low-energy phenomenology of quark and charged lepton masses and
  quark mixing. In addition, the GS anomaly cancellation conditions
  are satisfied and conservation of $B_3$ is imposed. $x$, $y$, $z$
  and $b$ are integers specified in Eqs.~(\ref{cond22}) and
  (\ref{parameters}). $\Delta^H$, $\Delta^L_{31}$, and $\zeta$ are
  integers as well but as yet unconstrained. $SU(5)$ invariance would
  require $y=1$ and $z=\Delta^L_{21}=\Delta^L_{31}=0$, but the latter
  is not compatible with the second condition in Eq.~(\ref{B_3condsa}).
}
\end{center}
\end{table}

Phenomenologically, the conservation of $B_3$ renders the proton
stable. For the proton to decay we need a baryon-number violating
operator. This in turn requires the parameter $\mathcal{C}$ in
Eq.~(\ref{n-eq1}) to be non-zero. On the other hand, $\mathcal{C}$
must be an integer multiple of three in the case of $B_3$ conservation
[see Eq.~(\ref{zw}) in App.~\ref{bpb3-app}, with $\iota_B = 0$].
Hence, only operators with $|\mathcal{C}| = 3,6,9,...$ are $B_3$
conserving and baryon-number violating.  Comparing with
Eq.~(\ref{n-eq1}) we see that at least nine quark (or antiquark)
superfields are needed. Such a superpotential term, however, is
suppressed by a factor of $\frac{1}{{M_{\mathrm{grav}}^{\,6}}}$ and thus
negligible.

Our baryon triality conserving model is not compatible with grand
unified theories (GUTs). Unlike in the $M_p$ conserving model in
Ref.~\cite{Dreiner:2003yr}, it is impossible to choose the parameters
$x,y,z,\Delta^H,\zeta,b,\Delta ^L_{31}$ such that the $X$-charges of
Table~\ref{Table2} are $SU(5)$ invariant. This should be obvious,
since after symmetry breaking the trilinear GUT superpotential term $~
\bsym{\ol{5}}\,\bsym{ \ol{5}}\,\bsym{1\!\!\:0}~$ produces $LL\ol{E}$,
$LQ\ol{D}$ (both $B_3$ conserving), and $\ol{U}\ol{D}\ol{D}$ ($B_3$
violating). $SU(5)$ invariance requires $y=1$ and $z=\Delta^L_{21}=
\Delta^L_{31}=0$. However, the latter is not compatible with the 
second condition in Eq.~(\ref{B_3condsa}) with
$\zeta\in\mathbb{Z}$. For a review of models where horizontal
symmetries are combined with unification see Ref.~\cite{Kane:2005va}.





\cleqn
\section{\label{NeutrinoSector}The Neutrino 
Sector}
\subsection{\label{ExpNuRes}Experimental Results}
We now include the experimental constraints from the neutrino sector,
in particular from the solar \cite{Ahmed:2003kj,Smy:2003jf},
atmospheric \cite{Ashie:2005ik}, reactor \cite{Araki:2004mb}, and
accelerator \cite{Ahn:2002up} neutrino experiments.\footnote{We do not include
the result of the LSND experiment
\cite{Athanassopoulos:1996jb}.} We first need to
translate the data into a form, such that we can compare it to our
FN-models. Then we can use this to further constrain the $X$-charges.
   
In our $B_3$ conserving model, there are no right-handed neutrinos.
Hence in our phenomenological analysis of the data we only consider
Majorana mass terms for the left-handed neutrinos with a symmetric
mass matrix $\bsym{M^{(\nu)}}$ in the current eigenstate basis
$(\nu_e,\nu_\mu,\nu_\tau)$. The Takagi factorization of the matrix
$\bsym{M^{(\nu)}}$ is given by
\beqn 
\bsym{M^{(\nu)}_{\mathrm{diag}}} & = 
& {\bsym{U^{(\nu)}}}^\ast \cdot\bsym{M^{(\nu)}}
\cdot  {\bsym{U^{(\nu)}}}^\dagger,\label{nudiag}
\eeqn
where $\bsym{U^{(\nu)}}$ is a unitary matrix and $\bsym{M^{(\nu)}_{
    \mathrm{diag}}}=\bsym{\mathrm{diag}}(m_1,m_2, m_3)$. The neutrino
masses $m_1,\,m_2,\,m_3\,$ are the singular values of $\bsym{M^{(\nu)}}
$ (not the eigenvalues). They are most easily computed as the positive
square roots of the eigenvalues of $\bsym{M^{(\nu)}}^\dagger\bsym{M^{(
\nu)}}$. At this stage, we choose not to make any statement on the
relative size of the three masses. The corresponding singular vectors
are denoted $(\nu_1,\nu_2,\nu_3)$.

In order to determine the connection between the structure of the
original mass matrix $\bsym{ M^{(\nu)}}$ and the ordering of the
masses, we will need to fix the lepton mixing matrix. Experimentally,
we have access to the Maki-Nakagawa-Sakata (MNS) matrix
\cite{Maki:1962mu}, which is the product of the left-handed charged
lepton mixing matrix $\bsym{U^{(E ^{\phantom |} _L)}}$ and
$\bsym{U^{(\nu)}}^\dagger$. As we are working in the basis with a
diagonal charged lepton mass matrix, we have
$\bsym{U^{(E^{\phantom|}_L)}} = \mathds{1}_3$ and thus
\beqn
\bsym{U^{\mathrm{MNS}}} & \equiv & \bsym{U^{(E^{\phantom|}_L)}} 
\cdot {\bsym{U^{(\nu)}}}^\dagger = \; 
{\bsym{U^{(\nu)}}}^\dagger.\label{MNSdef}
\eeqn
In the standard parameterization \cite{Eidelman:2004wy}, $\bsym{U^
  {\mathrm{MNS}}}$ is given by
\beqn
\begin{pmatrix}1 &\phantom{-}0 &0 \\0 & 
\phantom{-}c_{23} & s_{23}\\ 0& -s_{23}& c_{23} \end{pmatrix} \! \cdot \! 
\begin{pmatrix}c_{13} & 0& s_{13}~e^{\mathrm{i}\delta}\\ 0& 1& 0\\ 
-s_{13}~e^{-\mathrm{i}\delta}& 0& c_{13} \end{pmatrix} \! \cdot \! 
\begin{pmatrix} \phantom{-}c_{12}& s_{12}&0 \\ -s_{12}& c_{12} & 0\\ 
\phantom{-}0& 0&1  
\end{pmatrix} \! \cdot  \begin{pmatrix}
e^{-\mathrm{i}\alpha_1/2} &0 &0 \\0 & e^{-\mathrm{i}\alpha_2/2} & 
0\\0 &0 &1  \end{pmatrix},~
\eeqn
with $c_{ij} \equiv \cos{\theta_{ij}}$ and $s_{ij}\equiv\sin{\theta
_{ij}}$. Here $i,j=1,2,3$ are generation indices and $\delta$ and
$\alpha_1$, $\alpha_2$ are the CP-violating Dirac and Majorana phases,
respectively. A global three-generation neutrino oscillation fit
assuming CP conservation (\ie\ $\delta=\alpha_1=\alpha_2=0$) yields
the $3\sigma$~CL allowed ranges \cite{{Gonzalez-Garcia:2000sq},
{Gonzalez-Garcia:2004jd}}
\beqn 
\phantom{|}\Delta
m^2_{21}\phantom{|} \equiv\phantom{|} m_2^2 - m_1^2\phantom{|} =
8.2^{+1.1}_{-0.9} \: \times 10^{-5} \: \mathrm{eV}^2, &\;\;\;&
\tan^2{\theta_{12}} = 0.39^{+0.21}_{-0.11} \:,\nonumber \\ |\Delta
m^2_{32}| \equiv |m_3^2 - m_2^2| = 2.2^{+1.4}_{-0.6} \:
\times 10^{-3} \: \mathrm{eV}^2, &\;\;\;& \tan^2{\theta_{23}}
= \:1.0\;^{+1.1}_{-0.5} \:,\nonumber \\
&& \;\!\sin^2{\,\theta_{13}} \le 0.041 \:.  
\label{exp-results}
\eeqn 
The neutrino mass eigenstates are conventionally labeled such that the
solar neutrino problem is predominantly solved by $\nu_1\lra\nu_2$
oscillations and the atmospheric neutrino problem by $\nu_2\lra\nu_3$
oscillations. Furthermore $\nu_1$ is defined as the neutrino which is
predominantly $\nu_e$. In this convention the sign of $\Delta m_{12}^
2$ is known to be positive from the solar neutrino data \cite{
Smy:2003jf}, whereas the sign of $\Delta m_{23}^2$ is unknown. There
are then two possible orderings of the masses \cite{Giunti:2004vv},
either $m_1< m_2< m_3$ or $m_3< m_1< m_2$.  Taking into account the
magnitudes of $\Delta m^2_{21}$ and $\Delta m^2_{32}$ we have the
following possible solutions
\beq \label{nu-solns}
\begin{array}{rll}
&m_1< m_2\ll m_3,\qquad & \mathrm{hierarchical}, \\ [3mm]
&m_3\ll m_1< m_2,& \mathrm{inverted\; hierarchical}, \\ [3mm]
\sqrt{|{m_i^2} -{ m_j^2}|}\ll&m_1\approx m_2\approx m_3, & 
\mathrm{degenerate,}
\end{array}
\eeq
where in the degenerate case $i,j=1,2,3$ can take on all values and we
again have the two possible mass orderings. We discuss these solutions
and their individual implications in the context of our FN scenario in
Sects.~\ref{massconstraints} and \ref{mixingconstraints}.

\begin{figure}[t] 
\epsfig{figure=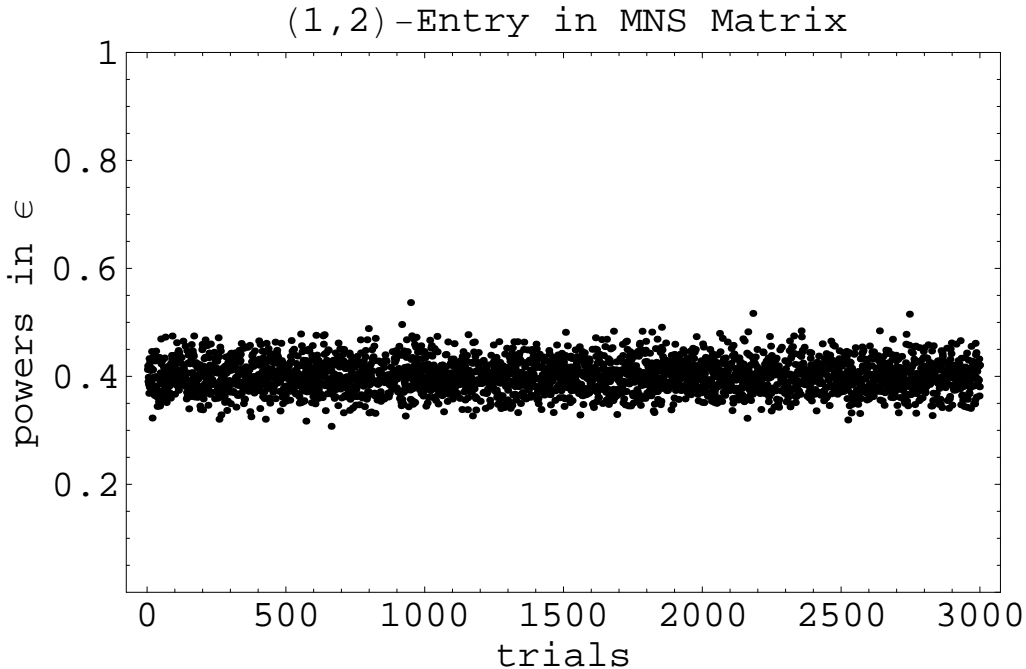,width=14.5cm} 
\caption{The powers in $\eps$ of the $(1,2)$-element for  an ensemble of 
$3000$ sets of mixing parameters obeying Gaussian statistics for 
$\theta_{12}$ and $\theta_{23}$, whereas $\theta_{13} 
\in [0^\circ,11.7^\circ]$ and $\delta \in [0,2\pi]$ are taken from 
an equal distribution.}  \label{MNSscatt12}
\end{figure}

The mixing angles together with their uncertainties can be translated
\cite{Gonzalez-Garcia:2004jd} into allowed ranges for the entries of
the MNS matrix in terms of the FN-parameter $\eps$. For the mixing
angles $\theta_{12}$ and $\theta_{23}$, we assume Gaussian errors in
their measured values.\footnote{Disregarding systematic effects,
  measured quantities follow a Gaussian distribution. Derived
  quantities such as the mixing angles $\theta_{ij}$ might however
  show a distorted statistical spread. Taking the central values of
  $\tan^2{\theta}_{ij}$ plus their $3 \sigma$ CL limits and
  translating these into corresponding angles $\theta_{ij}$, we found
  approximately symmetrical distributions for the mixing angles. Thus
  we are led to our simplifying assumption of Gaussian errors.}
Furthermore, assuming flat distributions for the unmeasured quantities
$\theta_{13} \in [0^\circ,11.7^\circ]$ and the Dirac phase $\delta \in
[0,2\pi]$, we calculate the scatter of the absolute values of the MNS
matrix elements. Figure~\ref{MNSscatt12} shows the powers in $\eps =
0.2$ of the $(1,2)$-element for an ensemble of $3000$ sets of mixing
parameters obeying the upper assumed statistics. From this we deduce an
FN $\eps$-structure (by definition the exponents must be integer) of
$\eps^0$ or $\eps^1$ for the (1,2)-element.

We employ a similar analysis of the other matrix elements. Due to the
unknown $\mathcal{O}(1)$ coefficients in FN models, we allow all
(integer) powers in $\eps$ within about $\pm 1$ of the center of the
scattering region. We then obtain the experimentally acceptable
$\eps$-structure for the MNS matrix
\beqn
\bsym{U_{\mathrm{exp}}^{\mathrm{MNS}}} & \sim & \begin{pmatrix}
  \eps^{0,1} & \eps^{0,1} & \eps^{0,1,2,...} \\ \eps^{0,1,2} &
  \eps^{0,1} & \eps^{0,1} \\ \eps^{0,1,2} & \eps^{0,1} & \eps^{0,1}
\end{pmatrix},
\label{MNSexpstr}
\eeqn
where multiple possibilities for the exponents are separated by
commas. The dots in the (1,3)-entry of Eq.~(\ref{MNSexpstr}) indicate
that arbitrarily high integer exponents are experimentally
allowed. Requiring an FN structure in the neutrino mass matrix however
excludes values beyond 2.  It should be mentioned again that this
calculation is done for $\eps = 0.2$. Varying $\eps$ within the
interval $[0.18 , 0.22]$ does not alter the allowed exponents in
Eq.~(\ref{MNSexpstr}).






\subsection{\label{nuMatrix}The Neutrino Mass Matrix}
In order to make use of the experimental information about the
neutrino sector, we need to specify the origin of the neutrino masses.
It has already been pointed out that $B_3$ invariance allows for
lepton-number violating $\not\!\!M_p$ interactions. Due to the
bilinear terms $\mu_iL^iH^U$ the neutrinos mix with the neutralinos,
which leads to {\it one} massive neutrino at tree level\footnote{
Strictly speaking, the distinction between neutrino and neutralino
mass eigenstates is no longer appropriate.  However, due to stringent
experimental constraints on the neutrino masses: $\mu_i/M_W\ll1$ and
the mixing between neutralinos and neutrinos is small
\cite{Allanach:2003eb}.}
\cite{Hall:1983id}. The inferred measured mass squared differences
$\Delta m^2_{21}$ and $\Delta m^2_{31}$ require at least {\it two}
massive neutrinos.  Therfore we must consider higher order
contributions to the neutrino mass matrix. In the following, we
concentrate on the effects of quark-squark and charged lepton-slepton
loop corrections \cite{{Grossman:1998py},{Grossman:1999hc},
{Davidson:2000uc}} due to the operators $LQ\overline{D}$ and
$LL\overline{E}$, respectively. (For a thorough analysis of all one-loop 
contributions to the neutrino mass matrix in the lepton-number violating but
$B_3$ conserving MSSM see Ref.~\cite{Dedes:2006ni}.)
The resulting effective neutrino mass
matrix in the flavor basis is given by
\beqn 
\bsym{M^{(\nu)}} &=& \bsym{M_{\mathrm{tree}}^{(\nu)}} +
\bsym{M_{{\lambda^\prime} \mbox{-}\mathrm{loop}}^{(\nu)}} +
\bsym{M_{\lambda \mbox{-} \mathrm{loop}}^{(\nu)}}\,,\label{MnuL} 
\eeqn
with 
\beqn 
\left(\bsym{M_{\mathrm{tree}}^{(\nu)}}\right)_{\,ij} \!&\! = \!&\!
\frac{{m_Z^2} ~ M_{\tilde{\gamma}} ~ \mu_0 ~ \cos^2{\beta}}{{m_Z^2} ~
  M_{\tilde{\gamma}} ~ \sin{2\beta} ~ - ~ M_1~ M_2 ~\mu_0} \cdot
\frac{\mu_i ~
  \mu_j}{{\mu_0^2}},\label{tree}\\
\left(\bsym{M_{{\lam^\prime} \mbox{-}\mathrm{loop}}^{(\nu)}}\right)
_{\,ij} \!&\!
\simeq \!&\! \frac{3}{32 \pi^2} \sum_{k,n} (\lam^\prime_{ikn} ~
\lam^\prime_{jnk} + \lam^\prime_{jkn} ~
\lam^\prime_{ink}) ~ m^d_k \sin{2\phi^{(\tilde{d})}_n}
~\ln\!\left[\Bigg{(}\frac{ m^{\tilde{d}}_{n_1} }{
    m^{\tilde{d}}_{n_2}}\Bigg{)}^{\!\!\!2}\right] , ~~~~
\label{lamploop} \\
\left(\bsym{M_{\,\lam \,\mbox{-}\mathrm{loop}}^{(\nu)}}\right)_{\,ij} 
\!&\!\simeq\!&\! 
\frac{1}{32 \pi^2} \sum_{k,n} \, (\,\lam_{ikn}~ \lam_{jnk} \, + 
\,\lam_{jkn} ~ \lam_{ink}\,) \;\, m^e_k \sin{2\phi^{(\tilde
{e})}_n} ~ \ln\!\left[\Bigg{(}\frac{  m^{\tilde{e}}_{n_1}   }{  
m^{\tilde{e}}_{n_2}}\Bigg{)}^{\!\!\!2}\right]  .
\;\;\;\;\;\;\;\;\;
\label{lamloop}
\eeqn
Here $m_Z$ is the $Z$-boson mass and $m^{d/e}_k$ denote the masses
of the down-type quarks/ charged leptons of generation $k=1,2,3$. $M_
1$ and $M_2$ are the soft supersymmetry breaking gaugino mass
parameters, which, together with the weak mixing angle $\theta_W$,
define the photino mass parameter $M_{\tilde {\gamma}}=M_1\cos^2{
\theta _W} + M_2 \sin^2{\theta_W}$. In addition, we have the down-type
squark/charged slepton masses $m^{\tilde{d}/\tilde{e}}_{ n_i}$ of
generation $n$. $i=1,2$ labels the two sfermion mass eigenstates in
each generation $n$. 
The $\phi^{(\tilde{d},\tilde{e})}_n$ are the mixing
angles in the sfermion sector for generation $n$. 
Explicitly (no summation over repeated indices),
\beqn
\tan{2 \phi^{(\tilde{d})}_n} &=& \frac{  2 m^d_n 
\Big|[\bsym{A_D}]_{0nn} - 
{\mu_0^\ast} \tan{\beta}\Big| }{ [\bsym{M_{\widetilde{Q}}^2}]_{nn} -
[\bsym{M_{\widetilde{D}}^2}]_{nn} -\frac{1}{24}(g_Y^2 - 3
g_W^2)({\upsilon_u^2}- {\upsilon_d^2}) }\,,
\label{angleSD}\\
\tan{2 \phi^{(\tilde{e})}_n} &=& \frac{ 2 m^e_n
  \Big|[\bsym{A_E}]_{0nn} - {\mu_0^\ast} \tan{\beta}\Big| }{
  [\bsym{M_{\widetilde{L}}^2}]_{nn} -
  [\bsym{M_{\widetilde{E}}^2}]_{nn} -\frac{1}{8}(3 g_Y^2 - g_W^2)
  ({\upsilon_u^2}-{\upsilon_d^2}) }\,,\label{angleSE} 
\eeqn 
where the $\bsym{A_{D,E}}$\  are the coefficients of the soft
supersymmetry breaking trilinear scalar interactions $[\bsym{A_{
D}}]_{\alpha jk}~ \lam^\prime_{\alpha jk}~\widetilde{L^\alpha
}\widetilde{Q^j} {\widetilde{{D^k}} }^* $ and $\frac{1}{2}[\bsym
{A_{E}}]_{\alpha\beta k}~\lam_{\alpha \beta k}~\widetilde{L^
\alpha}\widetilde{L^\beta}{ \widetilde{{E^k} }}^*$. Here 
$\widetilde{L^\alpha}$ refers to the scalar component of the chiral
superfield $L^\alpha$. The $[\bsym{M_ {\widetilde{...}  }^2}]_{ij}$
are the soft scalar masses squared. $g_Y$, $g_W$ are the $U(1)_Y$ and
$SU(2)_W$ gauge couplings, respectively. 

Assuming all soft supersymmetry breaking mass parameters are $
\mathcal{O}(m_{ 3/2}) > 100 \,\mathrm{GeV}$, and excluding accidental
cancellations, the denominators of Eqs.~(\ref{angleSD}) and
(\ref{angleSE}) are of order $m_{3/2}^2$. For the numerators we get
$2\,m^{d /e}_n m_{3/2}\cdot\mathcal{O}(1 +\eps\tan{\beta})$. Taking
into account the lower limit for $m_{3/2}$ of about $500\,\mathrm{GeV
}$, which originates from the combination of the experimental lower
bound on $\mu_0 \ge 100 \,\mathrm{GeV}$ and its $\eps$-structure $\mu_0
\sim m_{3/2}\cdot\eps$ in our model (see also App.~\ref{vevcond}), we
conclude that even for large $\tan{ \beta}\lesssim 50$ the 
left-right
mixing in (one generation of)  
the down squark and charged slepton sectors is small.  Thus
the sines in Eqs.~(\ref{lamploop}) and (\ref{lamloop}) can be
approximated by tangents.  Furthermore, the logarithms become
$\mathcal{O}(1)$ coefficients if the sfermion masses are
non-degenerate but not too different either, \textit{i.e.} $
\mathcal{O}(1) \lesssim\Big([m^{\tilde{d}/\tilde{e}}_{n_i}]^2
-[m^{\tilde{d}/\tilde{e}}_{n_j}]^2 \Big) \Big/ [m^{\tilde{d} /
\tilde{e}}_{n_j}]^2 \lesssim \mathcal{O}(10)$, where $[m^{\tilde
{d}/ \tilde{e}}_{n_i}] > [m^{\tilde{d}/\tilde{e}}_{n_j}]$. (Once again, $i$
and $j$
label the two mass eigenstates of a particular generation $n$.) We
consider these assumptions natural and apply the corresponding
simplifications to Eqs.~(\ref{lamploop}) and (\ref{lamloop}).
Inserting the $\;\not\!\!\!M_p$ parameters of the last line in
Table~\ref{sequence}, using the phenomenological constraints of
Table~\ref{Table2}, and keeping only leading terms, we obtain the FN
structure of the tree and the loop contributions to the neutrino mass
matrix
\beqn {M_{\mathrm{tree}}^{(\nu)}}_{\,ij}\! &\!\! \sim \!\!& \!
\frac{{m_Z^2} ~ M_{\tilde{\gamma}} ~ \mu_0 ~ \cos^2{\beta}}{{m_Z^2} ~
  M_{\tilde{\gamma}} ~ \sin{2\beta} ~ - ~ M_1 ~ M_2 ~ \mu_0} \cdot
\eps^{-2\Delta^H - \Delta^L_{i1}
  -\Delta^L_{j1} },\\
{M_{{\lambda^\prime} \mbox{-}\mathrm{loop}}^{(\nu)}}_{\,ij}\! &
\!\!\sim\!\! &\! \frac{3}{8 \pi^2} \frac{ {m_b^2} \;
  \Big|[\bsym{A_D}]_{033} - {\mu_0^\ast} \tan{\beta}\Big| \,
  \eps^{2x}}{ [\bsym{M_{\widetilde{Q}}^2}]_{33} -
  [\bsym{M_{\widetilde{D}}^2}]_{33} -\frac{1}{24}(g_Y^2 - 3
    g_W^2)({\upsilon_u^2}-{\upsilon_d^2}) } \cdot \eps^{-2\Delta^H -
  \Delta^L_{i1} -\Delta^L_{j1} } ,
\notag \\
{M_{\,\lambda \,\mbox{-}\mathrm{loop}}^{(\nu)}}_{\,ij} \!&\!\!
\sim\!\! &\! \frac{1}{8 \pi^2} \frac{ {m_\tau^2} \;
  |[\bsym{A_E}]_{033} - {\mu_0^\ast} \tan{\beta}| \, \eps^{2x}}{
  [\bsym{M_{\widetilde{L}}^2}]_{33} -
  [\bsym{M_{\widetilde{E}}^2}]_{33} -\frac{1}{8}(3 g_Y^2 -
  g_W^2)({\upsilon_u^2}-{\upsilon_d^2}) } \cdot \eps^{-2\Delta^H -
  \Delta^L_{i1} -\Delta^L_{j1}} \cdot f_{ij}\,.  \notag 
\eeqn 
Here we have replaced $m^d_3$ by $m_b$ and $m^e_3$ by
$m_\tau$.  The factors $f_{ij}=f_{ji}$ in the last term take care of
$\lambda_{ikn}$'s direct dependence on the charged lepton mass matrix
and its antisymmetry under interchange of the first two indices.
Depending on $i$ and $j$ the tau-stau loop may be forbidden by
symmetry and thus does not give the leading contribution. For
$i,j=1,2$ we find $f_{ij} \sim 1$, whereas $f_{23} \sim \eps^4$ and
$f_{13} \sim f_{33} \sim \eps^8$. See App.~\ref{symconsid} for
details.

Some remarks are in order at this point. Compared to the quark-squark
loop, the charged lepton-slepton loop does not contribute
significantly to the neutrino mass matrix. Therefore we neglect it in
our following discussion. There is a further source of neutrino
masses: The non-renormalizable but $B_3$ conserving superpotential
term $L^i H^U L^j H^U $. In our model, this effective term is
generated via the GM/KN mechanism and thus suppressed by a factor of
$\frac{m_{3/2}}{{M _{ \mathrm{grav}}^2}}\eps^{2z-2\Delta^H-\Delta^L_
{i1}-\Delta^L_{ j1}} $. Inserting the Higgs VEV $\upsilon_u$ for $H^U$
we find that the resulting neutrino mass scale is negligibly small
compared to the tree level contribution in Eq.~(\ref{tree}). The ratio
of the two is of the order $\frac{{m_{3/2}^2}}{{M_{\mathrm{grav }}^2}}
\,(1 + \tan^2{\beta})$. So even for large $\tan{\beta}$ it can be
safely discarded. Similarly, we find that the quark-squark loop
contribution of Eq.~(\ref{lamploop}) is significantly larger than the
mass scale of the non-renormalizable operators $L^i H^U L^j H^U$.






\subsection{\label{massconstraints}Constraints from 
Neutrino Masses}
In our model, we obtain one massive neutrino at tree level. A second
non-zero mass is supplied by the quark-squark loop. Notice that except
for an overall relative factor the $\eps$-structure of the tree-level
and one-loop matrices is exactly the same. However, they are not
aligned in the sense that one matrix is a (real or complex) multiple
of the other. The $\mu_i$ and the $\lam^\prime_{i33}$ have a
completely different origin, \ie\ the ${\cal O}(1)$ coefficients are
in general different. Adding the two terms, we therefore expect not
one but two non-zero masses. One neutrino remains massless
since $\bsym{M_{\mathrm{ tree}}^{( \nu)}}$ and
$\bsym{M_{{\lam^\prime}\mbox{ -} \mathrm{loop}} ^{(\nu)}}$ are both
rank one matrices.\footnote{ For the loop contribution this statement
relies on the fact that the $\not\!\!\!M_p$ coupling constants are
generated via the canonicalization of the K\"ahler potential and thus
proportional to the down-type quark mass matrix, 
\textit{cf.}  Eq.~(\ref{rpvbyCK}).} 
Hence a degenerate neutrino scenario is excluded. Notice that this
remains true even if we include the charged lepton-slepton loop
contribution: The resulting third non-zero mass is smaller by a factor
of $\frac{{m_\tau^2}}{3 {m_b^2}}\approx \frac{1}{15}$ compared to the
quark-squark loop mass. This is inconsistent with the degenerate
neutrino mass solution of Eq.~\ref{nu-solns}.

In order to see whether our model is compatible with the hierarchical
or the inverse-hierarchical neutrino solutions of Eq.~\ref{nu-solns},
we calculate the relative factor $\frac{m^{\mathrm{tree}}}{m^{\mathrm
    {loop}}}$ between the overall scales of the tree and the loop mass
matrix. This factor must not come out larger than the experimental
ratio of the atmospheric and solar neutrino mass scales, which is
approximately $5$. First, we do a rough estimate for $\tan{\beta}
\lesssim 2$, that is $\,x=2,3$ (thus $\cos{\beta} \gtrsim 0.5$), where
we assume all soft breaking parameters, even the gaugino masses $M_1$
and $M_2$, to be of the same order $\mathcal{O}(m_{3/2})$. Neglecting
the first term in the denominator of the tree level as well as the
second term in the numerator of the loop level overall mass scale, we
arrive at $\frac{m^{ \mathrm{tree}}}{m^{\mathrm{loop}}} \sim \frac{8
  \pi^2}{3} \cos^2{\!\beta} \: \frac{{m_Z^2}}{{m_b^2}} \, \eps^{-2x}$.
This is much too large for $x \ge 2$, so we are restricted to the
cases with $x=0,1$, {\it i.e.} $\tan{\beta}\gtrsim 8$.  We can then
approximate $\cos{\beta}$ by $\cot{\!\beta} \sim \eps^{-x}
\frac{m_b}{m_t}$.  Neglecting again the first term in the denominator
of the tree level overall mass scale, we get
\beqn
\frac{m^{\mathrm{tree}}}{m^{\mathrm{loop}}} & \sim & \eps^{-4x}~ 
\frac{8 \pi^2~{m_Z^2}}{3{m_t^2}}\!  \cdot\!  \frac{M_{\tilde{\gamma}}}{
M_1~M_2}\! 
\cdot\! \frac{ [\bsym{M_{\widetilde{Q}}^2}]_{33} - [\bsym{
M_{\widetilde{D}}^2}]_{33} -
\frac{1}{24}(g_Y^2 - 3 g_W^2)({\upsilon_u^2}-{\upsilon_d^2}) }
{\Big|[\bsym{A_D}]_{033} - {\mu_0^\ast} \, ~ \eps^{x} \,~ 
\frac{m_t}{m_b} \Big|}\,. \notag \\ \label{x=0}
\eeqn
Note that we have replaced $\tan{\beta}$ in the denominator of the
last factor. The second factor, $\frac{8\pi^2~{m_Z^2}}{3{m_t^2}}
\approx7$. Taking $x=1$ requires the product of the last two factors
to yield a tiny fraction of their natural value of about~$1$.  Such a
scenario, where there is either fine tuning in the scalar masses or
the gaugino masses are about $1000$ times larger than the scalar
masses is very unnatural. We therefore reject this case and focus on
$x=0$. This together with $z=1$ numerically determines the expansion
parameter $\eps \equiv \frac{\langle A \rangle}{M_{ \mathrm{grav}}} =
0.186$, see Ref.~\cite{Dreiner:2003yr}.\footnote{The highest exponent of this expansion parameter 
occurs in the up-type quark mass
matrix. Since $\left( \frac{0.186}{0.22} \right)^8 \approx \frac{1}{4}$, it is unproblematic to 
work with $\eps=0.186$.} 
Notice that
with $x=0$ it seems reasonable to assume that the denominator of the
last term in Eq.~(\ref{x=0}) is now dominated by the second term.
Taking the gaugino masses at a common scale $M_{1/2}$, the scalar mass
parameters all of $\mathcal{O}(m_{3/2})$, we can simplify
Eq.~(\ref{x=0})
\beqn
\frac{m^{\mathrm{tree}}}{m^{\mathrm{loop}}} & \sim & \frac{8
  \pi^2{m_Z^2} m_b}{3{m_t^3}} \cdot \eps^{-z} \cdot
\frac{m_{3/2}}{M_{1/2}}\; \sim \; \mathcal{O}(1) \:
\frac{m_{3/2}}{M_{1/2}}\,.
\label{tree/loop}
\eeqn
Thus if we choose supersymmetric parameter points for which the scalar
quark masses are bigger than the gaugino mass parameters by factors of
about two to five ($m_{3/2}\approx5M_{1/2}$) \cite{{Ghodbane:2002kg},
{Allanach:2002nj}}, we can accommodate a hierarchical neutrino mass
scenario. The tree level contribution then provides for one relatively
heavy neutrino while the second neutrino remains light. On the other
hand, an inverse hierarchy is possible just as well. Then the tree and
the quark-squark loop mass matrices must have the same order of
magnitude, thus generating two relatively heavy neutrinos while the
third neutrino remains light.  Due to our ignorance of the soft
breaking sector and the arbitrariness of all $\mathcal{O}(1)$
coefficients, our $B_3$-conserving FN models allow both, the
hierarchical and the inverse-hierarchical neutrino scenario.

In both cases however, the mass of the heaviest neutrino is given by
the atmospheric neutrino mass scale $\sqrt{|\Delta m_{32}^2|}$. Thus
the integer parameter $\Delta^H$ can be determined. Equating the
eigenvalue of the tree level neutrino mass matrix, which is
proportional to $\sum_i \frac{{\mu_i^2}}{{\mu_0^2}}$, with $\sqrt{|
  \Delta m_{32}^2|}$ and putting $M_{1/2} =\mathcal{O}(m_{3/2})$
yields
\beqn
- 2 \Delta^H &\sim& \frac{1}{\ln{\eps}}  
\cdot \ln{\frac{{m_t^2} \, m_{3/2} \, \sqrt{|\Delta m_{32}^2|} }{ 
{m_b^2} \,{m_Z^2}}}\,.
\eeqn
Here we made use of the ordering $X_{L^3} \leq X_{L^2} \leq X_{L^1}$,
so that $\sum_{i} \eps^{-2 \Delta^L_{i1}} \sim 1$. Inserting $\eps =
0.186$, $m_t = 175 \,\mathrm{GeV}$, $m_b = 4.2 \,\mathrm{GeV}$, $m_Z =
91.2 \,\mathrm{GeV}$, $\sqrt{|\Delta m_{32}^2|} = 0.047 \,\mathrm{eV}$,
and $1000 \,\mathrm{GeV} \geq m_{3/2} \geq 100 \,\mathrm{GeV}$ we obtain
\beqn
- 2 \Delta^H & \in & [\,11.0\,,\,12.3\,]\,.
\label{interval}
\eeqn
Here the lower bound corresponds to $m_{3/2} = 1000 \,\mathrm{GeV}$ and 
the upper one to $m_{3/2} = 100 \,\mathrm{GeV}$. Since $\Delta^H$ 
is integer, we end up with the single option
\beqn
\Delta^H & = & -~6\,.
\label{deltaH}
\eeqn
At the end of App.~\ref{vevcond}, we argue that the sequence of basis
transformations in Sect.~\ref{bases} generates $\;\not\!\!M_p$
coupling constants which are to some extent larger than expected.
Taking this feature into account, the interval in Eq.~(\ref{interval})
is shifted slightly to higher values.  For $\mu_i \sim \eps^{- 0.5}
\cdot m_{3/2} \, \eps^{-X_{L^i} - X_{H^U}}$, where the first factor
quantifies such a systematic effect, this shift is about one unit. So
the solution given in Eq.~(\ref{deltaH}) remains stable.






\subsection{\label{mixingconstraints}Constraints from 
Neutrino Mixing}
We now turn to the conditions on the $X$-charges imposed by the MNS
matrix. The effective neutrino mass matrix of Eq.~(\ref{MnuL}) is
diagonalized by the unitary transformation ${{U}^{(\nu')}}
_{ij}\sim\eps^{|X_{L^i} - X_{L^j}|}$. This transforms the current 
eigenstate basis into the mass eigenstate basis $(\nu_1^{\,\prime},
\nu_2^{\,\prime},\nu_3^{\,\prime})$ of $\bsym{M^{(\nu)}}^\dagger
\bsym{M^{(\nu)}}$. In the latter basis we denote the diagonal entries
of the mass matrix as $(m_1',m_2',m_3')$, with relative values $m_3'
\ll m_2'\lesssim m_1'$ (for details see App.~\ref{diagonalizeM}). 
It is important to note that $\bsym{{U}^{(\nu^{\,\prime})}}$ is 
different from $\bsym{U^{(\nu)}}$ defined in
Eq.~(\ref{nudiag}). In
order to compare with the possible solutions in Eq.~(\ref{nu-solns})
and the data of Eq.~(\ref{exp-results}) it is more convenient to
reorder the basis $(\nu_1^{\,\prime},\nu_2^{\,\prime},\nu_3^{\,\prime
})$ into a new basis $(\nu_1,\nu_2,\nu_3)$, with the corresponding
masses $(m_1, m_2,m_3)$ in the order of the hierarchical or inverted
hierarchical solution. We can then fix the mixing angles so that
$(\nu_1,\nu_2)$ solve the solar neutrino problem. We summarize the
bases choices in the following table
\begin{table}[ht]
\hspace{3.cm} \begin{tabular}{|c|c|c|c|}
\hline  
       \multicolumn{2}{|c|}{Mass Ordering} & \ Hierarchy\ \  &  
$\phantom{\Big|}$Inverse Hierarchy$\phantom{\Big|}$ \\
\hline \ Heaviest\ \    & $\;\nu_1^{\,\prime},\,m_1'\;$ & $\nu_3,\,m_3$ & 
$\phantom{\Big|}\nu_2,\,m_2\phantom{\Big|}$ \\
\hline Medium     & $\nu_2^{\,\prime},\,m_2'$ & $\nu_2,\,m_2$ & 
$\phantom{\Big|}\nu_1,\,m_1\phantom{\Big|}$ \\
\hline Lightest   & $\;\nu_3^{\,\prime},\,m_3'\;$ & $\nu_1,\,m_1$ & 
$\phantom{\Big|}\nu_3,\,m_3\phantom{\Big|}$\\ \hline
\end{tabular}
\caption{Options for the mass ordering of the neutrinos.} 
\end{table}

For the hierarchical scenario, $m_1$ must be the lightest and $m_3$
the heaviest neutrino mass. We must therefore exchange the first and
third states in the primed basis to obtain the relevant unprimed
basis. The new diagonalization matrix is then given by
\beqn
\bsym{{U}^{(\nu,\,\mathrm{h.})}} &\equiv & \bsym{T^{\mathrm{(h.)}}} 
\cdot \bsym{{U}^{(\nu^{\,\prime})}}\,,\label{MNShier}
\eeqn
where
\beqn
\bsym{T^{\mathrm{(h.)}}} &\equiv & \begin{pmatrix} 0&0&1 \\ 0&1&0 \\ 
1&0&0 \end{pmatrix}\,.\label{transpoH.}
\eeqn
Here the superscript \textbf{h.} refers to the hierarchical
solution. Combining Eqs.~(\ref{MNSdef}), (\ref{MNSexpstr}), 
(\ref{MNShier}), and (\ref{transpoH.}) we get
\beqn
\eps^{|L^i-L^j|} 
~\sim ~ \big[ \bsym {U^{(\nu^{\,\prime})}}\big]_{ij} 
&=& \big[ \bsym {T^{(\mathrm{h.})} \cdot U^{(\nu,\, \mathrm{h.})}}\big]_{ij}
\label{epsHier} \\
&=& \big[ \bsym {T^{(\mathrm{h.})} \cdot {U^{\mathrm{MNS}}}}^\dagger \big]_{ij}
~\sim ~ \begin{pmatrix}    
\eps^{0,1,2,...} & \eps^{0,1} & \eps^{0,1} \\  
\eps^{0,1} & \eps^{0,1} & \eps^{0,1} \\ \eps^{0,1} & \eps^{0,1,2} & 
\eps^{0,1,2}   \end{pmatrix}_{\!\!ij}\,, \notag
\eeqn
which restricts $\Delta^L_{i1}\equiv X_{L^i}-X_{L^1}$, for
$i=2,3$. All acceptable combinations which also comply with the
ordering $\Delta^ L_{31} \le \Delta^L_{21} \le 0$ are listed in
Table~\ref{deltas}. As we impose conservation of $B_3$, the second
condition in Eq.~(\ref{B_3condsa}) has to be satisfied, \ie $\!\, $
$\Delta^L_{31} +\Delta^L_{21}-z\not= 0\,{\rm mod}\,3$. The last column
of Table~\ref{deltas} shows which cases are compatible with $B_3$
conservation for $z=1$.
\begin{table}[t]
\hspace{1.4cm} \begin{tabular}{||c|c||c|c|c||} 
\hline \hline  $\Delta^L_{31}$ & $\Delta^L_{21}$  & Hierarchy & 
Inverse Hierarchy & Conservation of $B_3$\\ 
\hline \hline  $-1 $ & $-1 $ & yes & no & no \\
\hline   $-1 $ & $ \phantom{-}0 $ & yes & yes & yes \\
\hline   $ \phantom{-}0 $ & $ \phantom{-}0 $ & yes & yes & yes \\
\hline \hline
\end{tabular} \caption{\small All combinations of $\Delta^L_{i1}$ which are 
compatible with the experimental MNS matrix for the hierarchical and
the inverse-hierarchical neutrino scenario. In addition, the condition
of $B_3$ conservation on the $X$-charges as stated in
Eq.~(\ref{B_3condsa}) is checked.}\label{deltas}
\end{table}

In the case of an inverse hierarchy, $m_3\ll m_1\lesssim m_2$, we
need to interchange the first two states of the primed basis to obtain
the relevant unprimed basis. We have 
\beqn
\bsym{T^{\mathrm{(i.h.)}}} &\equiv & \begin{pmatrix} 0&1&0 \\ 1&0&0 \\ 
0&0&1 \end{pmatrix}.
\eeqn
With this we find 
\beqn
\eps^{|L^i-L^j|} 
~\sim ~ \big[ \bsym {U^{(\nu^{\,\prime})}}\big]_{ij} 
&=& \big[ \bsym {T^{(\mathrm{i.h.})} \cdot U^{(\nu,\,
    \mathrm{i.h.})}}\big]_{ij} \label{epsInv} \\
&=& \big[ \bsym {T^{(\mathrm{i.h.})} \cdot {U^{\mathrm{MNS}}}}^\dagger
\big]_{ij} 
~\sim ~\begin{pmatrix}\eps^{0,1} & \eps^{0,1} & \eps^{0,1} \\  
\eps^{0,1} & \eps^{0,1,2} & \eps^{0,1,2} \\ \eps^{0,1,2,...} & \eps^{0,1} & 
\eps^{0,1}   \end{pmatrix}_{\!\!ij} \,, \notag
\eeqn
for the inverse-hierarchical neutrino scenario. 
Again, the allowed $\Delta^L_{i1}$ are given in Table~\ref{deltas}. In
addition to the constraints arising from the experimental MNS matrix,
we now have to ensure that the ratio $\frac{m_1^{\,\prime}}{m_2^{\,
\prime}}$ of the two heavy masses is of order one. As shown in 
App.~\ref{diagonalizeM}, $m_2^{\,\prime}$ is not only determined by
the scale of the second largest contribution to the neutrino mass
matrix, but it is additionally suppressed by a factor of $\eps^{-2
\Delta^L_{21}}$, \textit{cf.}  Eq.~(\ref{frakm_2}). For this reason 
$\Delta^L_{21}= -1$ is forbidden in the case of an inverse hierarchy.





\cleqn
\section{\label{viableSets}Viable $\bsym{X}$-Charge
 Assignments}
In summary, we have fixed almost all parameters determining the FN
charges by imposing conservation of $B_3$, requiring GS anomaly
cancellation and finally taking into account the phenomenological
constraints of the low-energy fermionic mass spectrum, including the
neutrinos. Starting with Table~\ref{Table2}, we need $z=1$ if the
bilinear superpotential terms are to be generated via the
Giudice-Masiero/Kim-Nilles mechanism. Then we have $x=0$ due to the
upper limit on $\frac{m^{\mathrm{tree}}}{m^{\mathrm {loop}}}$, which
is given by the ratio of the atmospheric and the solar neutrino mass
scales. $\Delta^H$ is fixed through the absolute neutrino mass scale.
As degenerate neutrinos are excluded, this corresponds to the
atmospheric mass scale. Hence we find $\Delta^H = -6$.  Finally, the
constraints coming from the MNS mixing matrix together with the
requirement of $B_3$ conservation yield $\Delta^L_{21} = 0$ and
$\Delta^L_{31}= -1,0$ (see Table~\ref{deltas}). So, in the end we are
left with only the choice of
\begin{equation}
y=-1,0,1\,,\qquad \mathrm{and} \qquad 3\zeta + b \equiv
\Delta^L_{31}+\Delta^L_{21}-z = \Delta^L_{31} - 1 = -2,-1\,.
\end{equation}
This leads to six sets of viable $X$-charge assignments displayed in
Table~\ref{allsets}. All sets are compatible with either a
hierarchical or an inverse-hierarchical neutrino scenario, depending
on the ratio $\frac{m^{\mathrm{tree}}}{m^{\mathrm{loop}}}$ [\textit{cf.}
Eq.~(\ref{tree/loop})] and unknown $\mathcal{O}(1)$ coefficients in
$\bsym{M^{(\nu)} }$. Taking the smallness of the $(1,3)$-element of
the MNS matrix in Eq.~(\ref{MNSexpstr}) [corresponding to the
$(1,1)$-entry of Eq.~(\ref{epsHier}) in the hierarchical, and the
$(3,1)$-entry of Eq.~(\ref{epsInv}) in the inverse-hierarchical case]
as a crucial criterion, we prefer the inverse-hierarchical cases with
$\Delta^L_{31}=-1$. It is only there, that the FN prediction for this
entry is of $\mathcal{O}(\eps)$. In all other cases we have to assume
an unattractively small ``$\mathcal{O}(1)$ coefficient''.
\begin{table}[tp]
\hspace{0.5cm} 
\begin{tabular}{||c|c|c||c|c|c|c|c|c|c|c||} 
\hline \hline \multicolumn{3}{||c||}{$\phantom{\Big|}$  
{\bf Input}$\phantom{\Big|}$  } & \multicolumn{8}{|c||}{\bf Output} \\
\hline \hline $\bsym{\Delta^L_{31}}$ & $\bsym{3\zeta + b }$ 
& $\bsym{y}$ & 
$\bsym{X_{H^D}}$ & $\bsym{X_{H^U}}$&  
$\bsym{i}$ & $\bsym{X_{Q^i}}$ & $\bsym{X_{\ol{U^i}}}$ & 
$\bsym{X_{\ol{D^i}}}$ & 
$\bsym{X_{L^i}}$ & \rule[-3mm]{0mm}{8mm}
$\bsym{X_{\ol{E^i}}}$ \\ 
\hline \hline &&&&&
$1$ & $\frac{467}{105}$ & $\frac{722}{105}$ & $-\frac{97}{35}$ & 
$-\frac{386}{105}$ & \rule[-3mm]{0mm}{8mm}$\frac{667}{105}$ \\
\cline{6-11} $-1\phantom{-}$ & $-2$ & $-1\phantom{-}$ & 
$\frac{244}{105}$ & $-\frac{349}{105}$ &
$2$ & $\frac{467}{105}$ & $\frac{302}{105}$ & $-\frac{167}{35}$ 
& $-\frac{386}{105}$ & \rule[-3mm]{0mm}{8mm}$\frac{352}{105}$ \\ 
\cline{6-11} &&&&&
$3$ & $\frac{257}{105}$ & $\frac{92}{105}$ & $-\frac{167}{35}$ & 
$-\frac{491}{105}$ & \rule[-3mm]{0mm}{8mm}$\frac{247}{105}$ \\
\hline \hline &&&&&
$1$ & $\frac{177}{35}$ & $\frac{676}{105}$ & $-\frac{373}{105}$ & 
$-\frac{368}{105}$ & \rule[-3mm]{0mm}{8mm}$\frac{631}{105}$ \\
\cline{6-11} $-1\phantom{-}$ & $-2$ & $0$ & $\frac{262}{105}$ & 
$-\frac{367}{105}$ &
$2$ & $\frac{142}{35}$ & $\frac{361}{105}$ & $-\frac{478}{105}$ & 
$-\frac{368}{105}$ & \rule[-3mm]{0mm}{8mm}$\frac{316}{105}$ \\ 
\cline{6-11} &&&&&
$3$ & $\frac{72}{35}$ & $\frac{151}{105}$ & $-\frac{478}{105}$ & 
$-\frac{473}{105}$ & \rule[-3mm]{0mm}{8mm}$\frac{211}{105}$ \\
\hline \hline  &&&&&
$1$ & $\frac{17}{3}$ & $6$ & $-\frac{13}{3}$ & $-\frac{10}{3}$ 
& \rule[-3mm]{0mm}{8mm}$\frac{17}{3}$ \\
\cline{6-11} $-1\phantom{-}$ & $-2$ & $1$ & $\frac{8}{3}$ & 
$-\frac{11}{3}$ &
$2$ & $\frac{11}{3}$ & $4$ & $-\frac{13}{3}$ & $-\frac{10}{3}
$ & \rule[-3mm]{0mm}{8mm}$\frac{8}{3}$ \\ 
\cline{6-11} &&&&&
$3$ & $\frac{5}{3}$ & $2$ & $-\frac{13}{3}$ & $-\frac{13}{3}$ 
& \rule[-3mm]{0mm}{8mm}$\frac{5}{3}$ \\
\hline \hline &&&&&
$1$ & $\frac{458}{105}$ & $\frac{241}{35}$ & $-\frac{274}{105}$ 
& $-\frac{394}{105}$ & \rule[-3mm]{0mm}{8mm}$\frac{683}{105}$ \\
\cline{6-11} $0$ & $-1$ & $-1\phantom{-}$ & $\frac{236}{105}$ & 
$-\frac{341}{105}$ &
$2$ & $\frac{458}{105}$ & $\frac{101}{35}$ & $-\frac{484}{105}$ 
& $-\frac{394}{105}$ & \rule[-3mm]{0mm}{8mm}$\frac{368}{105}$ \\ 
\cline{6-11} &&&&&
$3$ & $\frac{248}{105}$ & $\frac{31}{35}$ & $-\frac{484}{105}$ 
& $-\frac{394}{105}$ & \rule[-3mm]{0mm}{8mm}$\frac{158}{105}$ \\
\hline \hline  &&&&&
$1$ & $\frac{174}{35}$ & $\frac{677}{105}$ & $-\frac{356}{105}$ 
& $-\frac{376}{105}$ & \rule[-3mm]{0mm}{8mm}$\frac{647}{105}$ \\
\cline{6-11} $0$ & $-1$ & $0$ & $\frac{254}{105}$ & $-\frac{359}{105}$ &
$2$ & $\frac{139}{35}$ & $\frac{362}{105}$ & $-\frac{461}{105}$ 
& $-\frac{376}{105}$ & \rule[-3mm]{0mm}{8mm}$\frac{332}{105}$ \\ 
\cline{6-11} &&&&&
$3$ & $\frac{69}{35}$ & $\frac{152}{105}$ & $-\frac{461}{105}$ 
& $-\frac{376}{105}$ & \rule[-3mm]{0mm}{8mm}$\frac{122}{105}$ \\
\hline \hline  &&&&&
$1$ & $\frac{586}{105}$ & $\frac{631}{105}$ & $-\frac{146}{35}$ 
& $-\frac{358}{105}$ & \rule[-3mm]{0mm}{8mm}$\frac{611}{105}$ \\
\cline{6-11} $0$ & $-1$ & $1$ & $\frac{272}{105}$ & $-\frac{377}{105}$ &
$2$ & $\frac{376}{105}$ & $\frac{421}{105}$ & $-\frac{146}{35}$ 
& $-\frac{358}{105}$ & \rule[-3mm]{0mm}{8mm}$\frac{296}{105}$ \\ 
\cline{6-11} &&&&&
$3$ & $\frac{166}{105}$ & $\frac{211}{105}$ & $-\frac{146}{35}$ 
& $-\frac{358}{105}$ & \rule[-3mm]{0mm}{8mm}$\frac{86}{105}$ \\
\hline \hline
\end{tabular} \caption{\small All six sets of viable $X$-charge assignments, 
  where $z=1$, $x=0$ (\ie $\,$large $\tan{\beta}$), and $\Delta^H =
  -6$. The other input parameters of Table~\ref{Table2}, namely
  $\Delta^L_{31}$, $3\zeta + b$, and $y$, differentiate between the
  various possible scenarios. All of them are compatible with
  hierarchical and inverse-hierarchical neutrino masses, depending on
  the ratio $\frac{m^ {\mathrm{tree}}}{m^{\mathrm{loop}}}$ and unknown
  $\mathcal{O}(1)$ coefficients in $\bsym{M^{(\nu)}}$. The former
  depends on the parameters of supersymmetry breaking. Here we assume
  gravity mediation so that all soft breaking mass parameters are of
  $\mathcal{O}(m _{3/2})$, with $m_{3/2} \in [100
  \mathrm{\,GeV} , 1000 \mathrm{\,GeV}]$. In order to determine the
  structure of the sneutrino VEVs, we have assumed an FN structure for
  $b_\alpha$ and $[\bsym{M^2_{\widetilde{L}}}]_{\alpha \beta}$, see
  App.~{\ref{vevcond}}.}\label{allsets}
\end{table}
Remarkably, there exists one set where all FN charges are multiples of
one third. This salient charge assignment is obtained for $3\zeta + b
= -2$ (or equivalently $\Delta^L_{31} = -1$) and $y=1$.  However, as
$y\neq 0$ the CKM matrix is not optimal (\textit{cf.} App. 
\ref{top-down}),  
but nonetheless acceptable due to the possibility of mildly adjusting the
unknown $\mathcal{O}(1)$ coefficients.

All other sets contain highly fractional
$X$-charges (just like the sets in Ref.~\cite{Dreiner:2003yr}) and are
thus ``esthetically disfavored''.  However, requiring that the FN
scenario is in agreement with the very tight experimental bounds on
exotic processes usually leads to highly fractional $X$-charge
assignments \cite{Dreiner:2003yr, {Dreiner:2003hw}}. Thus the
six models presented in this section are so-to-speak in good 
company; our ignorance
of the origin of the $U(1)_X$ gauge symmetry does not allow us to exclude
models just because of unpleasant $X$-charges. Due to our experience 
with hypercharge, it is natural to hope for  "nice", \emph{i.e.} not 
too fractional, $X$-charges. 
But actually, in string models ({\it e.g.} \cite{Lebedev:2006kn}) 
the anomalous $U(1)$-charges can very well be 
highly fractional \cite{akin}.
It is therefore not clear at 
all whether one should expect "nice" charges or not.

In the
manner of Ref.~\cite{Dreiner:2003hw}, we checked that the
$\not\!\!M_p$ coupling constants which are produced by the six sets of
$X$-charges are all in agreement with the very tight experimental
bounds \cite{Allanach:1999ic}, unless there is an unnatural adding-up
among the $\mathcal{O}(1)$ coefficients.  In App.~\ref{top-down}, we
give an explicit example of how the physics at the high energy scale,
constrained by the third $X$-charge assignment in Table~\ref{allsets},
boils down to a viable low-energy phenomenology.
 
Finally, one could raise the question: Is it possible to construct a
scenario where no hidden sector fields are needed to cancel the
$\mathcal{A}_{CCX}$ and $\mathcal{A}_{GGX}$ anomalies? Explicitly
\cite{Dreiner:2003yr}
\beqn
\mathcal{A}_{GGX} &=&2X_{H^U}+2X_{H^D}+
\sum_{i}\Big(6{X_{Q^i}}+3{X_{\ol{U^i}}}+
3{X_{\ol{D^i}}}+2{X_{L^i}}+{X_{\ol{E^i}}}\Big)
\nonumber\\
&~&\qquad +X_A+
\mcal{A}_{GGX}^{\mathrm{hidden~sector}}\,,\\ [3mm]
\mcal{A}_{CCX}&=&\frac{1}{2}\Big[\sum_{i}
\Big(2~X_{Q^i}+X_{\ol{U^i}}+X_{\ol{D^i}}\Big) \Big].  
\eeqn 
Inserting the relations of Table~\ref{Table2} with $z=1$ and $x=0$
yields
\beqn
\mathcal{A}_{GGX} = (3\zeta + b) +3\Delta^H + 68 +
\mathcal{A}_{GGX}^{\mathrm{hidden~sector}} ~~~&\mathrm{and}&~~~
\mcal{A}_{CCX}=\frac{21}{2}.  
\eeqn 
Anomaly cancellation \`a la Green-Schwarz requires
\cite{Dreiner:2003yr} $\frac{\mathcal{A}_{CCX}}{k_C}=\frac{\mathcal
  {A}_{GGX}}{24}$, where $k_C$ is the {\it positive integer}
Ka\v{c}-Moody level of $SU(3)_C$.  Assuming that the hidden sector
fields are uncharged under $U(1)_X$, \ie\ $\mathcal{A}_{GGX}^{
  \mathrm{hidden~sector}} = 0$, we arrive at the condition
\beqn 
\frac{2\cdot 2\cdot 3\cdot 3\cdot 7}{k_C} &=& (3\zeta
+ b) + 3\Delta^H + 68\,.\label{kac} 
\eeqn 
As both sides of this equation have to be integer, $k_C$ is restricted
to a product of a subset of the primes in the numerator on the
left. With $3\zeta +b = -2, -1$ and $\Delta^H = -6$ the right-hand
side of Eq.~(\ref{kac}) is either $48$ or $49$. This is not attainable
with the left-hand side, even if we allow a variation of $\pm1$ in
$\Delta^H$. This shows that $U(1)_X$-charged hidden sector fields are
necessary to cancel the gravity-gravity-$U(1)_X$ anomaly.





\section{\label{summarY}Summary, Conclusion  and
 Outlook}
We have constructed a minimalist and compact $U(1)_X$ Froggatt-Nielsen
scenario with the MSSM particle content plus one additional flavon
field $A$.  Furthermore, our model exhibits only two mass scales,
$M_{\mathrm{grav}}$ and $m_{3/2}$. A discrete symmetry is needed to
ensure a long-lived proton. Without enlarging the (low-energy)
fermionic particle content of the MSSM and excluding a GS mechanism,
there are only three discrete symmetries
\cite{{Ibanez:1991hv},{Ibanez:1991pr}, Dreiner:2005rd} which, besides
allowing for neutrino masses, can originate from an anomaly-free gauge
symmetry and thus do not experience violation by quantum gravity
effects.  These salient discrete symmetries are $M_p$, $B_3$, and
$P_6$.  Following the philosophy of \cite{Dreiner:2003yr}, where $M_p$
is a remnant of the continuous $U(1)_X$ symmetry, we have examined the
case with $B_3$ being generated by virtue of the $X$-charge
assignment.  This $\mathbb{Z}_{3}$-symmetry has some attractive
features: First, it phenomenologically stabilizes the proton. Second,
it allows bilinear and trilinear $\not\!\!M_p$ coupling constants, so
that neutrino masses are possible at the renormalizable level without
the need to introduce right-handed neutrinos.  Imposing the
restrictions of the measured fermionic mass spectrum and the GS
anomaly cancellation conditions we arrive at six phenomenologically
viable sets of $X$-charges presented in Table~\ref{allsets}.  All of
them feature large $\tan{\beta}$ ($\gtrsim~40$). Our ignorance about
the details of the soft supersymmetry breaking parameters does not
allow us to distinguish between models of normal and inverse neutrino
mass hierarchy. However, taking the smallness of
${U^{\mathrm{MNS}}}_{13}$ as a crucial criterion, we should prefer the
first three cases (\ie$\;$those with $\Delta^L_{31} = -1$) of
Table~\ref{allsets} and an inverse hierarchy. Doing so, our model {\it
  predicts} inverse-hierarchical neutrino masses. Of all six
possibilities, we find the third $X$-charge assignment of
Table~\ref{allsets} the most pleasing: All $X$-charges are integer
multiples of one third in this case. However, the other five models (with more
fractional $X$-charges) are 
phenomenologically possible just as well.

In constructing viable models of the fermionic mass spectrum we have
been guided by the principle of minimality and compactness. With only
the $U(1)_X$ symmetry and two mass scales at hand, we had to exclude
the choice $z=0$ right from the beginning as it does not
satisfactorily explain the origin of the $\mu$-parameter. However, the
quest for a dark matter candidate requires us to introduce at least
one additional particle like {\it e.g.} the axion, which in turn would
suggest the existence of a new global $U(1)$ symmetry
\cite{Peccei:1977hh}. Also superstring models often predict more than
just one $U(1)$. So it is tempting to assume that the $\mu$-term is
originally forbidden by such a symmetry. Effectively it may then be
generated via some mechanism (other than GM) at the phenomenologically
needed mass scale \cite{{Cvetic:1997ky},{Keith:1997zb},
  {Cleaver:1997nj}, {Martin:1999hc}, {Chamseddine:1996rc}}.
In that case, the possibility of $z=0$ should be considered and
investigated seriously.

Allowing for supersymmetric zeros in the leptonic mass matrices could be 
another direction of further study. 
In this paper, we have excluded the existence of supersymmetric zeros in 
all Yukawa mass matrices as 
we wanted to evade difficulties like those encountered with the CKM matrix. 
However, maybe they are a 
blessing for the MNS matrix. It would be interesting to examine if a small 
$\theta_{13}$ can 
naturally arise from supersymmetric zeros appearing in the charged lepton 
and/or neutrino mass matrix, 
see {\it e.g.} \cite{Psallidas:2005tr}.





\section*{Acknowledgments}

We are grateful for helpful discussions with and questions and/or
comments from D. Bartsch, S.~Davidson, \mbox{J.-M.}~Fr\`ere,
A.~Ibarra, J.~S.~Kim, 
S.~Nasri, P.~Ramond, 
G.~Weiglein and A.~Wingerter.
C.L. thanks the SPhT at the
CEA-Saclay and M.T. the Physikalisches Institut, Bonn, for
hospitality.  M.T. greatly appreciates that he was supported by a
fellowship within the postdoc-program of the German Academic
Exchange Service (Deutscher Akademischer Austauschdienst, DAAD) during
the early stage of this work, while he was funded by a Feodor Lynen
fellowship of the Alexander-von-Humboldt-Foundation at the end of this
project. This project is partially supported by the RTN European
Program MRTN-CT-2004-503369.  The work of H.M. was supported in part
by the DOE under the contract DE-AC03-76SF00098 and in part by NSF
grants PHY-0098840 and PHY-04-57315.


\begin{appendix}



\cleqn

~\\
\\
\noindent{\LARGE \bf Appendix}

\section{\label{bpb3-app} Baryon Parity vs. 
Baryon Triality}
In this appendix we would like to emphasize the differences between
baryon parity and baryon triality. Baryon parity, $B_p$, is defined
on the MSSM chiral superfields by \beqn \{H^D, H^U,L^i, \ol{E^i}\}
&\longrightarrow& \phantom{e^{2\pi i/2}~} \{H^D, H^U,L^i,
\ol{E^i}\}\,,\nonumber\\
\{Q^i,\ol{U^i},\ol{D^i}\} &\longrightarrow& e^{2\pi
  \mathrm{i}/2}~\{Q^i,\ol{U^i},\ol{D^i}\}\,.  \eeqn
The total $B_p$-charge of an operator $R$ [\textit{cf.} Eq.~(\ref{R})]
can then be written as
\beqn\label{bbpp} 
\sum_i n_{Q^i}~+~\sum_in_{\ol{U^i}}~ +~\sum_i
n_{\ol{D^i}}&=&2~\mathcal{I}_B^{\,\prime}+ ~\iota_B^{\,\prime}\,.
\eeqn
Here $\mathcal{I}_B^{\,\prime}$ is an integer, which can differ for
each operator. $\iota_B^{\, \prime}$ is fixed for all operators and is
0 or 1 if $B_p$ is conserved or broken, respectively. In order to
achieve $B_p$ by virtue of the $X $-charges, we need 
Eq.~(\ref{cond0}) as well as 
\begin{equation}
X_{H^D}-X_{L^1}=\mbox{ integer},\qquad \mbox{and}\qquad  3X_{Q^1}+X
_{L^1}=\mbox{integer}-\frac{b^{\,\prime}} {2}\,, 
\end{equation}
with $b^{\,\prime} =1$. Plugging this into Eq.~(\ref{x=i+f}), we get
\beq
X_{\mathrm{total }}=\mbox{integer}-\mathcal{C}\cdot\frac{b^{\,\prime}}{2}\,. 
\eeq
In analogy to Eq.~(\ref{alfa1}) we have $\iota_B^{\,\prime}+\mathcal{C}=2
\left(\sum_i n_{Q^i} -\mathcal{C}-\mathcal{I}_B^{\,\prime} \right)$
[which is obtained by adding Eqs.~(\ref{n-eq1}) and (\ref{bbpp})], leading to 
$\mathcal{C}=2 \cdot \mbox{integer}-\iota_B^{\,\prime}$, and thus the final
condition
\beq
X_{\mathrm{
total}}= \mbox{integer}+\frac{b^{\,\prime}\cdot\iota_B^{\,\prime}}{2}\,.
\eeq

The first main difference between $B_p$ and $B_3$ is that the former
is an anomalous discrete gauge symmetry whereas the latter is anomaly
free \cite{{Ibanez:1991hv},{Ibanez:1991pr},{Dreiner:2005rd}}. The next
question we would like to address is what is the lowest dimensional
operator of MSSM chiral superfields which is allowed by $B_p$ but
forbidden by $B_3$. It is easy to see that for the renormalizable
interactions the two symmetries act identically. This equality
persists for operators which are the product of four superfields. For
a more systematic approach consider that, since $B_p$ is a $\mathbb{Z}
_2$-symmetry, Eq.~(\ref{bbpp}) can be recast as
\beq 
\sum_i n_{Q^i}~-~\sum_in_{\ol{U^i}}~ -~\sum_i
n_{\ol{D^i}}~=~2~\mathcal{I}_B^{\,\prime}+ ~\iota_B^{\,\prime}\,.
\eeq 
This is to be compared with Eq.~(\ref{n-eq1}), which leads to
\beq\label{er}
3\mathcal{C}~=~2~\mathcal{I}_B^{\,\prime}+ ~\iota_B^{\,\prime}\,.  
\eeq
Solving Eqs.~(\ref{n-eq1})-(\ref{n-eq3}) for $\sum_in_{Q^i}\,$, $\sum
_in_{\ol{D^i}}\,$, and $\sum_in_{\ol{U^i}}$ and plugging the result
into the second line of Eq.~(\ref{rrpp}) we get
\beq\label{zw}
3\mathcal{W}~+~3\sum_{i}n_{\ol{E^i}}~-~4\,\mathcal{C}~=~3
\mathcal{I}_B~+~\iota_B\,.
\eeq 
Eqs.~(\ref{er}) and (\ref{zw}) imply that $B_p$ is conserved if $
\mathcal{C}$ is an integer multiple of two, whereas $B_3$ is conserved
if $4\,\mathcal{C}$ is an integer multiple of three. Thus the smallest
value for which baryon parity is conserved and baryon triality is
violated is $|\mathcal{C}|=2$. This in turn implies that the relevant
operator contains at least six quark superfields [see
Eq.~(\ref{n-eq1})]. An example is
\beq
\varepsilon^{abc} \varepsilon^{def}\, \ol{{{
U^{i}}_{a}} {{D^{j}}_{b}} {{D^{k}}_{c}} {{
U^{\ell}}_{d}} {{D^{m}}_{e}} {{D^{n}}_{f}}}\,,
\eeq
where $a,\ldots,f$ are $SU(3)_C$ indices and $i,\ldots,n$ are
generation indices. Such a non-renormalizable operator is highly
suppressed so that the effective low-energy phenomenology is identical
for $B_p$ and $B_3$.





\cleqn

\section{\label{2by2}Supersymmetric Zeros in 
Mass Matrices}

\begin{figure}[t!] \epsfig{figure=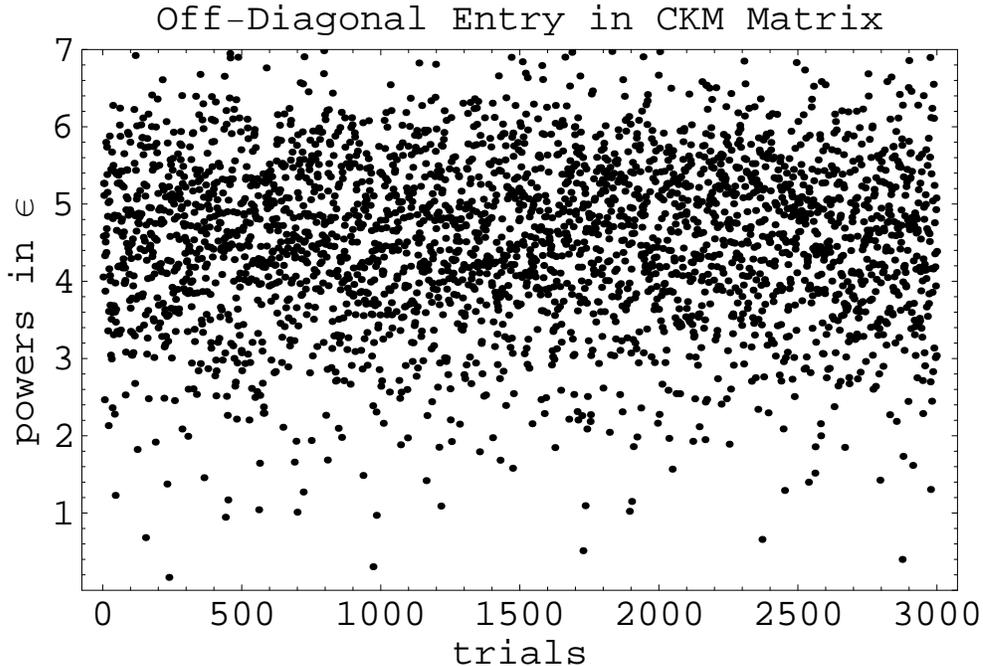,width=14.5cm} 
\caption{The powers in $\eps$ of the off-diagonal entry in the two generation 
CKM matrix for 3000 sets of randomly generated complex $\mathcal{O}(1)$ 
coefficients in the FN-generated Yukawa matrices and the kinetic K\"ahler 
potential terms.} \label{scatterCKM}
\end{figure}

\begin{figure}[t!] \epsfig{figure=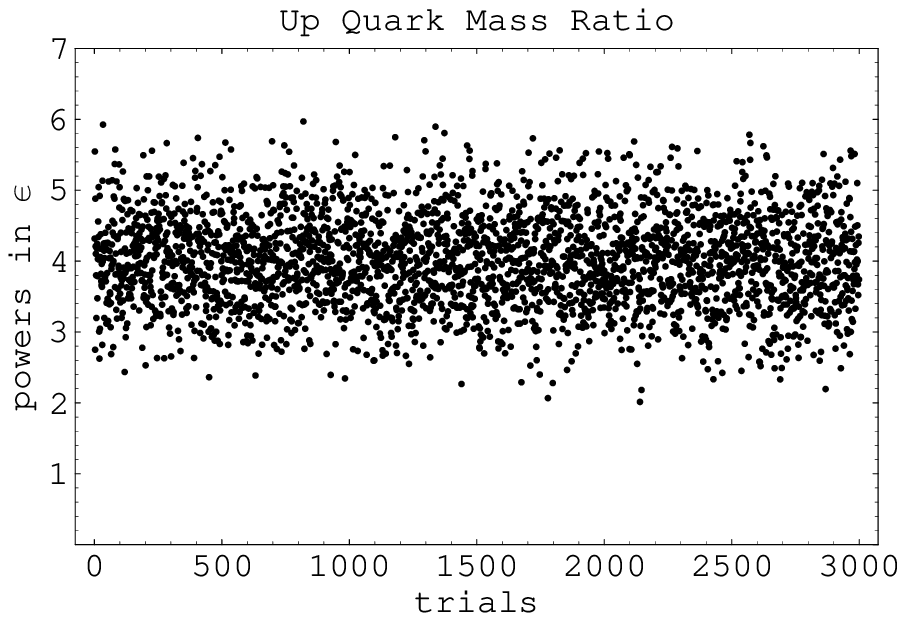,width=7.3cm} 
\epsfig{figure=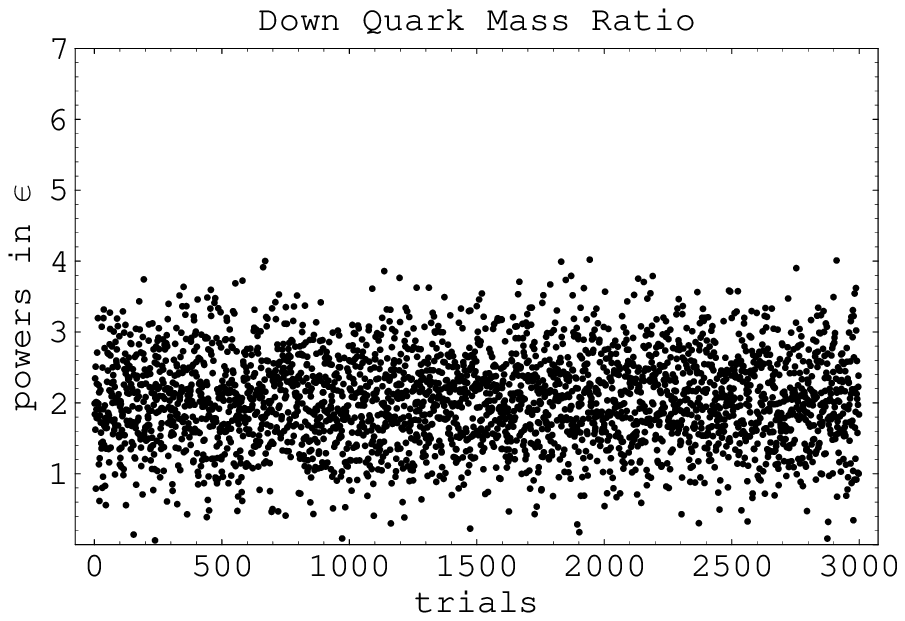,width=7.3cm} \caption{The powers in 
$\eps$ of the ratios $\frac{m_u}{m_c}$ and $\frac{m_d}{m_s}$.}  
\label{scattermass}
\end{figure}

The structure of the CKM matrix depends on the overall $X$-charges of
the operators in the quark mass matrices. Due to holomorphy of the
superpotential, terms with negative $X$-charge are forbidden. These
supersymmetric zeros are filled in by the canonicalization of the
K\"ahler potential \cite{Binetruy:1996xk}. Naively, the resulting mass
matrices might suggest a CKM matrix consistent with the experimentally
measured quark mixing.  However, if one allows for supersymmetric
zeros then things are more involved since the matrix canonicalizing
the kinetic terms of the quark doublet $Q$ affects both, the up- {\it
  and} the down-type quark mass matrices.  Diagonalizing these, we
therefore encounter cancellations in the CKM matrix (which is a
product of the two left-handed diagonalization matrices). These
cancellations spoil the naively expected nice results. Espinosa and
Ibarra \cite{{Espinosa:2004ya},{King:2004tx}} have investigated the
influence of supersymmetric zeros on the CKM matrix. In the following
we illustrate such a situation explicitly for \emph{two} {generations}
of quarks and the trilinear superpotential terms
\beq
W_3 = 
\left(\bsym{G^{(D)}_\mathrm{FN}}\right)_{ij} H^DQ^i \overline{D^j}+
\left(\bsym{G^{(U)}_\mathrm{FN}}\right)_{ij} H^UQ^i \overline{U^j}\,.
\eeq
Here the subscript ``$\mathrm{FN}$'' refers to the fact that the
K\"ahler potential is not canonicalized yet at this point. Now
consider an $X$-charge assignment with
\beqn 
X_{Q^2} \!&=&\!  X_{Q^1} + 1,\qquad X_{\ol{U^2}}\,=\,X_{\ol{U^1}}-5,
\qquad X_{\ol{D^2}}\,=\,X_{\ol{D^1}} - 3,\nonumber\\
X_{H^U} \!&=&\! 4  - X_{Q^1} - X_{\ol{U^1}},\qquad X_{H^D}
\,=\, 2  - X_{Q^1} - X_{\ol{D^1}}\,.
\eeqn
Then in the Froggatt-Nielsen scenario the Yukawa couplings come out to
be
\beqn
\bsym{G^{(U)}_\mathrm{FN}} \, \sim \, \begin{pmatrix}\eps^4 & 0 \\ 
\eps^5 & 1 \end{pmatrix} , & & \bsym{G^{(D)}_\mathrm{FN}} \, 
\sim \, \begin{pmatrix}\eps^2 & 0 \\ \eps^3 & 1 \end{pmatrix},
\eeqn
where the $(1,2)$-elements are supersymmetric zeros. The K\"ahler potential 
has to be canonicalized by matrices of the form
\beqn
{\bsym{C^{(Q)}}}^{-1}  \sim   \begin{pmatrix}1 & \eps \\ \eps & 1 
\end{pmatrix} , & {\bsym{C^{(\ol{U})}}}^{-1} \sim   
\begin{pmatrix} 1 & \eps^5 \\ \eps^5 & 1 \end{pmatrix} , & {
  \bsym{C^{(\ol{D})}}}^{-1} \sim   \begin{pmatrix} 1 & \eps^3 \\
  \eps^3 & 1 \end{pmatrix}\,.  
\eeqn 
These transformations change the Yukawa matrices to
\beqn
\bsym{G^{(U)}} \, \sim \, \begin{pmatrix}\eps^4 & \eps \\ \eps^5 & 1 
\end{pmatrix} ,
 & & \bsym{G^{(D)}} \, \sim \, \begin{pmatrix}\eps^2 & \eps \\ \eps^3 & 1 
\end{pmatrix}\,.
\eeqn
These are diagonalized by unitary matrices with the
textures\footnote{Note that most general unitary matrix with texture
  $\begin{pmatrix} 1 & \eps^a \\ \eps^a & 1\end{pmatrix}$ is
\beq
\bsym{U}=\begin{pmatrix} \xi & 0 \\ 0 & \tilde{\xi} \end{pmatrix} 
\cdot\begin{pmatrix} 1 &-\chi^\ast\eps^a\\\chi\eps^a&1\end{pmatrix}\,,\qquad 
\mathrm{with}\qquad  |\xi|^{-1} =|\tilde{\xi}|^ {-1}= \sqrt{1+|\chi|^2~ 
\eps^{2a}}\,, \nonumber
\eeq
and $a \in \mathbb{N}$. Applying this form to calculate the
off-diagonal elements of ${\bsym{U^{(U_L)}}}^\ast\!\cdot\! \bsym{G^{(U)}}
\!\cdot\!{\bsym{U^{(\ol{U})}}}^\dagger $ we readily find that $\bsym{G^
{(U)}}$ is diagonalized if $\chi$ is of $\mathcal{O}(1)$. The same holds 
for $\bsym{G^{(D)}}$. With our choice of $X$-charges the ratios of the 
quark masses are $\frac{m_u}{m_c} \sim \eps^4$ and $\frac{m_d}{m_s}\sim 
\eps^2$, respectively.}
\beqn
\bsym{U^{(U_L)}} \sim \begin{pmatrix} 1 & \eps \\ \eps & 1 
\end{pmatrix} , && \bsym{U^{(\ol{U})}} \sim \begin{pmatrix} 1 
& \eps^5 \\ \eps^5 & 1 \end{pmatrix} ,\nonumber\\
\bsym{U^{(D_L)}} \sim \begin{pmatrix} 1 & \eps \\ \eps & 1 
\end{pmatrix} , && \bsym{U^{(\ol{D})}} \sim \begin{pmatrix} 1 &
  \eps^3 \\ \eps^3 & 1 \end{pmatrix}.  
\eeqn 
If we naively neglect possible cancellations between
$\bsym{U^{(U_L)}}$ and $\bsym{U^{(D_L)}}$ we get
\beqn 
\bsym{U^\mathrm{CKM}} \equiv
\bsym{U^{(U_L)}}\cdot
{\bsym{U^{(D_L)}}}^\dagger & \sim & \begin{pmatrix} 1 & \eps \\
  \eps & 1 \end{pmatrix}\,, \label{naiveCKM} 
\eeqn 
which agrees well with the data. However, in order to show that this
line of reasoning is too simple, we numerically calculated the CKM
matrix for an ensemble of 3000 {\it
  Mathematica}$\,^\copyright\,$-randomly generated sets of complex
$\mathcal{O}(1)$ coefficients which remain undetermined by any FN
model and appear in both, the FN-generated Yukawa matrices and the kinetic
(Hermitian) K\"ahler potential terms.  Figure~\ref{scatterCKM} shows
the powers in $\eps=0.2$ of the off-diagonal element in the CKM
matrix. The corresponding quark mass ratios $\frac{m_u}{m_c}$ and
$\frac{m_d}{m_s}$ are depicted in Figure~\ref{scattermass}.

Obviously, with supersymmetric zeros the naive result of
Eq.~(\ref{naiveCKM}) is in gross disagreement with the numerically
calculated CKM matrix, where instead of $\eps$ we have $\eps^{4,5,\,
  \mbox{or}\;6}$. Cancellations between $\bsym{U^{(U_L)}}$ and $\bsym
{U ^{(D_L)}}$ render the off-diagonal entries from $\mathcal{O}(\eps)$
to $\mathcal{O}(\eps^5)$. However, the quark mass ratios come out
correct in the naive calculation, \textit{i.e.} $\frac{m_u}{m_c}\sim
\eps^4$ and $\frac{m_d}{m_s} \sim \eps^2$.





\cleqn

\section{\label{vevcond}The Structure of the Sneutrino VEVs}

In $\not\!\!\!M_p$ theories, 
the five neutral scalars $\tilde\nu_\alpha,\,h^U_0$
mix and we have the following minimization conditions for the scalar
potential \cite{Grossman:1998py}
\beqn \Big{(} {M_{H^U}^2} +
{\mu_\alpha^\ast} ~\mu_\alpha + \frac{g_W^2 + g_Y^2}{8}\:
(|\upsilon_u|^2 - |\upsilon_d|^2 ) \Big{)} \; \upsilon_u \; - \;
{b_\alpha^\ast} {\upsilon_\alpha^\ast} & \stackrel{!}{=} &
0\,,\label{min1} \\
\Big{(} [\bsym{M^2_{\widetilde{L}}}]_{\alpha \beta} +
{\mu_\alpha^\ast}~ \mu_\beta + \frac{g_W^2 + g_Y^2}{8}\:
(|\upsilon_d|^2 - |\upsilon_u|^2 ) \: \delta_{\alpha\beta} \Big{)} \;
\upsilon_\beta \; -{b_\alpha^\ast}  {\upsilon_u^\ast} \; &
\stackrel{!}{=} & 0\,,~~~~\label{min2} 
\eeqn 
with $|\upsilon_d| \equiv \sqrt{{\upsilon_\alpha^\ast}~\upsilon
  _\alpha}$. $M_{H^U}$ is the soft supersymmetry breaking mass for the
Higgs scalar $H^U$ and $b_\alpha$ is the soft supersymmetry breaking
bilinear mass parameter of the term $~b_\alpha \widetilde{L^\alpha}
H^U$. In an explicit $\not\!\!\!M_p$ mSUGRA model, the complete
scalar potential was numerically minimized in Ref.~\cite{
  Allanach:2003eb}. However within the context of our FN models, we
have no prediction for the soft supersymmetry breaking parameters. We
shall thus make the assumption that the hidden and observable sector
superpotentials separate and we get the alignment of the soft
supersymmetry breaking bilinear operator \cite{Allanach:2003eb}
\beqn 
b_\alpha & \propto & \mu_\alpha\,,
\label{sugra} 
\eeqn 
at the unification scale. In the context of our FN model, this implies
\beqn b_\alpha & \sim & B \; m_{3/2} \; \eps^{-
  X_{L^\alpha} - X_{H^U}}\,.
\label{softbilin}
\eeqn 
$B$ is a soft supersymmetry breaking mass parameter of $\mathcal{O}(m
_{3/2})$. The other crucial ingredient is the structure of the soft
supersymmetry breaking slepton mass squared $[\bsym{M^2_{\widetilde
{L}}]_{\alpha \beta}}\,$. For simplicity we take
\beqn 
[\bsym{M^2_{\widetilde{L}}}]_{\alpha \beta} & \sim & {m_{3/2}^2}
\; \eps^{| X_{L^\alpha} - X_{L^\beta}|}\,.\label{softscalar} 
\eeqn 
This might originate either directly from an FN structure of the
corresponding parent terms or via the CK transformation of a {\it
  diagonal} $[\bsym{M^2_{ \widetilde{L}}}]_{\alpha \beta}$ whose
eigenvalues are all of the same order but {\it not equal}. Soft
supersymmetry breaking parameters with the structure of
Eqs.~(\ref{softbilin}) and (\ref{softscalar}) have also been
considered in Ref.~\cite{Banks:1995by}.

Given Eqs.~(\ref{softbilin}) and (\ref{softscalar}), we can now solve
the minimization conditions Eqs.~(\ref{min1}) and (\ref{min2}) and
obtain an FN structure for the VEVs. To see this, we first simplify
Eq.~(\ref{min2}) by the observation that $\frac{g_W^2+g_Y^2}{8}
\:(|\upsilon_u|^2 - |\upsilon_d|^2 ) \: < \:\frac {1}{10}\,( 246\:
\mathrm{GeV} )^2$. On the other hand, the lower bound on the chargino
production cross section from LEP implies $\mu_0\ge100\,\mathrm{GeV}$
\cite{Abbiendi:2003sc}. In our case this translates into a lower bound
on $m_{3/2 }\,$. As we assume a GM/KN-generated $\mu_0\sim m_{3/2}\:
\eps^{-X_{L ^0}-X_{H^U}}\sim m_{ 3/2}\:\eps\,$, the lowest allowed 
value for $m_{ 3/2}$ is about $500\,\mathrm{GeV}$. Putting this
together we have
\beqn 
\frac{g_W^2 + g_Y^2}{8}\: (|\upsilon_u|^2 - |\upsilon_d|^2 )  & \ll &  
( 246 \: \mathrm{GeV} )^2  \: \lesssim \: m_{3/2}^2 \: \sim \: 
[\bsym{M^2_{\widetilde{L}}}]_{\alpha \alpha}  \:,
\eeqn
so that the cubic term of Eq.~(\ref{min2}) is negligible. Applying this 
approximation, we are left with a set of linear equations in the VEVs, 
which is solved with the {\it ansatz}
\beqn
\begin{pmatrix} \upsilon_u \\[3mm] \upsilon_\alpha \end{pmatrix}  & = &  
\begin{pmatrix} 
  N_u \; \frac{m_{3/2}}{B} \\[3mm] N_\alpha \; \eps^{\,- X_{L^\alpha} -
    X_{H^U}} \end{pmatrix} , 
\eeqn 
if the coefficients $N_u$ and $N_\alpha$ are of the same order. This
qualitative statement relies on the assumption commonly made in FN
models that the sum of several complex numbers with absolute value of
$\mathcal{O}(1)$ is again of $\mathcal{O}(1)$. The overall scale of
the VEVs is determined by the normalization requirement $\sqrt{|
  \upsilon_u|^2+|\upsilon_d|^2}=246\;\mathrm{GeV}$. Eq.~(\ref{min1})
does not constrain the sneutrino VEVs as the relevant terms are
negligibly small, {\it i.e.} $\frac{b_i ~\upsilon_i\phantom{|}}{b_0
  ~\upsilon_0}\sim \eps^{2\, (X_{L^0}-X_ {L^i})}$. Hence, we finally
end up with
\beqn \upsilon_\alpha & \propto & \eps^{-
  X_{L^\alpha}}.\label{vevstru} 
\eeqn 
The apparent alignment of $\upsilon_\alpha$ and $\mu_\alpha$ is with
respect to the power of $\eps$ and not exact. Both sets of $\mathcal{O
}(1)$ coefficients differ from each other due to the VEVs' dependence
on $[\bsym{M^2_{\widetilde{L}}}]_{\alpha\beta}$. Therefore, excluding
artificial exact alignment, the $\upsilon_i$ and $\mu_i$ ($i=1,2,3$)
cannot be rotated away simultaneously. This is important in order to
obtain a massive neutrino through mixing with the neutralinos.

Again, we have checked these results numerically. Except for a slight
tendency to have less $\eps$-suppression in the $\upsilon_i$ ($i=1,2,
3$) we found agreement with Eq.~(\ref{vevstru}). The systematic effect
of bigger $\upsilon_i$ is caused by the two-dimensional random walk.
Changing to a basis without sneutrino VEVs, this feature passes on to
other coupling constants with the generation structure $\eps^{ -
  X_{L^\alpha}}$. We take account of this by preferring higher
$\mathcal{O}(1)$ coefficients for those coupling constants which are
proportional to $\eps^{- X_{L^i}}$, namely $\mu_i$,
$\lambda^\prime_{ijk}$, and $\lambda_{ijk}\,$. Coupling constants
which -- after the canonicalization of the K\"ahler potential -- have
the structure $\eps^{ + X_{L^\alpha}}$ are affected differently by the
$\bsym{U^{\mathrm{VEVs}}}$ transformation. Their $\alpha=0$ component
gets somewhat enlarged. Thus the $\lambda_{0jk}$ remain unchanged.





\cleqn

\section{\label{symconsid}Symmetries in the 
$\bsym{\lambda}$-Loop Contribution to the 
Neutrino Mass Matrix }

The contribution of the charged lepton-slepton loop to the neutrino
mass matrix is given in Eq.~(\ref{lamloop}). In our model, the $\not\!
\!M_p$ parameters $\lambda_{ik n}$ are generated out of the charged
lepton mass matrix via the canonicalization of the K\"ahler potential.
This mechanism leads to Eq.~(\ref{rpvbyCKlept}), which can be written
as
\beqn 
\lambda_{ikn}&=&c_i\,\lambda_{0kn}-(i\leftrightarrow k)\,, 
\eeqn
with $c_i$ being some coefficient. Remember that we are working in the
charged lepton mass eigenstate basis, \ie\ $\lambda_{0kn}=\delta_{kn}
\,\lambda_{0kn}$. Using this structure of $\lambda_{ikn}$ we can
calculate the first term of the mass matrix $\bsym{M_{\,\lambda\,
    \mbox {-}\mathrm{loop}}^{( \nu)}}$ in Eq.~(\ref{lamloop})
symbolically
\beqn
\sum_{k,n} \lambda_{ikn} \, \lambda_{jn k} \, F^{(1)}_k
F^{(2)}_n &  =  & 
c_i \, c_j \, \Big({\lambda_{011}^2} \, F^{(1)}_1  F^{(2)}_1   + 
{\lambda_{022}^2} \,F^{(1)}_2  F^{(2)}_2 \nonumber \\
&&  \;\;+ \;{\lambda_{033}^2} \, F^{(1)}_3  F^{(2)}_3 + \lambda_{0ii} \, 
\lambda_{0jj} \, F^{(1)}_j  F^{(2)}_i \phantom{\sum_k}\nonumber \\
&&\;\;-\;{\lambda_{0ii}^2} \, F^{(1)}_i  F^{(2)}_i- {\lambda_{0jj}^2} 
\, F^{(1)}_j  F^{(2)}_j  \Big).
\label{leadcanc}
\eeqn
Here $F^{(1,2)}_k$ are functions of the charged lepton masses as
well as the charged slepton masses and mixing angles. For the purposes
of this appendix it suffices to know the ratios $F^{(1)}_k/F^{(1)}
_3$ and $F^{(2)}_k / F^{(2)}_3$.  Eq.~(\ref{lamloop}) together with
the simplifications made below yields
\beq
\frac{F^{(1)}_k}{F^{(1)}_3}~~=~~\frac{m^{(e)}_k}{m^{(e)}_3}~~
\sim~~\frac{F^{(2)}_k}{F^{(2)}_3}\,.
\eeq
Depending on $i$ and $j\,$, we can encounter exact cancellations of
seemingly dominating terms in Eq.~(\ref{leadcanc}). Applying the FN
structure of the charged lepton masses, $m_e : m_\mu :m_\tau \sim
\,\eps^{4+z} : \eps^{2} : 1$, and keeping only the leading (non-zero)
terms, we obtain
\beqn
\sum_{kn} \lambda_{ikn} \, \lambda_{jn k} \, F^{(1)}_k
F^{(2)}_n &  =  & c_i \, c_j \: {\lambda_{033}^2} \, F^{(1)}_3 
F^{(2)}_3  \, f_{ij}~, \label{lamloopF}
\eeqn
with $f_{ij} \sim 1$ for $i,j=1,2$, $f_{23} \sim f_{32} \sim \eps^4$
and $f_{13} \sim f_{31} \sim f_{33} \sim \eps^8$.  Adding the second
term of Eq.~(\ref{lamloop}) symmetrizes the mass matrix
$\bsym{M_{\lambda \mbox{-} \mathrm{loop}}^{(\nu)}}$. As our result in
Eq.~(\ref{lamloopF}) is already symmetric in $i$ and $j$ concerning
the magnitudes, we simply get a factor of two.

This shows that due to the $\lambda_{ikn}$'s direct dependence on
the charged lepton mass matrix and its antisymmetry under interchange
of the first two indices, the $(i,3)$- and the $(3,i)$-elements
($i=1,2, 3$) of $\bsym{M_{\lambda \mbox{-} \mathrm{loop}}^{(\nu)}}$
are highly suppressed.




\cleqn

\section{\label{diagonalizeM} Diagonalization of 
the Neutrino Mass Matrix}

The effective neutrino mass matrix $\bsym{M^{(\nu)}}$ is diagonalized
by the unitary matrix $\bsym{U^{(\nu)}}$ defined in
Eq.~(\ref{nudiag}).  In Sect.~\ref{nuMatrix}, we have seen that the
mass matrix has an $\eps$-structure
\beqn 
{M^{(\nu)}}_{ij}&\propto & \eps^{-X_{L_i}-X_{L_j}}\,.  
\eeqn 
Since $X_{L_1} \ge X_{L_2} \ge X_{L_3}$, we wish to find the unitary
matrix $\bsym{{U}^{(\nu^{\,\prime})}}$ such that $\bsym{M^{(\nu^{\,\prime})}_{
    \mathrm{diag}}}=\bsym{\mathrm{diag}}(m_1^\prime,m_2^\prime,
m_3^\prime)$ with the mass ordering $m_1^\prime \gtrsim m_2^\prime
\gtrsim m_3^\prime\,$. It is given as \cite{Hall:1993ni}
\beqn
{{U}^{(\nu^{\,\prime})}}_{ij} & \sim & \eps^{|X_{L_i}-X_{L_j}|},
\label{nutransfstruc}
\eeqn
as in the case of the down-type fermions, see Eq.~(\ref{massbasisQL}).
Unfortunately the two main contributions to the neutrino mass matrix,
Eqs.~(\ref{tree}) and (\ref{lamploop}), are both matrices of rank one.
They thus show an additional symmetry, which obscures the validity of
Eq.~(\ref{nutransfstruc}) in the model we consider. We wish to examine
this problem in more detail in this appendix. We focus on the
case with only the tree level and the quark-squark loop contributing
to the neutrino mass matrix, \ie\ with two independent contributions
leading to two massive neutrinos. The mass matrix can then be written
in the form
\beqn
{M^{(\nu)}}_{ij} &=& A~ a_i~ a_j + B~ b_i~ b_j,\label{genstructure}
\eeqn
where the upper case letters define the overall mass scale of each
term and the lower case letters give the generation structure $a_i\sim
b_i \sim \eps^{X_{L^1}-X_{L^i}}$. $a_i,\,b_i$ include in general
different factors of order 1. Since $X_{L^1}\ge X_{L^2}\ge X_{L^3} $,
we have $|a_1| \gtrsim |a_2| \gtrsim |a_3|$ and $|b_1|\gtrsim|b_2|
\gtrsim |b_3|$. In addition we take $A \gtrsim B$.\footnote{We do not
  specify which term is tree and which is loop level in order to stay
  general as long as possible.  So the treatment in this appendix is
  valid also for $\frac{m^\mathrm{tree}}{m^\mathrm{loop}} \lesssim
  1$.}

Notice that there are six degrees of freedom in
Eq.~(\ref{genstructure}). So at first glance 
Eq.~(\ref{nutransfstruc}) seems applicable. However, the mass
matrix $\bsym{M^{(\nu)}}$ in Eq.~(\ref{genstructure}) exhibits an
additional symmetry. It is of rank two, thus leaving one neutrino
massless. Therefore it is not obvious at all that
Eq.~(\ref{nutransfstruc}) correctly describes the structure of the
diagonalization matrix. We need to have a closer look at
Eq.~(\ref{genstructure}). In analogy to Eq.~(\ref{vevrot}), we perform
a unitary transformation which rotates away $a_2$ and $a_3$.  Thus
with
\beqn
\bsym{V}^\ast &\sim&  \begin{pmatrix} 1 & \eps^{X_{L^1}-X_{L^2}} & 
\eps^{X_{L^1}-X_{L^3}} \\
  \eps^{X_{L^1}-X_{L^2}} & 1 & \eps^{2X_{L^1}-X_{L^2}-X_{L^3}} \\
  \eps^{X_{L^1}-X_{L^3}} & \eps^{2X_{L^1}-X_{L^2}-X_{L^3}} & 1
\end{pmatrix}, 
\eeqn 
and $V^\ast_{ij} \,a_j \equiv a^\prime_i =\delta_{1i}\,a^\prime_i$ 
and $V^\ast_{ij} \, b_j \equiv b^{\,\prime}_i \sim \eps^{X_{L^1}-
X_{L^i}}$ we have 
\beqn
\bsym{M^{(\nu)}}^\prime \equiv \bsym{V}^\ast \cdot \bsym{M^{(\nu)}}
\cdot \bsym{V}^\dagger
=  B \begin{pmatrix} {\frac{A}{B}~a^\prime_1}^2  + {b^{\,\prime}_1}^2 &  
b^{\,\prime}_1 b^{\,\prime}_2 &  b^{\,\prime}_1 b^{\,\prime}_3 \\
  b^{\,\prime}_2 b^{\,\prime}_1 & b^{\,\prime}_2 b^{\,\prime}_2 &
  b^{\,\prime}_2
  b^{\,\prime}_3 \\
  b^{\,\prime}_3 b^{\,\prime}_1 & b^{\,\prime}_3  b^{\,\prime}_2 &
  b^{\,\prime}_3 b^{\,\prime}_3 \end{pmatrix}.  
\eeqn 
In the next step, we want to find the unitary matrix $\bsym{W}$ which
finally diagonalizes $\bsym{M^{(\nu)}}^\prime$, with the ordered mass
singular values. For this we consider
\beqn 
{{M^{(\nu)}}^\prime}_{ij}\: W^\dagger_{jk}
&=& W^T_{ik} \:m_k^{\,\prime}\,,\label{laststep} 
\eeqn 
where $m_1^{\,\prime} \ge m_2^{\,\prime}$ and $m_3^{\,\prime} = 0$. 
For $k=3$ we find that
\beqn
W^\dagger_{j3} &\sim & \frac{1}{b^{\,\prime}_2} \begin{pmatrix} 0 
\\ - b^{\,\prime}_3 \phantom{-}\\
  b^{\,\prime}_2 \end{pmatrix}_{\!\!\!j} \; \sim \; \begin{pmatrix} 0 \\
  \eps^{X_{L^2}-X_{L^3}} \\ 1 \end{pmatrix}_{\!\!\!j} 
\eeqn 
satisfies Eq.~(\ref{laststep}). Now consider $k=1$. For $i=1,2,3$ we
get the following order of magnitude relations
\beqn 
\left[ \frac{A}{B} \,\mathcal{O}(1) + \mathcal{O}( 1)
  \! \right] \! W^\dagger_{11} + \mathcal{O}(\eps^{X_{L^1}-X_{L^2}})
W^\dagger_{21} + \mathcal{O}(\eps^{X_{L^1}-X_{L^3}})
W^\dagger_{31} &=&
\frac{m_1^{\,\prime}}{B} W^T_{11},  \label{W1} 
\\
\mathcal{O}(\eps^{X_{L^1}-X_{L^2}}) W^\dagger_{11} +
\mathcal{O}(\eps^{2(X_{L^1}-X_{L^2})}) W^\dagger_{21} +
\mathcal{O}(\eps^{2 X_{L^1}-X_{L^2}-X_{L^3}}) W^\dagger_{31} &=&
\frac{m_1^{\,\prime}}{B} W^T_{21}, \label{W2}  \\
\mathcal{O}(\eps^{X_{L^1}-X_{L^3}}) W^\dagger_{11} +
\mathcal{O}(\eps^{2 X_{L^1}-X_{L^2}-X_{L^3}}) W^\dagger_{21} +
\mathcal{O}(\eps^{2(X_{L^1}-X_{L^3})}) W^\dagger_{31} &=&
\frac{m_1^{\,\prime}}{B} W^T_{31}\,.~~~~~\label{W3} 
\eeqn
Assuming no accidental cancellations among $\mathcal{O}(1)$
coefficients and keeping only leading terms\footnote{Notice that
  $\frac{m_1^{\,\prime}}{B} \ge 1$. Hence, considering for example the
  last equation, we can neglect the third term of the LHS compared to
  the RHS.}, we can determine the magnitude of $W^\dagger_{21}$ and
$W^\dagger_{31}$ from Eqs.~(\ref{W2}), (\ref{W3}):
\beqn 
W^\dagger_{21}
\sim \frac{B}{m_1^{\,\prime}}\: \eps^{X_{L^1}-X_{L^2}} \: W^T_{11}\,,
&\;\;\;\mathrm{and}\;\;\;& W^\dagger_{31} \sim
\frac{B}{m_1^{\,\prime}}\: \eps^{X_{L^1}-X_{L^3}} \: W^T_{11}.
\eeqn 
Eq.~(\ref{W1}) does not contain any information on the $\eps
$-structure of $W^\dagger_{j1}$. It simply states that $m_1^{\,
  \prime}$ is of $\mathcal{O}(A)$, the scale of the leading
contribution to the neutrino mass matrix. By means of normalization
arguments we conclude that
\beqn
W^\dagger_{j1}  &\sim & \begin{pmatrix} \phantom{\frac{B}{m_|}} 1 \\
  \frac{B^{\phantom{|}}}{{A}_{\phantom{|}}}\: \eps^{X_{L^1}-X_{L^2}} \\
  \frac{B^{\phantom{|}}}{A}\: \eps^{X_{L^1}-X_{L^3}}
\end{pmatrix}_{\!\!\!j}.  
\eeqn 
The remaining (second) column of the unitary matrix $\bsym{W}^\dagger$ is
obtained by finding the normalized vector which is orthogonal to
$W^\dagger_{j3}$ and $W^\dagger_{j1}$.  We get
\beqn
W^\dagger_{j2}  &\sim & \begin{pmatrix}    \frac{B}{A}\: 
\eps^{X_{L^1}-X_{L^2}} \\
 \phantom{\frac{B^|}{m_|}}\!\!\!\!\!\! 1 \\ \phantom{\frac{B^|}{A}} 
\!\!\!\!\! \eps^{X_{L^2}-X_{L^3}} 
\end{pmatrix}_{\!\!\!j}.  
\eeqn 
Inserting this into Eq.~(\ref{laststep}), we find the magnitude of the
second neutrino mass 
\beqn 
m_2^{\,\prime} &\sim& B \cdot \eps^{2(X_{L^1}-X_{L^2})}\,.
\label{frakm_2} 
\eeqn 
Notice that $m_2^{\,\prime}$ is not simply given by the scale $B$ of
the second largest contribution to the neutrino mass matrix but it is
additionally suppressed by a factor of $\eps^{2 (X_{L^1}-X_{L^2})}$.
This is of course only relevant if $X_{L^1} > X_{L^2}$.

We finally arrive at the diagonalization matrix $\bsym{{U}^
  {(\nu^{\,\prime})}}$ with the eigenvalues in the order $m_1^{\,\prime} \gtrsim
m_2^{\,\prime} >m_3^{\,\prime} = 0$:
\beqn 
{{U}^{(\nu^{\,\prime})}}_{ij} & \equiv & W_{ik} \: V_{kj} \; \sim
\; \eps^{|X_{L^i}-X_{L^j}|}\,.  
\eeqn 
The dependence on the factor $\frac{B}{A}$ which appears in $\bsym{W}$
drops out in leading order.



\cleqn

\section{\label{top-down}A Top-down Example}

Here we consider the $X$-charge assignment of Table~\ref{allsets} with
$\Delta^L_{31} = -1$, $3\zeta + b = -2$, and $y=1$. This is our
preferred scenario since the resulting $X$-charges are all multiples
of one third, \ie$\,$ they are not highly fractional.  With this
choice we obtain the following Yukawa matrices for the superpotential
terms $H^D Q^i \ol{D^j}$, $H^D L^i\ol{E^j}$, and $H^U Q^i \ol{U^j}$
without supersymmetric zeros:
\beq
\bsym{G^{(D)}_{\mathrm{FN}}} \sim \begin{pmatrix} \eps^4 & \eps^4 &
  \eps^4 \\ \eps^2 & \eps^2 & \eps^2 \\ 1~ & 1~ & 1~
\end{pmatrix}\,, ~~~~\bsym{G^{(E)}_{\mathrm{FN}}} \sim
\begin{pmatrix} \eps^5 & \eps^2 & \eps \\ \eps^5 & \eps^2 & \eps \\
  \eps^4 & \eps & 1 \end{pmatrix}\,,
~~~~\bsym{G^{(U)}_{\mathrm{FN}}} \sim \begin{pmatrix} \eps^8 &
  \eps^6 & \eps^4 \\ \eps^6 & \eps^4 & \eps^2 \\ \eps^4 & \eps^2 & 1~
\end{pmatrix}\,.~~~~ 
\eeq
The trilinear $\rpv$ terms $L^i Q^j \ol{D^k}$ and $L^i L^j \ol{E^k}$
are disallowed due to negative integer overall $X$-charges: $X_{L^1} +
X_{Q^1} + X_{\ol{D^1}} = -2$ and $X_{L^1} + X_{L^2} + X_{\ol{E^1}} =
-1$, respectively. For higher generational indices we obtain even
smaller total $X$-charges. Analogously, we have for the bilinear terms
$L^\alpha H^U$: $X_{L^0} + X_{H^U} = -1$ (corresponding to the
statement $z=1$) and even smaller for the three lepton doublets $L^i$
due to $X_{L^i} < X_{L^0}$. The $B_3$ violating terms $\ol{U^i D^j
  D^k} $ are forbidden by non-integer overall $X$-charge,
$X_{\ol{U^1}} + X_{\ol{D^1}} + X_{\ol{D^2}} = - \frac{8}{3}$.

The GM/KN mechanism, however, reintroduces the terms
disallowed by {\it negative integer} overall $X$-charge in the
effective superpotential. Thus we get
\beq
\frac{m_{3/2}}{M_{\mathrm{grav}}} \: \eps^{- (X_{L^i} + X_{Q^j} + 
X_{\ol{D^k}})} ~ L^i Q^j \ol{D^k},~~~~~~~ 
\frac{m_{3/2}}{M_{\mathrm{grav}}} \: \eps^{- (X_{L^i} + X_{L^j} + 
X_{\ol{E^k}})} ~ L^i L^j \ol{E^k}, 
\eeq
and
\beq
{\mu_\mathrm{FN}}_{\,\alpha} ~ L^\alpha H^U,  ~~~~ 
\mathrm{with}~~~~  {\mu_\mathrm{FN}}_{\,\alpha} \sim \, 
m_{3/2} ~ \eps^{- (X_{L^\alpha} + X_{H^U})}.
\eeq
Since $\frac{m_{3/2}}{M_{\mathrm{grav}}} = \mathcal{O}(\frac{10^3
  \,\mathrm{GeV}}{10^{18}\,\mathrm{GeV}}) = \mathcal{O}(10^{-15})$ the
GM/KN-generated trilinear terms are negligibly small. In contrast, the
bilinear terms including the MSSM $\mu$-term are phenomenologically
the correct order of magnitude. Of course, care has to be taken for
sufficient $\eps$-suppression of the $\rpv$ bilinears. In the scenario
considered here, we have
\beqn
{\mu_\mathrm{FN}}_{\, \alpha} &\sim & m_{3/2} ~ \begin{pmatrix} \eps~ 
\\ \eps^7 \\ \eps^7 \\ \eps^8  \end{pmatrix}_{\!\!\!\alpha}.
\eeqn

Next, we canonicalize the K\"ahler potential. The only CK
transformation that changes the $\eps$-structure of the above coupling
constants is the one connected to the superfields $L^\alpha$ [thus
\emph{e.g} $\bsym{G^{(U,D)}}\sim\bsym{G^{(U,D)}_{\mathrm
    {FN}}}$ concerning the CK transformations of $Q^i,\ol{U^i},\ol
{D^i}, H^U, H^D$]. The corresponding transformation
matrix takes the form [see Eq.~(\ref{defofckmatrix})]
\beqn
\bsym{C^{(L)}} &\sim& 
\begin{pmatrix} 
1 & \eps^6 & \eps^6 & \eps^7 \\
\eps^6 & 1 & 1 & \eps  \\ 
\eps^6 & 1 & 1 & \eps \\ 
\eps^7 & \eps & \eps & 1   
\end{pmatrix}.
\eeqn
The $\rpv$ coupling constants $\lambda^\prime_{i j k}$ of the
trilinear terms $L^i Q^j \ol{D^k}$ are now generated from 
${\lambda^\prime_{\mathrm{FN}}}_{\, 0 j k} \equiv {G^{(D)}_{
\mathrm{FN}}}_{\,jk}$ as shown in Eq.~(\ref{CKlqd}):
\beqn
\lambda^\prime_{\alpha j k} & = & [{\bsym{C^{(L)}}}^{-1}]_{0 \alpha}~ 
{G^{(D)}_{\mathrm{FN}}}_{\,jk} 
~~\sim~~ \begin{pmatrix}  1 \\ \eps^6 \\ \eps^6 \\   \eps^7      
\end{pmatrix}_{\!\!\!\!\alpha} {G^{(D)}_{\mathrm{FN}}}_{\,jk}\,.
\eeqn
Likewise, the $\rpv$ coupling constants ${\lambda}_{i j k}$ of the
trilinear terms $L^i L^j \ol{E^k}$ are generated from
$\bsym{G^{(E)}_{\mathrm{FN}}}$. An additional antisymmetrizing
term accounts for the antisymmetry of ${\lambda}_{i j k}$ in the first
two indices, see Eq.~(\ref{CKlle}).  The $\eps$-structure of the
bilinear coupling constants $\mu_\alpha$ is not affected by the
CK-transformation:
\beq 
\mu_\alpha ~ = ~ [{\bsym{C^{(L)}}}^{-1}]_{\beta \alpha}
~{\mu_\mathrm{FN}}_{\,\beta} ~~ \sim ~~ m_{3/2}
\begin{pmatrix} 1 & \eps^6 & \eps^6 & \eps^7   \\    \eps^6 & 1 & 1 & \eps \\ 
\eps^6 & 1 & 1 & \eps 
\\ \eps^7 & \eps & \eps & 1  
\end{pmatrix}_{\!\!\!\!\beta \alpha} \!\!\!\! \cdot \!  
\begin{pmatrix}  \eps \\ \eps^7 \\ \eps^7 \\   \eps^8      
\end{pmatrix}_{\!\!\!\!\beta} 
~~ \sim ~~  m_{3/2} \begin{pmatrix}\eps \\ \eps^7 \\ \eps^7 \\ \eps^8
\end{pmatrix}_{\!\!\!\!\alpha}.
\eeq

Neglecting renormalization flow effects, we now rotate away the
sneutrino VEVs \cite{Grossman:1998py}. To leading order in $\eps$ the
necessary unitary transformation is given in Eq.~(\ref{vevrot}). For
our $X$-charge assignment it reads 
\beqn 
\bsym{U^{\mathrm{VEVs}}} & \sim &
\begin{pmatrix} 1 & \eps^6 & \eps^6 &\eps^7 \\ \eps^6 & 1 & \eps^{12}
  & \eps^{13} \\ \eps^6 & \eps^{12} & 1 & \eps^{13}\\ \eps^7 &
  \eps^{13} & \eps^{13} & 1 \end{pmatrix}\,. 
\eeqn 
It is easy to see that this transformation does not change the
coupling constants $\lambda^\prime_{\alpha j k}$, ${\lambda}_{\alpha
  \beta k}$, and $\mu_\alpha$ in their flavor structure.

Having generated the above $\rpv$ couplings via the GM/KN mechanism and
the subsequent CK transformation, it is possible to have neutrino
masses without introducing right-handed neutrinos.
Assuming\footnote{Equally, we could have taken
  $\frac{m^{\mathrm{tree}}}{m^{\mathrm{loop}}} \lesssim 1$ leading to
  the same flavor structure of $\bsym{M^{(\nu)}}$. However, the
  prefactor of the neutrino mass matrix would then originate from
  Eq.~(\ref{lamploop}).}  $\frac{m^{\mathrm{tree}}}{m^{\mathrm{loop}}}
\gtrsim 1$, we obtain an effective Majorana neutrino mass matrix of
the structure [\emph{cf.}  Eq.~(\ref{tree}) for $x=0$ which
corresponds to $\tan{\beta} \gtrsim 40$, thus $\cos^2{\!\beta} \approx
\frac{1}{\tan^2{\!\beta}} \sim \frac{{m_b^2}}{{m_t^2}}$ and
$\sin{2\beta} = 2\, \sin{\beta} \cos{\beta} \approx
\frac{2}{\tan{\beta }} \ll 1$]
\beqn
\bsym{M^{(\nu)}} &\sim& \frac{{m_Z^2}~ {m_b^2}}{{m_t^2}} \cdot
\frac{M_{\tilde{\gamma}}}{M_1 ~M_2} \cdot \eps^{12} \cdot
\begin{pmatrix} 1 & 1 & \eps \\ 1 & 1 & \eps \\ \eps & \eps & \eps^2
\end{pmatrix}.\nonumber
\eeqn 
Differentiating between hierarchical and inverse-hierarchical neutrino
scenarios ($\!\!\!\,\,${\it{cf.}}$\!$ Sect.$\:$\ref{mixingconstraints}) 
we arrive at an MNS
mixing matrix with either
\beqn 
\bsym{U_{\mathrm{(h.)}}^{\mathrm{MNS}}} 
= \bsym{U^{(\nu^{\,\prime})}}^\dagger \cdot \bsym{T^{\mathrm{(h.)}}} 
\sim \begin{pmatrix} 1&1&\eps \\ 1&1&\eps \\ \eps  &\eps & 1 \end{pmatrix}
\cdot \bsym{T^{\mathrm{(h.)}}}  
\sim \begin{pmatrix} \eps&1&1 \\ \eps&1&1 \\ 1 &\eps & \eps \end{pmatrix},
\eeqn
or
\beqn
\bsym{U_{\mathrm{(i.h.)}}^{\mathrm{MNS}}} 
= \bsym{U^{(\nu^{\,\prime})}}^\dagger \cdot \bsym{T^{\mathrm{(i.h.)}}} 
\sim \begin{pmatrix} 1&1&\eps \\ 1&1&\eps \\ \eps  &\eps & 1 \end{pmatrix}
\cdot \bsym{T^{\mathrm{(i.h.)}}}  
\sim \begin{pmatrix} 1&1&\eps \\ 1&1&\eps \\ \eps & \eps &1\end{pmatrix}.
\eeqn
Both scenarios are compatible with Eq.~(\ref{MNSexpstr}). However, due
to the smallness of the $(1,3)$-element in the experimentally measured
MNS matrix, leptonic mixing would suggest an inverse hierarchy.  Then,
consistency with the neutrino mass single values (see
Sect.~\ref{massconstraints}) would require two masses of similar
magnitude, thus $\frac{m^{\mathrm{tree}}}{m^{\mathrm{loop}}} \sim
\mathcal{O}(1)$.

As for the CKM matrix we refer to Ref.~\cite{Dreiner:2003yr} and
state the result for the sake of completeness: 
\beqn
\bsym{U^{\mathrm{CKM}}} &\sim & \begin{pmatrix} 1 & \eps^2 & \eps^4 \\
  \eps^2 & 1 & \eps^2 \\ \eps^4 & \eps^2 & 1
 \end{pmatrix}.
\eeqn
The price we have to pay for nice, \ie$\,$ not too fractional,
$X$-charges is a not-so-nice CKM matrix [{\it e.g.} the (1,2)-element
is $\mathcal{O}(\eps^2)$ and not $\mathcal{O}(\eps)$].

\end{appendix}
\bibliographystyle{hunsrt}
\bibliography{references}
\end{document}